\documentclass[manuscript,screen]{acmart}
\AtBeginDocument{%
  }

\setcopyright{acmlicensed}
\copyrightyear{2025}
\acmYear{2025}
\acmDOI{XXXXXXX.XXXXXXX}
\acmConference[Conference acronym 'XX]{Make sure to enter the correct
  conference title from your rights confirmation email}{June 03--05,
  2025}{Woodstock, NY}
\acmISBN{978-1-4503-XXXX-X/2018/06}

\usepackage{longtable}

\newcommand\revision[1]{\textcolor{black}{#1}}

\begin{document}

\title{FATe of Bots: Ethical Considerations of Social Bot Detection}


\author{Lynnette Hui Xian Ng}
\affiliation{%
  \institution{Carnegie Mellon University}
  \city{Pittsburgh}
  \state{Pennsylvania}
  \country{USA}}
\email{lynnetteng@cmu.edu}

\author{Ethan Pan}
\affiliation{%
  \institution{University of Wisconsin-Madison}
  \city{Madison}
  \state{Wisconsin}
  \country{USA}}
\email{pyy122759996@gmail.com}

\author{Michael Miller Yoder}
\affiliation{%
  \institution{University of Pittsburgh}
  \city{Pittsburgh}
  \state{Pennsylvania}
  \country{USA}}
\email{mmyoder@pitt.edu}

\author{Kathleen M. Carley}
\affiliation{%
  \institution{Carnegie Mellon University}
  \city{Pittsburgh}
  \state{Pennsylvania}
  \country{USA}}
\email{kathleen.carley@cs.cmu.edu}

\renewcommand{\shortauthors}{Ng et al.}

\begin{abstract}
A growing suite of research illustrates the negative impact of social media bots in amplifying harmful information with widespread social implications. Social bot detection algorithms have been developed to help identify these bot agents efficiently. While such algorithms can help mitigate the harmful effects of social media bots, they operate within complex socio-technical systems that include users and organizations. As such, ethical considerations are key while developing and deploying these bot detection algorithms, especially at scales as massive as social media ecosystems. In this article, we examine the ethical implications for social bot detection systems through three pillars: training datasets, algorithm development, and the use of bot agents. We do so by surveying the training datasets of existing bot detection algorithms, evaluating existing bot detection datasets, and drawing on discussions of user experiences of people being detected as bots. This examination is grounded in the FATe framework, which examines Fairness, Accountability, and Transparency in consideration of tech ethics. We then elaborate on the challenges that researchers face in addressing ethical issues with bot detection and provide recommendations for research directions.
We aim for this preliminary discussion to inspire more responsible and equitable approaches towards improving the social media bot detection landscape.
\end{abstract}

 

\received{20 February 2007}
\received[revised]{12 March 2009}
\received[accepted]{5 June 2009}

\maketitle

\section{Introduction}
Social media platforms (e.g., Facebook, Instagram, Reddit, X) were built to connect people to each other and to information. In fact, a November 2023 Pew Research Center survey revealed that half of the American population consumes their news from social media sites \cite{walker_matsa_2023}. This statistic highlights how social media platforms shape information consumption and social interaction patterns \cite{olteanu2021facts}. As these platforms increasingly mediate how people access and engage with information, the processes of content creation and dissemination have also become progressively automated \cite{woolley2016automation}.

Information spread on social media platforms is not entirely organic. Information content generation and dissemination on social media platforms have been increasingly subjected to automated influences \cite{ng2025global}. Social media bots -- partially or fully automated accounts -- play a significant role in the inorganic information pipeline. They artificially amplify specific narratives, manipulate engagement metrics and distort perceptions of popularity or consensus by liking, retweeting and reposting content at scale \cite{ferrara_rise_2016,subrahmanian_darpa_2016}. In some cases, hybrid human-bot accounts called cyborgs further blur authenticity by combining automated posting with human oversight, making detection and accountability more complex \cite{chu_detecting_2012,ng2024cyborgs}.

Many studies illustrate the negative impact of bots in amplifying harmful information and reshaping society  \cite{ng_assembling_2024,bessi2016social,schuchard2021insights}.  For example, bots are believed to have impacted electoral processes in the United States and in Europe through spreading misinformed claims \cite{chang2021social,likarchuk2023manipulation}, and the collective social pressures of bots can flip opinions in contentious political or health issues \cite{macy2003polarization,lu2024agents}. In extreme cases, bots can incite offline violence, which can lead to protests and riots \cite{chen2022election2020,sharma2022characterizing,singh2020multidimensional}. 

Bot detection algorithms are used to identify bot agents in large-scale settings. The detection and analysis of social media bots falls within the field of social cybersecurity, which aims to protect our social cyber health \cite{carley2020social}. This involves detecting malicious bots that can erode public confidence and polarize societies, and also deploy prosocial bots to safeguard social trust and community stability \cite{starbird2019disinformation,ng2025dual}. \revision{To do so requires incorporating behavioral and social considerations about cyber actors (i.e., artificial intelligence-driven social media bots) into computational algorithms. Ethical considerations are a subset of these considerations \cite{national2019decadal}.}

Many of these bot detection algorithms are based on analyzing patterns of bot behavior through distilling hundreds to thousands of features for supervised machine learning algorithms \cite{sayyadiharikandeh2020detection,schuchard2021insights}, deep learning algorithms \cite{ng2022botbuster,kudugunta2018deep}, or graph-based link inference algorithms \cite{feng2021twibot,magelinski2020graph}. These algorithms rely heavily on their training dataset inputs. Balancing training datasets for fairness is a key factor to consider in the creation of an ethical bot detection algorithm. This involves creating training datasets that have equitable representations of each demographic so that algorithms developed from these datasets will not be disproportionately inaccurate by language, social media platform, or type of post.

Bot detection algorithms aim to spot bots to maintain a healthy social media ecosystem. When deployed in the wild, these bot detection algorithms inevitably have errors in classification. Such errors include false positives, where legitimate users are classified as malicious bots \cite{rauchfleisch2020false}. False positive classification can result in punitive actions such as the de-platforming of these users. Algorithm development thus needs to be refined to communicate and patch algorithm blind spots.

Bot detection techniques have also been disproportionately geared towards the study of malicious bots that harm social discourse \cite{ng_assembling_2024,chang2021social,lee2011seven}. Identifying malicious bots such as @CoronavirusCon3 \cite{CoronavirusCon3}, which shares conspiracy theories on X, is important in safeguarding our online information systems. But we must not forget the existence of ``good bots'', or bots that assist us humans in our daily activities. \revision{One} example is the news bot @COVID19digest1 \cite{covid19digest1} that shares news digests on X to update its followers on the current world health situation. Ethical bot detection algorithms should consider how bot agents are used before acting on the classification.

\begin{figure}[h]
    \centering
    \includegraphics[width=0.8\linewidth]{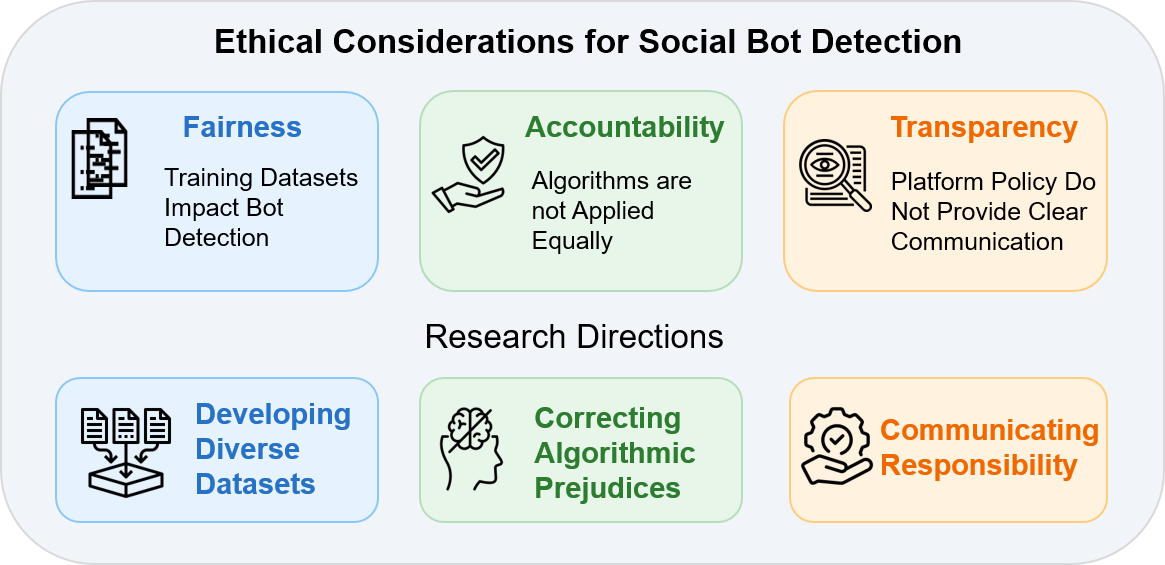}
    \caption{Summary of FATe framework applied to social bot detection in this article, including recommendations for respective research directions.}
    \label{fig:summary_diagram}
\end{figure}

With all these factors in mind, ethics must be considered alongside algorithmic progress when constructing social bot detection algorithms so that these algorithms can be deployed meaningfully towards a healthier online ecosystem. A common framework used to analyze ethics of machine learning algorithms in socio-technical systems is that of fairness, accountability, and transparency (sometimes referred to as FATe or FAccT).
We apply this framework to have three pillars in the design and deployment of a social bot detection system: the training dataset, the algorithm development, and the use of the bot agents. Our analysis examines these pillars through the lenses of training data asymmetry, algorithmic prejudice and engineering responsibility \cite{alameda2020fate,bogina2022educating}. \autoref{fig:summary_diagram} illustrates our application of FATe to social bot detection, along with corresponding proposed research directions.

This paper makes three contributions to social bot research:
\begin{enumerate}
    \item \textbf{Conceptually}, we extend the FATe framework from the machine learning domain to the new application domain of social bot detection, situating ethical analysis within socio-technical systems that involve human and automated actors.
    \item \textbf{Empirically}, we present a mixed-methods analysis that combines a survey of bot detection algorithms, fairness evaluation on a multilingual dataset and a qualitative study of user experiences from Reddit on how ethical blind spots materialize.
    \item \textbf{Practically,} we advance an agenda for ethical social cybersecurity that links dataset diversity, accountability mechanisms, and transparency practices to actionable strategies for building more trustworthy and equitable bot detection systems.
\end{enumerate}


This paper is organized as \revision{follows}: \autoref{sec:backgroundmotivation} motivates the need for ethical bot detection by providing a background to the landscape of social media bot detection and the social significance of bot detection, grounding the work in a socio-technical backdrop. To evaluate our ethical considerations, we use a mixed-methods analysis. \autoref{sec:analysis} elaborates on the quantitative and qualitative analysis methodologies used to evaluate each pillar. \autoref{sec:results_implications} presents the result of the analyses and elaborates on the implications of the results. In particular, we find that training datasets impact bot detection, algorithms are not applied equally\revision{,} and that platform \revision{policies} do not provide clear communication. Then, \autoref{sec:recommendations} describes recommendations for research directions: developing diverse datasets, correcting algorithmic prejudices and communicating responsibility. Finally, \autoref{sec:conclusion} elaborates on suggestions for ethical detection of social media bots and concludes the article. 

\section{Background Literature on Ethical Bot Detection}
\label{sec:backgroundmotivation}
\subsection{Cultural Considerations of Ethical Bot Detection}
In this work on ethical bot detection, we acknowledge that what is considered ethical varies by culture \cite{shao2019does}. For example, for cultures that value originality such as the American and European cultures, when a bot copies sets of words, the action might be thought of as plagiarism. In contrast, in East Asian cultures that value time-saving, this action might be considered as an efficient method of information dissemination. Further, the perception of deception is a central ethical concern in bot detection, and leads to considerations of accountability and transparency. This perception varies significantly across cultures. Cultural factors heavily influence what people perceive as harmful versus benign deception, and more so in the space of artificial intelligence bots \cite{chakraborti2019can}. In some cultures, small social lies by bots are acceptable, especially when they preserve social harmony, while in other cultures, complete transparency is the ethical imperative \cite{global2006world}. 

Bots operate in a world system, and not all countries and creators of bot detectors may subscribe to the same ethical system. This cultural contingency reflects the description of ``machine culture'', where the embedding of human values, habits and imaginaries into algorithmic systems in turn \revision{influences} human culture \cite{brinkmann2023machine}. Similarly, \cite{winner2020whale} argues that technologies are not neutral tools, but their usages embody and reproduce social orders and political values. In our context, bot detection algorithms are technologies which encode assumptions about authenticity, credibility and moral legitimacy that \revision{mirror} the cultural environments in which they are built, operated and deployed within. That is, the ``practices, institutions, and relations'' in which this bot technology operates gives it meaning \cite{suchman2007human}. Recognizing this interplay between cultural values and technological systems is essential for developing ethical approaches to bot detection. Such approaches should respect the diversity of users in the social media space yet maintain shared principles of transparency and accountability.

In this paper, we primarily take a Western, American-centric perspective. However, we acknowledge that as bot detection technologies become deployed across increasingly global platforms, future work must pay attention to these cross-cultural ethical considerations, to ensure that bot detection mechanisms respect cultural-specific ethical frameworks while still effectively identifying harmful automated behavior.

\subsection{Social Significance of Bot Detection}
Classifying social media users as bots or humans can have severe impacts on both individual users' experiences and entire online information ecosystems.
\autoref{tab:social_implications} summarizes the social significance of social media bot detection, listing both potential positive and negative social implications. 

Social bot detection allows for distinguishing inorganic and automated users from genuine human users, deepening the understanding of how these two types of accounts interact with each other, and showcasing how technology automation can be used to improve human lives through tasks like scheduling or aggregating posts \cite{ng2024cyborgs}. The identification of social bots, both good and bad, improves trust on online platforms, because users can better differentiate genuine human interactions from automated bot connections \cite{hajli2022social}. 
Automated accounts that propagate harmful content can be censored or removed from networks where they may spread misinformation or sow disorder.

\begin{table}
    \centering
    \begin{tabular}{p{7cm}p{7cm}}
        \hline
        \textbf{Positive implications} & \textbf{Negative implications} \\ \hline
        \begin{itemize}
            \item Leverage automated users to aid human information consumption (e.g. cyborgs)
            \item Improved trust in online platforms where users can better identify genuine human interactions
            \item Identify problematic bot agents and take countermeasures to affect their influence
            \item Understand the intertwined nexus of bot agents and humans
        \end{itemize}
         & 
         \begin{itemize}
             \item Suppression of the increasingly common use of AI technology in creating social agents
             \item With more visibility on bot algorithms, bad actors can write bots to evade detection
             \item Suppression of freedom of speech on social media 
         \end{itemize}
         \\ \hline 
    \end{tabular}
    \caption{Social Significance of Bot Detection}
    \label{tab:social_implications}
\end{table}

On the flip side, identifying and subsequently removing these bot accounts has been seen as a suppression of freedom of speech on social media, as well as suppressing the use of AI technology to create social agents \cite{vese2022governing}.
The publication and study of bot detection algorithms puts visibility on bot detection algorithms, which can provide bad actors with knowledge to program their bot agents to evade detection. For example, bots are known to change the use of their message artifacts such as using different hashtags and embedding messages in images to evade detection \cite{jacobs2024det}. 

These tensions highlight that bot detection operates within complex socio-technical systems in which infrastructures, institutions, and everyday usages intertwine \cite{pinch2012social,douglas2012social}. The design of algorithms and moderation policies govern digital publics. Decisions about how bots are classified or which bots are removed are not merely technical design choices but interventions in the social platform that determines whose voices are amplified and whose voices are constrained, thus defining the boundaries of authenticity and acceptable agentic participation in networked spaces \cite{gillespie2018custodians}.

\subsection{Fairness, Accountability, Transparency}
\label{sec:ethicaldetection}
Ethical challenges in machine learning (ML) and artificial intelligence (AI) tasks such as biometric systems, recommendation systems and language modeling have been studied at length \revision{\cite{buolamwini2017gender,milano2020recommender,bolukbasi2016man,bender_dangers_2021,ng2023digital}}. In doing so, sets of principles have been used to analyze such systems, often varying according to the application domain. Principles such as privacy and informed consent are particularly relevant in healthcare contexts \cite{javed_ethical_2024}, while ethical frameworks for cryptocurrency must take into account market manipulation and profit motives \cite{alibasic_developing_2023}, and ethical frameworks for robotics must consider implementation directly through encoded moral behaviors or through the system's ability to learn from interaction and experimentation \cite{wallach_moral_2008}.

For this article, we structure our ethical analysis around the core principles of \textit{fairness, accountability,} and \textit{transparency}, sometimes referred to as FATe or FAccT. These principles form a framework that is frequently invoked in the AI ethics community \cite{veale_fairer_2017,ahmad_fairness_2020,givens_centering_2020}. The FATe principles are commonly identified in surveys of AI ethics \cite{jobin_global_2019,hagendorff_ethics_2020} and used as a first step in ethical analysis \cite{mariotti_framework_2021}.
This framework is not without drawbacks. Prior work has critiqued its focus on the design rather than the use of systems \cite{gansky_counterfacctual_2022,widder_limits_2022,keyes_mulching_2019}.
Others note the framework's deficiency in accounting for institutional contexts that are vital determinants of social justice outcomes \cite{mittelstadt_principles_2019,laufer_four_2022,widder_its_2023}.
We keep these limitations in mind as we cautiously adopt FATe as a framework that we believe is fitting for an initial investigation of ethical issues in social bot detection.
In particular, we attempt to broaden our focus beyond the \textit{design} of social bot detection systems to include how they are \textit{used} in practice, embedded within institutions that may or may not be accountable to users.
We hope our work will stimulate a conversation about ethics in this research area, which may include additional ethical frameworks for analysis.

As machine learning algorithms, the data used and the knowledge embodied by social bot detection algorithms are necessarily ``partial, situated, and contextual'' \cite{leurs_feminist_2017,gray_feminist_2021}. In this article, we adopt the FATe framework as the first step to illuminate the ethical shortcomings and particularities in the case of social bot detection. For each principle, we identify ethical challenges through practical illustrations of the ethical issues and discuss recommendations to address those challenges.

We begin by first elucidating our interpretations of fairness, accountability, and transparency in the context of social bot detection. \autoref{tab:challenges_table} briefly summarizes the three principles of the framework and the recommendations we put forth in this article.

The term \textbf{fairness} has a multiplicity of definitions and implementations. Fairness in algorithms is generally defined as avoiding bias, prejudice, or favoritism towards groups or individuals \cite{mehrabi_survey_2021,jobin_global_2019}. In the context of bot detection, fairness is most clearly indicated in the geographic, linguistic, and platform distribution within training datasets. The distribution of training data affects the ability of bot detection algorithms to accurately evaluate posts from a wide variety of users.

The second principle is \textbf{accountability}, which refers to taking responsibility for outcomes of an ML system \cite{kohli_translation_2018}. Others note that accountability also includes the ``standards against which conduct and outcomes can be evaluated'' \cite{suzor_what_2019}, in order to hold actors, such as designers, developers, and institutions, to account \cite{gray_feminist_2021,jobin_global_2019}. Accountability in a bot detection system manifests largely in algorithmic development and system implementation.
Notably, it is crucial for bot classification operators to provide better explanations on the bot/human classification results and the action taken, for example through explainable AI techniques or better user interface indications.

Finally, \textbf{transparency} is the increase of ``interpretability or other acts of communication and disclosure'' \cite{jobin_global_2019}. Transparency is the ethical principle most commonly mentioned in AI ethics guideline surveys \cite{jobin_global_2019}. However, there is substantial variation in how transparency is interpreted across cases, from simply disclosing details about a system's operation to including representative groups in the system design \cite{kohli_translation_2018}. For the case of social bot detection, we select the definition of transparency as clearly communicating the definitions of an acceptable bot behavior and distinguishing between legitimate and malicious bots.

In the next sections, we detail these challenges for each principle, our methodology for analyzing the current landscape, and provide recommendations for future research.

\begin{table}
    \centering
    \begin{tabular}{p{7cm}@{\hskip 0.3cm} @{\hskip 0.3cm}p{6cm}}
        \multicolumn{2}{l}{\textbf{Fairness: Training Dataset}} \\
        \textit{Current Ethical Considerations} & \textit{Recommendation} \\ \hline
        Many datasets are gathered from X, based on political events, and are English-dominant & Develop datasets covering diverse contexts  \\ \hline
        & \\
        \multicolumn{2}{l}{\textbf{Accountability: Algorithm Development}} \\
        \textit{Current Ethical Considerations} & \textit{Recommendation} \\ \hline
        Algorithmic biases resulting in some demographic groups having a higher likelihood of being classified as a bot & Build a more accessible appeal process \\
        Take responsibility for actions taken as a result of the algorithmic classification & Multi-stakeholder discussion to develop process and solutions \\ \hline
        & \\
        \multicolumn{2}{l}{\textbf{Transparency: Bot Agent Usage}} \\
        \textit{Current Ethical Considerations} & \textit{Recommendation} \\ \hline
        Unclear classification of malicious bots & Explicitly define acceptable bot behavior  \\
        Lack of system transparency & Explain the evaluation of bot classification \\ \hline
        & \\
    \end{tabular}
    \caption{Ethical Considerations and Recommendations towards ethical bot detection algorithms}
    \label{tab:challenges_table}
\end{table}

\section{Methodology}
\label{sec:analysis}
This section describes the set of mixed-methods \revision{analyses} we employ to analyze \revision{the} bot landscape for our ethical analysis.

\subsection{\revision{Fairness: Literature Survey and Empirical Evaluation of Bot Detection Algorithms}}
To analyze fairness of the construction of bot detection algorithms, we used two approaches. The first approach is a literature survey of relevant papers of bot detection algorithms (\autoref{sec:lit_survey}). The second approach is an empirical evaluation of a bot detection algorithm for its performance in English vs. Non-English tweets (\autoref{sec:empirical_bot}).

\subsubsection{\revision{Literature Survey of Bot Detection Algorithms}}
\label{sec:lit_survey}
We first begin our analysis by examining the landscape of bot detection algorithms. This literature survey was conducted using a snowball method. We \revision{began} with studying the seminal paper ``The Rise of the Social Bot'' \cite{ferrara_rise_2016}. This seminal paper put forth a position that the rapid proliferation of automated agents on social media has transformed online discourse, and motivated the scientific community to systematically investigate how social bots influence online discussions. This paper was chosen because it is platform-agnostic and served as a conceptual anchor for subsequent computational approaches to bot detection. We then extracted the papers that cited the seminal paper from Google Scholar, and kept the papers that developed bot detection algorithms and \revision{were} written in English. We excluded papers that applied bot detection methods for an analysis, and papers that mentioned bot detection in passing. For each of those papers, we then collected the papers that cited them as a reference, snowballing the paper search. We stopped when the set of papers \revision{began} to repeat itself. Each inclusion and exclusion of papers were independently validated by two members of the team. In total, this survey returned 51 papers from the years of 2015 to 2024, which are presented in \autoref{tab:bot-detection}.

For each of the relevant papers that we kept, we annotated the platform that bot detection algorithm was created for, the event (if any) that the algorithm was premised upon, the language of the posts that were used to train the algorithm and a brief description of the type of machine learning algorithm. We then calculated some descriptive statistics of these data collected \revision{(i.e. proportion of algorithms targeting each social media platform, distribution of language targeted, distribution of events targeted, distribution of algorithm type or approach)}. We note that this snowball sampling may bias our dataset via how the papers are associated with each other, but this bias itself reflects the interdependent nature of bot detection research, where datasets are repeatedly reused and extended.

\subsubsection{\revision{Fairness: Empirical Evaluation of a Bot Detection Algorithm for Performance in English vs. Non-English Tweets}}
\label{sec:empirical_bot}
For an empirical \revision{understanding} of fairness, we evaluated how a popular multilingual bot detection system performs on different user groups that are segmented by language. These groups are: (1) users that primarily tweet in English and (2) users that primarily tweet in non-English language. The social bot detection algorithm that we used is the BotBuster algorithm, \revision{which was trained in 2023} on 87,056 users on primarily English language tweets \cite{ng2022botbuster}. The algorithm was updated in 2025 to include a multilingual word vector module to handle the diverse nature of language of social media \cite{ng2025social}. We used the updated multilingual version of the algorithm and applied \revision{it} across different datasets from the OSOME bot repository\footnote{\url{https://botometer.osome.iu.edu/bot-repository/datasets.html}}, \revision{one of the largest collections of social media user datasets human-flagged as ``bot'' or ``human''. This} repository contains 19 manually labeled datasets as of May 2024. The labels produced by the BotBuster algorithm are then compared with the manual labels from the OSOME repository. Each dataset was curated with different events and keyword focus.

Within each dataset, we compared the performance of the BotBuster algorithm on users that write primarily in English and users that write primarily in non-English languages (i.e., languages other than English). A user is considered to write in the English language if at least 80\% of their tweets are written in the English language. The language of the tweet is taken from the metadata in the tweet; that is, the language of the tweet is based on what the platform X would have inferred the language that the tweet was written in. We then compared the accuracy metrics \revision{of BotBuster's predictions} between the users that primarily tweeted in English and the users that primarily tweeted in non-English languages. A fair algorithm would perform the same on both types of users across all datasets. The results of this empirical evaluation \revision{are} presented in \autoref{tab:english_perc}.

\subsection{\revision{Accountability: Case Study of Reddit Users Being Mislabeled as Bots}}
\label{sec:reddit}
To illustrate the issue of accountability in bot classification, we draw on user-posted anecdotes from Reddit threads about users being incorrectly labeled as bots by social media platforms. 
In May 2024, we searched subreddits that correspond to three of the most active social media platforms (r/Instagram, r/Facebook, r/Twitter) for the phrase ``I'm a bot'', commonly mentioned in posts about social media users being misclassified as bots.
Subreddit threads that surfaced in this search were typically titled, ``Facebook thinks im a bot'', ``Twitter thinks i'm a bot'', etc.
The authors deductively coded the top 20 resulting threads for each subreddit (60 in total) for three categories: 1) whether the post describes misclassification as a bot (is relevant to our study), 2) whether the user was able to appeal their misclassification, and 3) any consequences the user mentions from their misclassification as a bot.
This coding enhances the understanding of accountability by surfacing how the classification errors of bot detection algorithms are actually experienced by users, and the availability (or lack of) recourse mechanisms, as narrated by online users.

This is not intended as an exhaustive study of user experiences being mislabeled as bots, but simply a set of motivating anecdotes for illustrating a real-world ethical consequence of social bot detection.
\autoref{tab:reddit_summary} presents the summary of our examination, and the full table is in the Appendix \autoref{tab:reddit_table}.

\subsection{\revision{Transparency: Qualitative Analysis of Reddit User Anecdotes and Platform Policy Analysis}}
\subsubsection{Qualitative Analysis of Reddit User Anecdotes}
\revision{Using the data of Reddit user anecdotes that we collected and annotated in the previous section, we annotated the experiences of the users in terms of whether they were able to appeal their misclassification as a bot, and whether they understood the reason for being misclassified. These stories, which motivate our analysis, reflect the pillar of Transparency by assessing the extent to which the platform moderation decisions are accompanied by meaningful explanations and opportunities for users to correct the misclassification, or are they instead left to infer the causes of misclassification because of opaque system behavior.}

\subsubsection{Policy Analysis of Platform's Actions for Bot Accounts}
To evaluate the transparency of social media platforms, we compared the current platform policies of several social media platforms on acceptable or unacceptable automation, or bot usage. These platform policies are manually extracted from the platform websites themselves in 2025. \autoref{tab:platform_policies} presents the summary of platform policies that we have consolidated.

\revision{In the next section, \autoref{sec:results_implications}, we present the results of our methods and explore the implications of our findings by describing ethical considerations and research directions for each portion of FATe (Fairness, Accountability, and Transparency) as applied to bot detection.}

\section{\revision{Results and Implications}}
\label{sec:results_implications}
\subsection{Fairness: Training Datasets \revision{Impact Bot Detection}}
\label{sec:dataset}
\subsubsection{\revision{Current Bot Detection Algorithms Lack Dataset Diversity}}
Bot detection systems rely on patterns in training data to identify bots from humans. In our first study of fairness, we performed a literature survey (see \autoref{sec:lit_survey}) of bot detection algorithms. \autoref{tab:bot-detection} presents a list of current bot detection tools, containing 51 papers from the years of 2015 to 2024. 

While bot detection algorithms have evolved over time from heuristic-based algorithms \cite{beskow2018bot,ng2022botbuster,sayyadiharikandeh2020detection} to unsupervised time-based algorithms \cite{chavoshi2016debot,mazza2019rtbust} to neural network-based algorithms \cite{liu_botmoe_2023,hui_botslayer_2019} to large language model-based algorithms \cite{feng2024does,zhou2024lgb}, the dataset signatures used to train the algorithms remain the same.
There is a notable lack of diversity in these training datasets with respect to language, event and source platform.
94.2\% ($n=48$) of these algorithms are trained on datasets collected from X, and only 5\% ($n=3$) are trained on other social media platforms like Reddit, Instagram, or TikTok. Further, 84.3\% ($n=43$) datasets contain posts that are authored primarily in the English language, showing the language imbalance of the curated datasets.

The homogeneity of training datasets used to construct bot detection algorithms creates asymmetry in training data. Most bot detection algorithms like Botometer \cite{yang2019arming} and BotBuster, a multilingual bot detection system trained on diverse social media datasets across Twitter, Reddit, and Instagram \cite{ng2022botbuster}, rely on supervised machine learning algorithms, in which the composition of the training data deeply affects the fairness of the algorithm.
Even bot detection algorithms based on Large Language Models (LLMs) do not overcome these limitations, as it is known that LLMs have inherent biases towards language, ethnicity, and geographical population groups when under zero-shot prompting scenarios \cite{kotek2024protected,manvi2024large,malik2024textit}.
How can bot detection algorithms accurately classify accounts from communities whose data does not appear in training datasets? This phenomenon is known as representation, or exclusion, bias, and affects the fairness of bot detection classification \cite{mehrabi_survey_2021}.

Current bot detection training datasets also lack topical diversity.
Much of the analysis on bot activity is event-based, so much of the training data is restricted to political or health events. The event of interest determines the data collection technique, which then affects the set of accounts and content retrieved. This event-based collection strategy is typically performed by selecting a set of keywords, or hashtags. While this strategy is appropriate for collecting relevant discourse on an event, such selection limits the search towards accounts that are interested and active in that topic, building a topic-specific training dataset. For example, collecting data via the hashtag \#ReleaseTheSnyderCut, which promoted a fan-driven-campaign for the Justice League film, would overwhelmingly surface accounts tied to that movement. Warner Bros later reported that a significant amount of support for this campaign was fueled by bot accounts \cite{siegel_2022}. 

A common collection strategy is snowball sampling, which begins with a seed account and recursively collects followers. This approach is typically used in social network studies to capture communities of interacting accounts, allowing researchers to map the propagation of information through the connected clusters of users. For example, \cite{ng2024exploratory} employed this method to collect Telegram accounts \revision{linked} to the ``Disinformation Dozen'', a group of influential online users that are known to start the spread of coronavirus-related disinformation \cite{nogara2022disinformation}. By collecting accounts that are in the channels where the Disinformation Dozen had posted, and the people who had shared those posts, the study mapped that in Telegram, bots were more effective in sustaining conversations while humans were spreading the information \cite{ng2024exploratory}. Such network-based collection enables information diffusion analysis but also risks over-representing highly connected user clusters while overlooking peripheral and less active users, potentially biasing the resultant training dataset.

Both the event-based and snowball sampling data collection strategies inevitably create isolated groups of accounts, resulting in algorithms overfitting for one type of account and underfitting for other types. This method limits the resultant set of accounts to users in the same social group, particularly the group that follows the seed account. That is, if the seed account is a political leader, the collected data will contain a large number of political advocates. Algorithms trained on this data will be adept at spotting political bots but not other types of bots, like health bots.

Other methods involve manual identification and vendor acquisition. Manual identification often involves coincidental spotting of accounts, such as through honeypots \cite{lee2011seven,yang2019arming} or abnormal behavioral characteristics \cite{danaditya2022curious}, which narrows the scope of search to visibly identifiable bot accounts. Vendors of bot accounts, who sometimes advertise these accounts as a service to boost influence on social media, usually create and maintain specialized types of bot accounts, such as fake follower bots \cite{yang2019arming}. Acquiring bot accounts through such vendors to include in datasets thus also narrows the scope of accounts algorithms learn from to the type of automated accounts that are marketable.

Finally, an essential part of this diversity is sampling data across platforms. Currently, there has been an almost exclusive focus on X, formerly Twitter, in data collection. Although there are many bot detection systems developed by research groups, only 16.7\% (n=7) of the systems surveyed \cite{knauth_language-agnostic_2019,chavoshi_debot_2016,mazza_rtbust_2019,sayyadiharikandeh_detection_2020,wang_you_nodate,wang_social_nodate,ferreira_dos_santos_uncovering_2019}, have the capability to classify beyond the Twitter/X platform. This lack of attention to other social media platforms not only leaves these platforms vulnerable to the activities of social media bots but also means that an account originating from other sites will not be accurately classified by current algorithms. 

\subsubsection{\revision{Empirical Performance of Bot Detection Algorithm Fares Better on English than Non-English Tweets}}
One of the largest \revision{collections} of bot datasets is the OSOME Bot Repository,\footnote{\url{https://botometer.osome.iu.edu/bot-repository/datasets.html}}, which contains 19 manually labeled datasets as of May 2024. Out of the posts across datasets in the OSOME repository analyzed in \cite{ng2022botbuster}, only about 40\% of the users predominantly write in English, as determined by X's tweet language classifier.
However, the vast majority of published bot detection systems focus exclusively on English (see \autoref{tab:bot-detection}).
\autoref{tab:english_perc} presents the results from an empirical investigation into the performance of a popular multilingual bot detection system, BotBuster, across users who \revision{primarily tweet in the} English language versus non-English languages across a number of datasets in OSOME. The language of each tweet is derived from the metadata annotated by the X platform. Users that tweet at least 80\% of the tweets in the English language are then classified as ``users who tweet primarily in English'', and the reverse for users that tweet primarily in non-English languages. Then, we compare the accuracy of the BotBuster algorithm in identifying bots from users that primarily tweet in English and non-English. That is, we report the accuracy of the generic BotBuster system in classifying the user as a bot or not, comparing with the manually annotated gold labels.

\begin{table}
    \centering
    \begin{tabular}{p{3cm}p{2.5cm}p{2.5cm}p{2.5cm}p{2.5cm}}
    \hline
        \textbf{Dataset} & \textbf{Users that \newline primarily write in English (\%)} & \textbf{Accuracy \newline identifying bots from users that primarily write in English (\%)} & \textbf{Users that \newline primarily write in Non-English (\%)} & \textbf{Accuracy \newline identifying bots from users that primarily write in Non-English (\%)} \\ \hline 
        astroturf \cite{sayyadiharikandeh2020detection} & 37.2 & 98.8 & 62.8 & 98.6 \\ \hline 
        botometer-feedback-2019 \cite{yang2019arming} & 47.5 & 99.2 & 52.5 & 98.6 \\ \hline
        botwiki-2019 \cite{yang2020scalable}  & 43.1 & 97.2  & 56.9 & 95.9 \\ \hline
        cresci-rtbust-2019 \cite{mazza2019rtbust} & 9.5 & 98.8 & 59.0 & 92.9 \\ \hline
        cresci-stock-2018 \cite{cresci2018fake} & 41.0 & 98.8 & 59.0 & 92.9 \\ \hline
        gilani-2017 \cite{diesner2017proceedings} & 27.0 & 97.9 & 73.0 & 97.4 \\ \hline
        midterm-2018 \cite{yang2020scalable} & 56.8 & 99.6 & 43.2 & 98.6 \\ \hline
        varol-2018 \cite{varol2017online} & 50.4 & 98.2 & 49.6 & 97.7 \\ \hline
        verified-2019 \cite{yang2020scalable} & 50.3 & 99.9 & 49.7 & 99.9 \\ \hline
        Average & 40.3 & 98.7 & 56.2 & 96.9  \\ \hline  
    \end{tabular}
    \caption{Difference in bot detection performance for users who primarily write in English and non-English languages (as determined by 80\% of their tweet language). \revision{A \textbf{fair} algorithm would perform equally well on both groups.} Refer to the Appendix \autoref{tab:english_perc_full} for the full statistics.}
    \label{tab:english_perc}
\end{table}

Overall, the BotBuster bot detection algorithm performs slightly better for users who primarily tweet in English than for users who do not. Even this is a generous evaluation estimate of the performance of bot detection in non-English languages because many of the non-English posts are written in high-resource European languages rather than low-resource languages. Given that most benchmark datasets and bot detection models disproportionately represent English-language content, it is not surprising that non-English languages are often overlooked, and the performance of the algorithm on non-English users is lower. While there are some language agnostic algorithms based on sender information only \cite{chavoshi_debot_2016,chavoshi_temporal_2017}, a vast majority of bot detection algorithms consider the texts as an input.
Constructing algorithms specific to different languages is important for the understanding of the social cyber geography landscape of information dissemination \cite{ng2025social}.


Dependence, or overdependence, on current training datasets means that accounts that primarily post in languages other than English and accounts that post about topics outside the scope of politics and finance are more likely to be misclassified due to under-representation: either they are humans that are misclassified as bots or they are undetected bots. There is a need for more research into bot detection datasets and the development of bot detection systems from social media platforms other than X, such as Facebook, Instagram, TikTok, and Bluesky.

\subsection{Accountability: \revision{Algorithms} Are Not Applied Equally}
\label{sec:algorithm}
To illustrate the consequences on users in the event of false positive bot classification, we \revision{qualitatively} analyzed subreddits related to three popular social media \revision{platforms} for threads where users seek help and express their frustration on being classified as a bot. \autoref{tab:reddit_summary} presents the summary of our examination, and the full table is in the Appendix \autoref{tab:reddit_table}. From our manual and deductive examination of the top 60 reddit threads, we find that 78\% ($n=32$) of the users expressed frustration on being incorrectly labeled as bots (``Anyone else run into this issue? How am I supposed to prove I’m real if I can’t do \textit{anything}?'', ``I'm thinking maybe because I'm liking too many posts and comments Instagram thinks I'm a spamming bot???'').

\begin{table}[h]
    \centering
    \begin{tabular}{p{7cm}r r  r }
    \revision{\textbf{Category}} & \revision{\textbf{Number of posts (\textit{n})}} & \textbf{Percentage (\%)} \\ \hline
    \hline
    \revision{\textbf{Relevant anecdodes (\revision{total, out of 60 reviewed)}}} & \revision{41} & \revision{100.00} \\ \hline
    Expresses frustration & \revision{32} & \revision{78.05} \\ \hline 
    \hline
    \revision{\textbf{Mentions consequences (total)}} & \revision{30} & \revision{100.00} \\ \hline 
       No access to account / login restrictions & \revision{17} & 56.66 \\ \hline 
       Messaging restrictions & \revision{2} & 6.67 \\ \hline 
       Interaction restrictions & \revision{3} & 10.00 \\ \hline 
       Posting restrictions & \revision{6} & 20.00 \\ \hline 
       Others & \revision{2} & 6.67 \\ \hline 
    \hline
    \textbf{Mentions appeal actions (total)} & \revision{9} & 100.00 \\ \hline 
        Cannot appeal & \revision{5} & 55.56 \\ \hline 
        No response & \revision{4} & 44.44 \\ \hline 
    \end{tabular}
    \caption{Summary of user-posted anecdotes of their accounts being flagged as a bot on Instagram, Facebook and X. \revision{Users expressed frustration on being incorrectly labeled as a bot, and are commonly not able to regain access to their account. There must be more \textbf{accountability} in the deployed algorithm to indicate to users the reason for being moderated.}}
    \label{tab:reddit_summary}
\end{table}

After receiving notifications from platforms that their accounts are suspected bots, these users are commonly locked out, have trouble logging in, are restricted in common actions like commenting, liking and messaging, or are forced to provide detailed personal information to prove that they are human. Many times, the appeal process is extremely complicated, with users echoing frustration at the complexity of the process (``Showed a popup that said `Facebook doesn't allow fake accounts' and then told me to jump through flaming hoops to open my account again, such as attaching my phone numbers and uploading a photo of myself, seriously not worth the trouble.''). In 55.56\% of these cases, there is no way to appeal, and in 44.44\% of these cases, there is no response after the user appealed. One user even mentioned that they had been waiting for months for a reply from Facebook after appealing their account restriction (``I've been waiting for 7 months and nothing...'').  Such unresolved moderation outcomes can have broader implications for procedural fairness and digital rights, undermining users' trust in automated governance. Therefore, there must be some accountability in the deployed algorithm to build a healthy digital ecosystem.


In the complex socio-technical environment of social media, bot detection algorithms can have a broad impact across different scales of users, ranging from individual users to organizations and governments. For example, labeling and deplatforming a government-run bot that broadcasts the number of malaria cases reduces critical information on the disease from reaching the masses, while doing the same for a state-sponsored bot that aims to destabilize politics can restore order. In both scenarios, the impact of the algorithm results in the loss of information transmission to a large community, but it is not easy to define a positive and negative impact. Banning and deplatforming can also have acute individual effects on hobbyists or small business owners who rely on social media platforms. From our Reddit qualitative analysis, some users noted that their account restrictions mean they cannot connect with friends, grow their influence or share their hobbies (e.g., ``I just want to enjoy sharing some art, but I feel like i'm being treated like a bot.'', ``I wouldn't really care, except it's how all my friends interact with one another, and I'm getting left out of the conversation because of this.''). These examples illustrate how algorithmic biases can cascade into social outcomes, and affect certain user groups or contexts, underscoring the need to design detection systems that are both socially aware and technically robust.

In order for different groups of users to be evaluated fairly by bot detection algorithms, generalizable algorithms are needed. Training generalizable bot detection algorithms requires projecting multi-domain, multi-lingual and multi-country datasets onto the same feature space so that algorithms can collectively make sense of them. This introduces a number of open challenges. For machine learning-based approaches, researchers must evaluate the most effective techniques to combine representations across domains. This involves comparing model architectures of different aggregation strategies, e.g., concatenation and summation \cite{smith2016deep}.
There are also choices to be made about at which stage the domain differences are applied, i.e., early, late or slow fusion \cite{karpathy2014large}. Incorporating data from multiple languages further requires researchers to make decisions about building joint or separate embedding spaces for each language \cite{ormazabal2019analyzing,feng_language-agnostic_2022}. 

\revision{Accountability in bot detection is further complicated by cultural variation. That is, how social media platforms are used, interpreted and embedded in the everyday life of different cultures. Norms around posting frequency, identity presentation, device sharing and social media usage differ widely across regions, particularly between Western and Eastern contexts \cite{alsaleh2019cross,zhao2011cultural}. In parts of Southeast Asia and Africa, social media use is often collective \cite{lim2023activist} and economically motivated \cite{olanrewaju2018influence}, which differs from the conversational patterns of the Western usage. Such differences will make the users appear anomalous or bot-like when evaluated against models trained on more Western-centric assumptions. These cultural contexts need to be factored into bot detection systems to avoid the risk of systematically misclassifying legitimate users, which then creates an accountability gap where users are penalized for their behavior that matches their local norms. This issue is reflected within some of the Reddit user anecdotes where users struggle to reconcile their everyday platform use with the opaque bot accusations which they cannot successfully contest.}

\revision{Lastly, a key accountability challenge in bot detection arises from the fact that social media companies frequently adapt their platform algorithms to comply with country-specific laws and constraints. Content moderation and algorithmic enforcement practices can vary substantially across countries due to state-platform relationships and regulatory risk exposure, resulting in uneven application of the bot detection results across regions. For example, Germany's Network Enforcement Act (NetzDG) mandates the rapid takedown of illegal content, which leads to more aggressive automated enforcement compared to other regions \cite{wagner2020regulating,kasakowskij2020network}. In India, platforms have to comply with intermediary liability rules and government takedown requests, which shapes how automated systems flag political speech and coordinated activity \cite{dubey2025digital,mukherjee2019jio}.  
These regionally tailored algorithmic adjustments can introduce systematic biases, where identical behaviors may be flagged as suspicious in one country but tolerated in another. Such differences complicate the efforts to ensure consistent accountability on bot classification. However, these regional differences are rarely disclosed to users, so the individuals affected by misclassification are often unable to determine whether the enforcement outcomes stem from technical error, platform policy choice or legal compliance.}

The detection and measurement of biases in algorithms is crucial to avoiding unwanted discrimination across groups in bot detection. To discover biases, bot detection systems \revision{should} be evaluated on diverse datasets that reflect differences in user attribute such as country, language and gender. Comparative performance analyses across these user groups can reveal systematic disparities in classification accuracy and highlight where algorithmic fairness interventions are needed.

\subsection{Transparency: \revision{Platform Policies Lack Clear Communication}}
\label{sec:botusage}
Transparency in bot detection refers to the communication of acceptable behavior, especially acceptable bot behavior. Transparency also extends to taking into account and possibly communicating the uncertainty of the bot detection algorithm. 

From our examination of Reddit threads (\autoref{tab:reddit_summary} and \autoref{tab:reddit_table} in the Appendix), many users indicate that they encounter stress and frustration when their accounts are flagged as bots (``annoying'', ``please help!!''). This leads to speculation among users of bot-like indicators of their behaviors, such as too much liking, mak/ing too many follow requests, or posting too frequently. One user related that ``I was just scrolling memes'' before he was deplatformed, expressing confusion at the platform's policies that associated his behavior with bots. The added friction of self-evaluating whether online actions will be flagged as inauthentic before performing an action like creating a post could take a toll on user engagement (``facebook will penalize you for scrolling too fast. Be slow or else.''). Some posts mentioned that the stress of receiving warnings about suspected automated activity reduced their engagement on the platform. One user had mentioned that they had even paid for the premium status on X for a few months and had blocked bots ``immediately upon seeing them'', but had instead received a notice that they own account was being labeled as a harmful account with a restricted reach (``I think they think I'm a bot or that I'm promoting bots. [...] I block them immediately upon seeing them, but I'm effectively shadow-banned and wonder how to get this lifted''). However, many of these discussions are just user speculation through lived experiences, and there \revision{is} no confirmation from the \revision{platforms} themselves.

Social media platforms should be mindful that their moderation activities can cause stress and emotional toll on their user base, and therefore should better communicate their standards and practices. Communication strategies should be carefully selected to alert users if their actions led a bot detection system to suspect that they are a bot, and a warning system may be employed to allow for time for users to amend their actions rather than deplatforming users on their first content violation.

When deployed with live users on social media platforms, bot detection algorithms need to have system transparency. These algorithms typically provide a probability between 0 and 1 that represents the likelihood that a user is a bot. That probability itself can have a margin of error. Oftentimes, this probability gets converted to binary bot/human labels that are attached to the users without taking into account the uncertainty quantification. Any user that has a bot probability above the threshold is classified as a bot, and a user with a probability below that threshold is classified as a human. Many academic systems select a threshold of 0.7 \cite{danaditya2022curious,jacobs_tracking_2023} or 0.5 \cite{chang2021social,rauchfleisch2020false}. In some cases, differing threshold values are used for the same bot detection algorithm \revision{depending on the intented use of the algorithm}. Different threshold values affect the proportion of bots detected. The stricter 0.7 threshold means that lesser users are classified as bots, while the more relaxed 0.5 threshold means that more users are classified as bots, which could mean more false positives. Retributive actions taken by social media platforms based on a binary \revision{classification that masks} uncertainty can lead to inaccurate conclusions. 

Beyond the issue of thresholding, another challenge lies in the opacity of how these probability scores are generated in the first place. Most supervised machine learning classifiers offer little justification for their output probabilities, leaving users and researchers unable to discern why an account was labeled as a bot. There has been progress in explainable AI that allows cracking open the machine learning box, such as through the use of game-theoretic approaches \cite{NIPS2017_7062} or identifying saliency of words in a sentence \cite{ding2019saliency}, providing insights into which features are important for the tested machine learning algorithm. Understanding which features are most important in a machine learning model is essential to identify potential sources of algorithmic biases. For example, \cite{ng_assembling_2024} revealed that in a random forest bot detection algorithms, the entropy or randomness of the characters in the user name or screen name contributes greatly to the bot classification result. Such ideas should be incorporated into bot detection algorithms to provide algorithmic transparency, thus providing a better understanding as to which feature(s) contributed to an account being classified as a bot.

Existing research-based bot detection systems have varying degrees of clarity around factors that go into their probability scores.
Botometer \cite{sayyadiharikandeh2020detection} breaks down a bot classification result by the contribution of certain features (i.e., complete automation probability) to the final probability score. BotBuster \cite{ng2022botbuster} works on a mixture-of-experts model, aggregating separate feature-based probability scores to form the final bot classification. TwiBot uses a graph-based approach to identify bot classes based on homophily of user interactions \cite{feng2021twibot}.
These techniques illuminate some algorithm decisions, pointing to the features that exhibit bot-like behavior.
However, this transparency, even limited, does not seem to extend to many deployed systems.

Currently, social media platforms do put up guidelines on unacceptable human-based behavior. \autoref{tab:platform_policies} curates the acceptable policies of bot usage for several popular Western platforms. While it is laudable that platforms do put out policies to define acceptable behavior (e.g. malicious activity, spamming), however, they usually only state the negative behaviors \revision{and} seldom publicize specific acceptable bot behavior. Further, it is difficult to infer acceptable usages of automation from the acceptable usages of the platforms from humans. The Rules and Policies of X state that ``You may not target others with abuse or harassment, or encourage other people to do so'', and elaborate on the behaviors that violate such policies \cite{xrules}, but they do not talk about the scenarios of acceptable uses.

\begin{table}[h]
    \centering
    \begin{tabular}{p{6cm}p{6cm}}
    \hline
    \textbf{Platform} & \textbf{Bot usage policy} \\ \hline 
    X & Bots are forbidden to engage in spamming, platform manipulation, aggressive following/ unfollowing, misleading or deceptive behavior. Bots can be used for helpful, informative or entertainment purposes if they follow the rules. \footnote{\url{https://help.x.com/en/rules-and-policies/x-automation}} \\ \hline 
    Facebook & For Messenger bots: responsiveness time requirement, no deception or misleading behavior, content restrictions, opt-out option for users\footnote{\url{https://developers.facebook.com/docs/messenger-platform/policy/responsiveness/}}  \\ \hline 
    Instagram & Cannot impersonate others, do anything unlawful, misleading or \revision{fraudulent}, create accounts or access or collect information in an automated way without the service's permission \footnote{\url{https://help.instagram.com/termsofuse}} \\ \hline 
    TikTok & Strictly \revision{prohibits} using bots or scripts to write fake reviews or comments or increase likes or shares; permissible automation includes post scheduling with TikTok's native scheduler, comment moderation and automated ad management with TikTok Ads Manager\footnote{\url{https://www.tiktok.com/community-guidelines}} \\ 
    \hline 
    \end{tabular}
    \caption{Summary of current platform policies of allowed bot agent usage. \revision{A \textbf{Transparent} system would clearly set forth the policy of which automation can be allowed.}}
    \label{tab:platform_policies}
\end{table}

Transparency operates together with fairness and accountability. The value of transparency comes from the increase in fairness and accessibility \cite{suzor_what_2019}. Having a better understanding of the decision-making process of the algorithm enables analysis of the classification results of groups of accounts, providing insights into which groups are disproportionately treated unfairly by the algorithm, which can lead to more targeted data collection. With more transparency around the behavior of a system, including its performance, better accountability can be assumed by the appropriate stakeholders. If the algorithm classifies celebrities that use the assistance of automated posting as bots, as in the case of cyborgs \cite{ng2024cyborgs}, social media companies should assume the task of not blocking those legitimate accounts.

\section{\revision{Recommendations for Research Directions}}
\label{sec:recommendations}
\revision{From our analysis, we propose the following research directions that the field could consider for a more ethical bot detection ecosystem.}

\subsection{\revision{Fairness}: Developing Diverse Datasets}
Bot detection algorithms should not solely rely on the set of readily available English-dominant X datasets. Bot detection systems need to evolve to capture a diverse set of users. A wider scope of user account collection and annotation, particularly across languages, events, and social media platforms, is needed to increase the diversity of datasets. Content selection needs to produce a diverse set of data to expose the developed bot detection algorithm to a wide variety of account types, languages, and topics in order to sufficiently represent the diverse social media population.

Ideal bot detection datasets would contain a proportionate amount of bot/human annotations from accounts that span across a range of user demographics, including differences across cultures, languages, ethnicities, genders, and so forth. To construct such a dataset, sampling techniques also need to evolve to identify representative, or at least more diverse, samples of social media populations and sub-populations for data collection and algorithm verification to maintain algorithmic fairness.
Techniques from preferential sampling \cite{kamiran_classifying_2009,kamiran_data_2012} could be useful. Diverse datasets could also be created from aggregating smaller datasets, for this method has been shown to improve the overall classifier and out-of-domain classification performance \cite{ng2022my}.
Datasets from currently under-represented social media accounts outside of English tweets are particularly needed to counteract the current bias.

Beyond dataset composition, fairness considerations also extend to how ground truth labels are generated and validated. Many existing bot detection datasets rely on platform-provided labels (i.e., ``spam'', ``verified''), suspended accounts, or heuristics that may themselves encode systematic biases against certain user behaviors, languages, or regions. Without careful scrutiny, such labeling practices risk conflating atypical but legitimate user activity with automation, disproportionately affecting users from marginalized or less-studied communities. Incorporating multi-stage annotation pipelines, human-in-the-loop verification, and culturally informed labeling guidelines can help mitigate these risks and improve the fairness and robustness of bot detection systems across diverse social media contexts.

\subsection{\revision{Accountability}: Correcting Algorithmic Prejudices}
To maintain accountability, bot detection algorithms need to be refined to correct for algorithmic prejudices. These algorithms need to be able to handle a wide variety of social media users. Since the availability of bot data by different languages varies considerably, approaches using multi-task learning to maximize accuracies of each demographic group may be useful \cite{oneto_taking_2019,dwork_decoupled_2018}. Rigorous testing is required to ensure that the algorithm maintains its accuracy across varied languages, cultures and types of social media users. Finally, these algorithms should account for the changing nature of social media bots \cite{ng2022botbuster,yang2023anatomy}, and algorithmic testing should also examine the stability of correctness metrics over time \cite{huang_stable_2019}.

Accountability is a complex issue due to the multi-stakeholder nature of the social media ecosystem: researchers, social media companies and even governments have their interests intertwined in this problem. Yet the responsibility for addressing malicious automation remains ill-defined, with no clear consensus on who should bear the primary ethical and operational burden. Would the onus be on researchers to develop new ways for detecting malicious bots? Would it be the social media companies' job to perform massive operations of take-downs of inauthentic accounts? Would it be regulatory bodies (i.e., government) to monitor and enact laws against bot accounts? 

As a first step, researchers should investigate whether any group of users have a higher likelihood of being classified as a bot compared to another user group. This can be done through systematic fairness auditing of detection models, using methods like subgroup evaluation, counterfactual fairness testing and group fairness metrics. Subgroup evaluation can reveal how false positive and false negative rates differ across demographic or behavioral groups, prompting researchers to focus their data collection or algorithmic design efforts on those groups \cite{mitchell2019model,raji2020closing}. Counterfactual fairness testing involve creating controlled variations of user profiles that differ only in certain attributes like language to assess whether classification outcomes remain stable across both control and testing populations \cite{kusner2017counterfactual,wexler2019if}.
Group fairness metrics such as equalized odds, which compares true positive and false positive rates across groups \cite{hardt_equality_2016,mehrabi_survey_2021}, may be particularly useful here. Studies in computer vision show that facial analysis algorithms are skewed towards lighter-skinned people and struggle to detect dark-skinned people \cite{buolamwini2017gender}. Social bot studies should draw inspiration from image algorithm studies and identify strata of users that are more likely to be classified as bots and strata for which classification yields a disproportionate amount of false positives and negatives. Biases could be investigated across salient large-scale demographic categories such as gender, race, and ethnicity, but there also may be social media-specific categories of users relevant to bot detection, such as frequent posters, hobbyists, or professional or organizational accounts.

Platforms and governments have taken some measures. X periodically removes accounts participating in information operations and releases selected sets on the X Transparency Center. The Singapore government introduced a Foreign Interference Bill in September 2021 to give the government authority to order social media platforms and Internet providers to prevent identified harmful content from being seen within the country. This bill comes years after a 2018 incident where the government requested the removal of posts from an Australian-based political activist, to which Facebook released a statement that it ``cannot be relied upon to filter falsehoods or protect Singapore from a false information campaign'' \cite{kim_aravindan_2018}. As each entity has its separate goals, one critical challenge to overcome as a whole is to communicate and harmonize the objectives of each group and collectively take responsibility for the problem of malicious social media bots.

Another part of accountability lies in taking responsibility for the actions taken as a result of the bot detection algorithms. One of the most prominent real-world outcomes of bot detection algorithms is mis-classification. False positives occur when a user is determined as a bot despite not being a bot. An essential part of responsibility for a system, i.e. accountability, involves deciding when and how to correct an error. 

In the case where legitimate users are wrongly classified as bots, social media platforms can put systems in place for them to prove their legitimate status and retain control of their social media accounts. Such appeal processes must not only be available but also accessible for average users \cite{miller_beyond_2019,gray_feminist_2021}. Unfortunately, as seen in the Reddit examples we found, some users are unable to regain their lost accounts and networks despite concerted efforts \cite{berger2015isis}.
This issue reflects a usability and access problem, but also could reflect a lack of institutional incentives to provide users who have been wrongfully labeled as bots with an accessible recourse to regain their accounts.

\subsection{\revision{Transparency}: Communicating Responsibility}
To increase transparency and reduce user frustration, social media platforms need to clearly communicate the definitions of acceptable bot behavior. To do so, they must first acknowledge that there exist good bots, such as moderation bots on Reddit \cite{kiene2020uses}, and provide guidelines on the types of bots and behaviors that are allowed on the platform. X vaguely acknowledges this by writing ``may post automated posts based on sources of outside information -- such as an RSS feed, weather data, etc''. However, it is subjective what a ``good'' bot is, or how ``good'' a bot is, but such decisions could be made in tandem with research institutions. A research direction is to profile the uses and analyze the impact of different types of automation. Further, more human-centered research needs to be conducted to aid the platforms in differentiating between humans having bot-like behavior and actual bots. Such research will help avoid false positives where users are incorrectly flagged as bots while performing bot-like behaviors, and also provide appropriate recourse to regain accounts.

In terms of system transparency, while it is desirable to communicate the decision path of bot detection algorithms, because doing so enables greater user understanding and trust in algorithmic governance, bot operators can take advantage of this fact to eventually bypass the algorithms, rendering them ineffective. This occurs because revealing the internal logic or key features used for bot detection can enable adversarial actors to identify exploitable \revision{weaknesses}, reverse-engineer feature weights or adapt their behaviors or design new bots that deliberately avoid these detection cues \cite{cresci2017paradigm}. Therefore, techniques need to be devised to communicate the reasons behind labeling an account as a bot, yet not give away the secrets of the algorithms so they cannot be manipulated for nefarious use. One method to be strategic in this communication is to leverage meta-explanations and extract examples by providing aggregated model comparisons and results, and avoid any explanation through feature importance attributions \cite{speith2022review}. Research in human-computer interaction and communication studies could assist in evaluating the best communication strategies that satisfy users' needs to understand why a specific account was flagged as a bot/human through broader patterns, without being specific in how the algorithm works, which could aid algorithmic evasion.

\section{Conclusion}
\label{sec:conclusion}
Social media bot detection algorithms do not operate in isolation, but instead operate in complex socio-technical environments alongside human users. The classifications they produce and the decisions enacted from those results can have an impact on other human users. While the current line of research to improve bot detection algorithms in response to the changing social media landscape is a vital advancement, it is equally important that these systems be designed with attention to human and societal values such as fairness, accountability, transparency, and also give respect to digital autonomy \cite{10.1145/3531146.3533097}. Embedding ethical reflection into each stage of system development ensures that detection efforts strengthen, rather than undermine, the trust, equity and the integrity of the online discourse.

This article summarizes ethical considerations in developing and deploying social bot detection in terms of the \textbf{fairness} of the training dataset, the \textbf{accountability} of the algorithmic development, and the \textbf{transparency} of the bot agent usage. With respect to each of these three pillars, we have proposed and elaborated on research directions, namely: developing diverse datasets, correcting algorithmic prejudices and engineering responsibility.
Though we find this framework productive as a first step in considering the ethics of bot detection, future work should examine the design and, crucially, usage and institutional issues we describe in more detail.
For example, how do policies around bot identification and subsequent action taken against bot accounts align with institutional notions of ``good'' and ``bad'' behavior on different platforms? How might appeal processes or transparency standards be implemented without enabling system evasion?

This article is a primer for examining ethical issues, their challenges, and potential research directions within the development of bot detection algorithms. The pursuit of accurate bot detection must be \revision{balanced} with a civic duty to uphold fairness and user rights.
We hope that this article inspires researchers, social media companies, and regulatory bodies alike towards pursuing an ethical balance in their quest to maintain the health of our social media systems through bot detection algorithms by protecting platform integrity but also reinforcing the ethical foundations of digital society.

\begin{acks}
This material is based upon work supported by the Knight Foundation, Office of Naval Research (Bothunter, N000141812108) and Scalable Technologies for Social Cybersecurity/ARMY (W911NF20D0002). The views and conclusions contained in this document are those of the authors and should not be interpreted as representing the official policies, either expressed or implied, of the Knight Foundation, Office of Naval Research, the US Army, or the U.S. Government.
\end{acks}

\bibliographystyle{ACM-Reference-Format}
\bibliography{references}


\begin{thebibliography}{170}


\ifx \showCODEN    \undefined \def \showCODEN     #1{\unskip}     \fi
\ifx \showISBNx    \undefined \def \showISBNx     #1{\unskip}     \fi
\ifx \showISBNxiii \undefined \def \showISBNxiii  #1{\unskip}     \fi
\ifx \showISSN     \undefined \def \showISSN      #1{\unskip}     \fi
\ifx \showLCCN     \undefined \def \showLCCN      #1{\unskip}     \fi
\ifx \shownote     \undefined \def \shownote      #1{#1}          \fi
\ifx \showarticletitle \undefined \def \showarticletitle #1{#1}   \fi
\ifx \showURL      \undefined \def \showURL       {\relax}        \fi
\providecommand\bibfield[2]{#2}
\providecommand\bibinfo[2]{#2}
\providecommand\natexlab[1]{#1}
\providecommand\showeprint[2][]{arXiv:#2}

\bibitem[Ahmad et~al\mbox{.}(2020)]%
        {ahmad_fairness_2020}
\bibfield{author}{\bibinfo{person}{Muhammad~Aurangzeb Ahmad}, \bibinfo{person}{Ankur Teredesai}, {and} \bibinfo{person}{Carly Eckert}.} \bibinfo{year}{2020}\natexlab{}.
\newblock \showarticletitle{Fairness, accountability, transparency in {AI} at scale: lessons from national programs}. In \bibinfo{booktitle}{\emph{Proceedings of the 2020 {Conference} on {Fairness}, {Accountability}, and {Transparency}}} \emph{(\bibinfo{series}{{FAT}* '20})}. \bibinfo{publisher}{Association for Computing Machinery}, \bibinfo{address}{New York, NY, USA}, \bibinfo{pages}{690}.
\newblock
\href{https://doi.org/10.1145/3351095.3375690}{doi:\nolinkurl{10.1145/3351095.3375690}}


\bibitem[Alameda-Pineda et~al\mbox{.}(2020)]%
        {alameda2020fate}
\bibfield{author}{\bibinfo{person}{Xavier Alameda-Pineda}, \bibinfo{person}{Miriam Redi}, \bibinfo{person}{Jahna Otterbacher}, \bibinfo{person}{Nicu Sebe}, {and} \bibinfo{person}{Shih-Fu Chang}.} \bibinfo{year}{2020}\natexlab{}.
\newblock \showarticletitle{FATE/MM 20: 2nd international workshop on fairness, accountability, transparency and ethics in multimedia}. In \bibinfo{booktitle}{\emph{Proceedings of the 28th ACM International Conference on Multimedia}}. \bibinfo{pages}{4761--4762}.
\newblock


\bibitem[Ali~Alhosseini et~al\mbox{.}(2019)]%
        {ali_alhosseini_detect_2019}
\bibfield{author}{\bibinfo{person}{Seyed Ali~Alhosseini}, \bibinfo{person}{Raad Bin~Tareaf}, \bibinfo{person}{Pejman Najafi}, {and} \bibinfo{person}{Christoph Meinel}.} \bibinfo{year}{2019}\natexlab{}.
\newblock \showarticletitle{Detect Me If You Can: Spam Bot Detection Using Inductive Representation Learning}. In \bibinfo{booktitle}{\emph{Companion Proceedings of The 2019 World Wide Web Conference}} (San Francisco {USA}, 2019-05-13). \bibinfo{publisher}{{ACM}}, \bibinfo{pages}{148--153}.
\newblock
\showISBNx{978-1-4503-6675-5}
\href{https://doi.org/10.1145/3308560.3316504}{doi:\nolinkurl{10.1145/3308560.3316504}}


\bibitem[Alibašić(2023)]%
        {alibasic_developing_2023}
\bibfield{author}{\bibinfo{person}{Haris Alibašić}.} \bibinfo{year}{2023}\natexlab{}.
\newblock \showarticletitle{Developing an {Ethical} {Framework} for {Responsible} {Artificial} {Intelligence} ({AI}) and {Machine} {Learning} ({ML}) {Applications} in {Cryptocurrency} {Trading}: {A} {Consequentialism} {Ethics} {Analysis}}.
\newblock \bibinfo{journal}{\emph{FinTech}} \bibinfo{volume}{2}, \bibinfo{number}{3} (\bibinfo{date}{July} \bibinfo{year}{2023}), \bibinfo{pages}{430--443}.
\newblock
\href{https://doi.org/10.3390/fintech2030024}{doi:\nolinkurl{10.3390/fintech2030024}}


\bibitem[Alsaleh et~al\mbox{.}(2019)]%
        {alsaleh2019cross}
\bibfield{author}{\bibinfo{person}{Dhoha~A Alsaleh}, \bibinfo{person}{Michael~T Elliott}, \bibinfo{person}{Frank~Q Fu}, {and} \bibinfo{person}{Ramendra Thakur}.} \bibinfo{year}{2019}\natexlab{}.
\newblock \showarticletitle{Cross-cultural differences in the adoption of social media}.
\newblock \bibinfo{journal}{\emph{Journal of Research in Interactive Marketing}} \bibinfo{volume}{13}, \bibinfo{number}{1} (\bibinfo{year}{2019}), \bibinfo{pages}{119--140}.
\newblock


\bibitem[Bender et~al\mbox{.}(2021)]%
        {bender_dangers_2021}
\bibfield{author}{\bibinfo{person}{Emily~M. Bender}, \bibinfo{person}{Timnit Gebru}, \bibinfo{person}{Angelina McMillan-Major}, {and} \bibinfo{person}{Shmargaret Shmitchell}.} \bibinfo{year}{2021}\natexlab{}.
\newblock \showarticletitle{On the {Dangers} of {Stochastic} {Parrots}: {Can} {Language} {Models} {Be} {Too} {Big}?}. In \bibinfo{booktitle}{\emph{Proceedings of the 2021 {ACM} {Conference} on {Fairness}, {Accountability}, and {Transparency}}} \emph{(\bibinfo{series}{{FAccT} '21})}. \bibinfo{publisher}{Association for Computing Machinery}, \bibinfo{address}{New York, NY, USA}, \bibinfo{pages}{610--623}.
\newblock
\href{https://doi.org/10.1145/3442188.3445922}{doi:\nolinkurl{10.1145/3442188.3445922}}


\bibitem[Benevenuto et~al\mbox{.}({[n.\,d.]})]%
        {benevenuto_detecting_nodate}
\bibfield{author}{\bibinfo{person}{Fabrıcio Benevenuto}, \bibinfo{person}{Gabriel Magno}, \bibinfo{person}{Tiago Rodrigues}, {and} \bibinfo{person}{Virgılio Almeida}.} \bibinfo{year}{[n.\,d.]}\natexlab{}.
\newblock \showarticletitle{Detecting Spammers on Twitter}.
\newblock  (\bibinfo{year}{[n.\,d.]}).
\newblock


\bibitem[Berger and Morgan(2015)]%
        {berger2015isis}
\bibfield{author}{\bibinfo{person}{Jonathon~M Berger} {and} \bibinfo{person}{Jonathon Morgan}.} \bibinfo{year}{2015}\natexlab{}.
\newblock \showarticletitle{The ISIS Twitter Census: Defining and describing the population of ISIS supporters on Twitter}.
\newblock  (\bibinfo{year}{2015}).
\newblock


\bibitem[Beskow and Carley(2019)]%
        {beskow_its_2019}
\bibfield{author}{\bibinfo{person}{David~M. Beskow} {and} \bibinfo{person}{Kathleen~M. Carley}.} \bibinfo{year}{2019}\natexlab{}.
\newblock \showarticletitle{Its All in a Name: Detecting and Labeling Bots by Their Name}.
\newblock  \bibinfo{volume}{25}, \bibinfo{number}{1} (\bibinfo{year}{2019}), \bibinfo{pages}{24--35}.
\newblock
\showISSN{1381-298X, 1572-9346}
\href{https://doi.org/10.1007/s10588-018-09290-1}{doi:\nolinkurl{10.1007/s10588-018-09290-1}}
\showeprint[arxiv]{1812.05932 [cs]}


\bibitem[Beskow and Carley(2018a)]%
        {beskow2018bot}
\bibfield{author}{\bibinfo{person}{David~M Beskow} {and} \bibinfo{person}{Kathleen~M Carley}.} \bibinfo{year}{2018}\natexlab{a}.
\newblock \showarticletitle{Bot-hunter: a tiered approach to detecting \& characterizing automated activity on twitter}. In \bibinfo{booktitle}{\emph{Conference paper. SBP-BRiMS: International conference on social computing, behavioral-cultural modeling and prediction and behavior representation in modeling and simulation}}, Vol.~\bibinfo{volume}{3}.
\newblock


\bibitem[Beskow and Carley(2018b)]%
        {beskow_bot-hunter_nodate}
\bibfield{author}{\bibinfo{person}{David~M Beskow} {and} \bibinfo{person}{Kathleen~M Carley}.} \bibinfo{year}{2018}\natexlab{b}.
\newblock \showarticletitle{Bot-hunter: a tiered approach to detecting \& characterizing automated activity on twitter}. In \bibinfo{booktitle}{\emph{Conference paper. SBP-BRiMS: International conference on social computing, behavioral-cultural modeling and prediction and behavior representation in modeling and simulation}}, Vol.~\bibinfo{volume}{3}.
\newblock


\bibitem[Bessi and Ferrara(2016)]%
        {bessi2016social}
\bibfield{author}{\bibinfo{person}{Alessandro Bessi} {and} \bibinfo{person}{Emilio Ferrara}.} \bibinfo{year}{2016}\natexlab{}.
\newblock \showarticletitle{Social bots distort the 2016 US Presidential election online discussion}.
\newblock \bibinfo{journal}{\emph{First monday}} \bibinfo{volume}{21}, \bibinfo{number}{11-7} (\bibinfo{year}{2016}).
\newblock


\bibitem[Beutel et~al\mbox{.}(2013)]%
        {beutel_copycatch_2013}
\bibfield{author}{\bibinfo{person}{Alex Beutel}, \bibinfo{person}{Wanhong Xu}, \bibinfo{person}{Venkatesan Guruswami}, \bibinfo{person}{Christopher Palow}, {and} \bibinfo{person}{Christos Faloutsos}.} \bibinfo{year}{2013}\natexlab{}.
\newblock \showarticletitle{{CopyCatch}: stopping group attacks by spotting lockstep behavior in social networks}. In \bibinfo{booktitle}{\emph{Proceedings of the 22nd international conference on World Wide Web}} (Rio de Janeiro Brazil, 2013-05-13). \bibinfo{publisher}{{ACM}}, \bibinfo{pages}{119--130}.
\newblock
\showISBNx{978-1-4503-2035-1}
\href{https://doi.org/10.1145/2488388.2488400}{doi:\nolinkurl{10.1145/2488388.2488400}}


\bibitem[Bogina et~al\mbox{.}(2022)]%
        {bogina2022educating}
\bibfield{author}{\bibinfo{person}{Veronika Bogina}, \bibinfo{person}{Alan Hartman}, \bibinfo{person}{Tsvi Kuflik}, {and} \bibinfo{person}{Avital Shulner-Tal}.} \bibinfo{year}{2022}\natexlab{}.
\newblock \showarticletitle{Educating Software and AI Stakeholders About Algorithmic Fairness, Accountability, Transparency and Ethics}.
\newblock \bibinfo{journal}{\emph{International Journal of Artificial Intelligence in Education}} \bibinfo{volume}{32}, \bibinfo{number}{3} (\bibinfo{year}{2022}), \bibinfo{pages}{808--833}.
\newblock


\bibitem[Bolukbasi et~al\mbox{.}(2016)]%
        {bolukbasi2016man}
\bibfield{author}{\bibinfo{person}{Tolga Bolukbasi}, \bibinfo{person}{Kai-Wei Chang}, \bibinfo{person}{James~Y Zou}, \bibinfo{person}{Venkatesh Saligrama}, {and} \bibinfo{person}{Adam~T Kalai}.} \bibinfo{year}{2016}\natexlab{}.
\newblock \showarticletitle{Man is to computer programmer as woman is to homemaker? debiasing word embeddings}. In \bibinfo{booktitle}{\emph{30th {Conference} on {Neural} {Information} {Processing} {Systems} ({NIPS} 2016)}}. \bibinfo{pages}{4349--4357}.
\newblock


\bibitem[Boshmaf et~al\mbox{.}(2013)]%
        {boshmaf_design_2013}
\bibfield{author}{\bibinfo{person}{Yazan Boshmaf}, \bibinfo{person}{Ildar Muslukhov}, \bibinfo{person}{Konstantin Beznosov}, {and} \bibinfo{person}{Matei Ripeanu}.} \bibinfo{year}{2013}\natexlab{}.
\newblock \showarticletitle{Design and analysis of a social botnet}.
\newblock  \bibinfo{volume}{57}, \bibinfo{number}{2} (\bibinfo{year}{2013}), \bibinfo{pages}{556--578}.
\newblock
\showISSN{13891286}
\href{https://doi.org/10.1016/j.comnet.2012.06.006}{doi:\nolinkurl{10.1016/j.comnet.2012.06.006}}


\bibitem[Brinkmann et~al\mbox{.}(2023)]%
        {brinkmann2023machine}
\bibfield{author}{\bibinfo{person}{Levin Brinkmann}, \bibinfo{person}{Fabian Baumann}, \bibinfo{person}{Jean-Fran{\c{c}}ois Bonnefon}, \bibinfo{person}{Maxime Derex}, \bibinfo{person}{Thomas~F M{\"u}ller}, \bibinfo{person}{Anne-Marie Nussberger}, \bibinfo{person}{Agnieszka Czaplicka}, \bibinfo{person}{Alberto Acerbi}, \bibinfo{person}{Thomas~L Griffiths}, \bibinfo{person}{Joseph Henrich}, {et~al\mbox{.}}} \bibinfo{year}{2023}\natexlab{}.
\newblock \showarticletitle{Machine culture}.
\newblock \bibinfo{journal}{\emph{Nature Human Behaviour}} \bibinfo{volume}{7}, \bibinfo{number}{11} (\bibinfo{year}{2023}), \bibinfo{pages}{1855--1868}.
\newblock


\bibitem[Buolamwini(2017)]%
        {buolamwini2017gender}
\bibfield{author}{\bibinfo{person}{Joy~Adowaa Buolamwini}.} \bibinfo{year}{2017}\natexlab{}.
\newblock \emph{\bibinfo{title}{Gender shades: intersectional phenotypic and demographic evaluation of face datasets and gender classifiers}}.
\newblock \bibinfo{thesistype}{Ph.\,D. Dissertation}. \bibinfo{school}{Massachusetts Institute of Technology}.
\newblock


\bibitem[Cai et~al\mbox{.}(2017)]%
        {cai_behavior_2017}
\bibfield{author}{\bibinfo{person}{Chiyu Cai}, \bibinfo{person}{Linjing Li}, {and} \bibinfo{person}{Daniel Zengi}.} \bibinfo{year}{2017}\natexlab{}.
\newblock \showarticletitle{Behavior enhanced deep bot detection in social media}. In \bibinfo{booktitle}{\emph{2017 {IEEE} International Conference on Intelligence and Security Informatics ({ISI})}} (2017-07). \bibinfo{pages}{128--130}.
\newblock
\href{https://doi.org/10.1109/ISI.2017.8004887}{doi:\nolinkurl{10.1109/ISI.2017.8004887}}


\bibitem[Cao et~al\mbox{.}(2014)]%
        {cao_uncovering_2014}
\bibfield{author}{\bibinfo{person}{Qiang Cao}, \bibinfo{person}{Xiaowei Yang}, \bibinfo{person}{Jieqi Yu}, {and} \bibinfo{person}{Christopher Palow}.} \bibinfo{year}{2014}\natexlab{}.
\newblock \showarticletitle{Uncovering Large Groups of Active Malicious Accounts in Online Social Networks}. In \bibinfo{booktitle}{\emph{Proceedings of the 2014 {ACM} {SIGSAC} Conference on Computer and Communications Security}} (Scottsdale Arizona {USA}, 2014-11-03). \bibinfo{publisher}{{ACM}}, \bibinfo{pages}{477--488}.
\newblock
\showISBNx{978-1-4503-2957-6}
\href{https://doi.org/10.1145/2660267.2660269}{doi:\nolinkurl{10.1145/2660267.2660269}}


\bibitem[Carley(2020)]%
        {carley2020social}
\bibfield{author}{\bibinfo{person}{Kathleen~M Carley}.} \bibinfo{year}{2020}\natexlab{}.
\newblock \showarticletitle{Social cybersecurity: an emerging science}.
\newblock \bibinfo{journal}{\emph{Computational and mathematical organization theory}} \bibinfo{volume}{26}, \bibinfo{number}{4} (\bibinfo{year}{2020}), \bibinfo{pages}{365--381}.
\newblock


\bibitem[Chakraborti and Kambhampati(2019)]%
        {chakraborti2019can}
\bibfield{author}{\bibinfo{person}{Tathagata Chakraborti} {and} \bibinfo{person}{Subbarao Kambhampati}.} \bibinfo{year}{2019}\natexlab{}.
\newblock \showarticletitle{(When) can AI bots lie?}. In \bibinfo{booktitle}{\emph{Proceedings of the 2019 AAAI/ACM Conference on AI, Ethics, and Society}}. \bibinfo{pages}{53--59}.
\newblock


\bibitem[Chang et~al\mbox{.}(2021)]%
        {chang2021social}
\bibfield{author}{\bibinfo{person}{Ho-Chun~Herbert Chang}, \bibinfo{person}{Emily Chen}, \bibinfo{person}{Meiqing Zhang}, \bibinfo{person}{Goran Muric}, {and} \bibinfo{person}{Emilio Ferrara}.} \bibinfo{year}{2021}\natexlab{}.
\newblock \showarticletitle{Social bots and social media manipulation in 2020: the year in review}.
\newblock In \bibinfo{booktitle}{\emph{Handbook of Computational Social Science, Volume 1}}. \bibinfo{publisher}{Routledge}, \bibinfo{pages}{304--323}.
\newblock


\bibitem[Chavoshi et~al\mbox{.}(2016)]%
        {chavoshi_debot_2016}
\bibfield{author}{\bibinfo{person}{Nikan Chavoshi}, \bibinfo{person}{Hossein Hamooni}, {and} \bibinfo{person}{Abdullah Mueen}.} \bibinfo{year}{2016}\natexlab{}.
\newblock \showarticletitle{{DeBot}: Twitter Bot Detection via Warped Correlation}. In \bibinfo{booktitle}{\emph{2016 {IEEE} 16th International Conference on Data Mining ({ICDM})}} (2016-12). \bibinfo{pages}{817--822}.
\newblock
\href{https://doi.org/10.1109/ICDM.2016.0096}{doi:\nolinkurl{10.1109/ICDM.2016.0096}}
\newblock
\shownote{{ISSN}: 2374-8486}.


\bibitem[Chavoshi et~al\mbox{.}(2017)]%
        {chavoshi_temporal_2017}
\bibfield{author}{\bibinfo{person}{Nikan Chavoshi}, \bibinfo{person}{Hossein Hamooni}, {and} \bibinfo{person}{Abdullah Mueen}.} \bibinfo{year}{2017}\natexlab{}.
\newblock \showarticletitle{Temporal Patterns in Bot Activities}. In \bibinfo{booktitle}{\emph{Proceedings of the 26th International Conference on World Wide Web Companion - {WWW} '17 Companion}} (Perth, Australia, 2017). \bibinfo{publisher}{{ACM} Press}, \bibinfo{pages}{1601--1606}.
\newblock
\showISBNx{978-1-4503-4914-7}
\href{https://doi.org/10.1145/3041021.3051114}{doi:\nolinkurl{10.1145/3041021.3051114}}


\bibitem[Chavoshi et~al\mbox{.}(2016)]%
        {chavoshi2016debot}
\bibfield{author}{\bibinfo{person}{Nikan Chavoshi}, \bibinfo{person}{Hossein Hamooni}, {and} \bibinfo{person}{Abdullah Mueen}.} \bibinfo{year}{2016}\natexlab{}.
\newblock \showarticletitle{Debot: Twitter bot detection via warped correlation.}. In \bibinfo{booktitle}{\emph{Icdm}}, Vol.~\bibinfo{volume}{18}. \bibinfo{pages}{28--65}.
\newblock


\bibitem[Chen et~al\mbox{.}(2022)]%
        {chen2022election2020}
\bibfield{author}{\bibinfo{person}{Emily Chen}, \bibinfo{person}{Ashok Deb}, {and} \bibinfo{person}{Emilio Ferrara}.} \bibinfo{year}{2022}\natexlab{}.
\newblock \showarticletitle{\# Election2020: The first public Twitter dataset on the 2020 US Presidential election}.
\newblock \bibinfo{journal}{\emph{Journal of Computational Social Science}} \bibinfo{volume}{5}, \bibinfo{number}{1} (\bibinfo{year}{2022}), \bibinfo{pages}{1--18}.
\newblock


\bibitem[Chen et~al\mbox{.}(2024)]%
        {chen_cacl_2024}
\bibfield{author}{\bibinfo{person}{Sirry Chen}, \bibinfo{person}{Shuo Feng}, \bibinfo{person}{Songsong Liang}, \bibinfo{person}{Chen-Chen Zong}, \bibinfo{person}{Jing Li}, {and} \bibinfo{person}{Piji Li}.} \bibinfo{year}{2024}\natexlab{}.
\newblock \bibinfo{title}{{CACL}: Community-Aware Heterogeneous Graph Contrastive Learning for Social Media Bot Detection}.
\newblock
\href{https://doi.org/10.48550/arXiv.2405.10558}{doi:\nolinkurl{10.48550/arXiv.2405.10558}}
\showeprint[arxiv]{2405.10558 [cs]}


\bibitem[Chu et~al\mbox{.}(2012)]%
        {chu_detecting_2012}
\bibfield{author}{\bibinfo{person}{Zi Chu}, \bibinfo{person}{Steven Gianvecchio}, \bibinfo{person}{Haining Wang}, {and} \bibinfo{person}{Sushil Jajodia}.} \bibinfo{year}{2012}\natexlab{}.
\newblock \showarticletitle{Detecting Automation of Twitter Accounts: Are You a Human, Bot, or Cyborg?}
\newblock  \bibinfo{volume}{9}, \bibinfo{number}{6} (\bibinfo{year}{2012}), \bibinfo{pages}{811--824}.
\newblock
\showISSN{1941-0018}
\href{https://doi.org/10.1109/TDSC.2012.75}{doi:\nolinkurl{10.1109/TDSC.2012.75}}
\newblock
\shownote{Conference Name: {IEEE} Transactions on Dependable and Secure Computing}.


\bibitem[{Coronavirus Conspiracy}(2025)]%
        {CoronavirusCon3}
\bibfield{author}{\bibinfo{person}{{Coronavirus Conspiracy}}.} \bibinfo{year}{2025}\natexlab{}.
\newblock \bibinfo{title}{CoronavirusCon3 on X}.
\newblock \bibinfo{howpublished}{\url{https://x.com/CoronavirusCon3}}.
\newblock
\newblock
\shownote{Retrieved April 2025}.


\bibitem[{CORONAVIRUSUPDATE}(2025)]%
        {covid19digest1}
\bibfield{author}{\bibinfo{person}{{CORONAVIRUSUPDATE}}.} \bibinfo{year}{2025}\natexlab{}.
\newblock \bibinfo{title}{CoronavirusCon3 on X}.
\newblock \bibinfo{howpublished}{\url{https://x.com/COVID19digest1}}.
\newblock
\newblock
\shownote{Retrieved April 2025}.


\bibitem[Cresci et~al\mbox{.}(2017)]%
        {cresci2017paradigm}
\bibfield{author}{\bibinfo{person}{Stefano Cresci}, \bibinfo{person}{Roberto Di~Pietro}, \bibinfo{person}{Marinella Petrocchi}, \bibinfo{person}{Angelo Spognardi}, {and} \bibinfo{person}{Maurizio Tesconi}.} \bibinfo{year}{2017}\natexlab{}.
\newblock \showarticletitle{The paradigm-shift of social spambots: Evidence, theories, and tools for the arms race}. In \bibinfo{booktitle}{\emph{Proceedings of the 26th international conference on world wide web companion}}. \bibinfo{pages}{963--972}.
\newblock


\bibitem[Cresci et~al\mbox{.}(2018)]%
        {cresci2018fake}
\bibfield{author}{\bibinfo{person}{Stefano Cresci}, \bibinfo{person}{Fabrizio Lillo}, \bibinfo{person}{Daniele Regoli}, \bibinfo{person}{Serena Tardelli}, {and} \bibinfo{person}{Maurizio Tesconi}.} \bibinfo{year}{2018}\natexlab{}.
\newblock \showarticletitle{FAKE: Evidence of spam and bot activity in stock microblogs on Twitter}. In \bibinfo{booktitle}{\emph{Twelfth international AAAI conference on web and social media}}.
\newblock


\bibitem[Danaditya et~al\mbox{.}(2022)]%
        {danaditya2022curious}
\bibfield{author}{\bibinfo{person}{Adya Danaditya}, \bibinfo{person}{Lynnette Hui~Xian Ng}, {and} \bibinfo{person}{Kathleen~M Carley}.} \bibinfo{year}{2022}\natexlab{}.
\newblock \showarticletitle{From curious hashtags to polarized effect: profiling coordinated actions in indonesian twitter discourse}.
\newblock \bibinfo{journal}{\emph{Social Network Analysis and Mining}} \bibinfo{volume}{12}, \bibinfo{number}{1} (\bibinfo{year}{2022}), \bibinfo{pages}{105}.
\newblock


\bibitem[Dehghan et~al\mbox{.}(2023)]%
        {dehghan_detecting_2023}
\bibfield{author}{\bibinfo{person}{Ashkan Dehghan}, \bibinfo{person}{Kinga Siuta}, \bibinfo{person}{Agata Skorupka}, \bibinfo{person}{Akshat Dubey}, \bibinfo{person}{Andrei Betlen}, \bibinfo{person}{David Miller}, \bibinfo{person}{Wei Xu}, \bibinfo{person}{Bogumił Kamiński}, {and} \bibinfo{person}{Paweł Prałat}.} \bibinfo{year}{2023}\natexlab{}.
\newblock \showarticletitle{Detecting bots in social-networks using node and structural embeddings}.
\newblock  \bibinfo{volume}{10}, \bibinfo{number}{1} (\bibinfo{year}{2023}), \bibinfo{pages}{119}.
\newblock
\showISSN{2196-1115}
\href{https://doi.org/10.1186/s40537-023-00796-3}{doi:\nolinkurl{10.1186/s40537-023-00796-3}}


\bibitem[Dickerson et~al\mbox{.}(2014)]%
        {dickerson_using_2014}
\bibfield{author}{\bibinfo{person}{John~P. Dickerson}, \bibinfo{person}{Vadim Kagan}, {and} \bibinfo{person}{V.S. Subrahmanian}.} \bibinfo{year}{2014}\natexlab{}.
\newblock \showarticletitle{Using sentiment to detect bots on Twitter: Are humans more opinionated than bots?}. In \bibinfo{booktitle}{\emph{2014 {IEEE}/{ACM} International Conference on Advances in Social Networks Analysis and Mining ({ASONAM} 2014)}} (2014-08). \bibinfo{pages}{620--627}.
\newblock
\href{https://doi.org/10.1109/ASONAM.2014.6921650}{doi:\nolinkurl{10.1109/ASONAM.2014.6921650}}


\bibitem[Diesner et~al\mbox{.}(2017)]%
        {diesner2017proceedings}
\bibfield{author}{\bibinfo{person}{Jana Diesner}, \bibinfo{person}{Elena Ferrari}, {and} \bibinfo{person}{Guandong Xu}.} \bibinfo{year}{2017}\natexlab{}.
\newblock \bibinfo{booktitle}{\emph{Proceedings of the 2017 IEEE/ACM International Conference on Advances in Social Networks Analysis and Mining 2017}}.
\newblock


\bibitem[Ding et~al\mbox{.}(2019)]%
        {ding2019saliency}
\bibfield{author}{\bibinfo{person}{Shuoyang Ding}, \bibinfo{person}{Hainan Xu}, {and} \bibinfo{person}{Philipp Koehn}.} \bibinfo{year}{2019}\natexlab{}.
\newblock \showarticletitle{Saliency-driven Word Alignment Interpretation for Neural Machine Translation}. In \bibinfo{booktitle}{\emph{Proceedings of the Fourth Conference on Machine Translation (Volume 1: Research Papers)}}. \bibinfo{pages}{1--12}.
\newblock


\bibitem[Douglas(2012)]%
        {douglas2012social}
\bibfield{author}{\bibinfo{person}{Deborah~G Douglas}.} \bibinfo{year}{2012}\natexlab{}.
\newblock \bibinfo{booktitle}{\emph{The social construction of technological systems, anniversary edition: New directions in the sociology and history of technology}}.
\newblock \bibinfo{publisher}{MIT press}.
\newblock


\bibitem[Dubey and Ghosh(2025)]%
        {dubey2025digital}
\bibfield{author}{\bibinfo{person}{Riya Dubey} {and} \bibinfo{person}{Ashima Ghosh}.} \bibinfo{year}{2025}\natexlab{}.
\newblock \showarticletitle{Digital Governance and Free Speech in India by Navigating Algorithmic Censorship Policy Frameworks and Public Dissent}.
\newblock \bibinfo{journal}{\emph{Ianna Journal of Interdisciplinary Studies, ISSN (O): 2735-9891, ISSN (P): 2735-9883}} \bibinfo{volume}{7}, \bibinfo{number}{1} (\bibinfo{year}{2025}), \bibinfo{pages}{454--470}.
\newblock


\bibitem[Dukic et~al\mbox{.}(2020)]%
        {dukic_are_2020}
\bibfield{author}{\bibinfo{person}{David Dukic}, \bibinfo{person}{Dominik Keca}, {and} \bibinfo{person}{Dominik Stipic}.} \bibinfo{year}{2020}\natexlab{}.
\newblock \showarticletitle{Are You Human? Detecting Bots on Twitter Using {BERT}}. In \bibinfo{booktitle}{\emph{2020 {IEEE} 7th International Conference on Data Science and Advanced Analytics ({DSAA})}} (sydney, Australia, 2020-10). \bibinfo{publisher}{{IEEE}}, \bibinfo{pages}{631--636}.
\newblock
\showISBNx{978-1-72818-206-3}
\href{https://doi.org/10.1109/DSAA49011.2020.00089}{doi:\nolinkurl{10.1109/DSAA49011.2020.00089}}


\bibitem[Dwork et~al\mbox{.}(2018)]%
        {dwork_decoupled_2018}
\bibfield{author}{\bibinfo{person}{Cynthia Dwork}, \bibinfo{person}{Nicole Immorlica}, \bibinfo{person}{Adam~Tauman Kalai}, {and} \bibinfo{person}{Max Leiserson}.} \bibinfo{year}{2018}\natexlab{}.
\newblock \showarticletitle{Decoupled {Classifiers} for {Group}-{Fair} and {Efficient} {Machine} {Learning}}. In \bibinfo{booktitle}{\emph{Proceedings of the 1st {Conference} on {Fairness}, {Accountability} and {Transparency}}}. \bibinfo{publisher}{PMLR}, \bibinfo{pages}{119--133}.
\newblock
\urldef\tempurl%
\url{https://proceedings.mlr.press/v81/dwork18a.html}
\showURL{%
\tempurl}


\bibitem[Efthimion et~al\mbox{.}(2018)]%
        {efthimion_supervised_2018}
\bibfield{author}{\bibinfo{person}{Phillip~George Efthimion}, \bibinfo{person}{Scott Payne}, {and} \bibinfo{person}{Nicholas Proferes}.} \bibinfo{year}{2018}\natexlab{}.
\newblock \showarticletitle{Supervised Machine Learning Bot Detection Techniques to Identify Social Twitter Bots}.
\newblock  \bibinfo{volume}{1}, \bibinfo{number}{2} (\bibinfo{year}{2018}).
\newblock


\bibitem[Eshraqi et~al\mbox{.}(2015)]%
        {eshraqi_detecting_2015}
\bibfield{author}{\bibinfo{person}{Nasim Eshraqi}, \bibinfo{person}{Mehrdad Jalali}, {and} \bibinfo{person}{Mohammad~Hossein Moattar}.} \bibinfo{year}{2015}\natexlab{}.
\newblock \showarticletitle{Detecting spam tweets in Twitter using a data stream clustering algorithm}. In \bibinfo{booktitle}{\emph{2015 International Congress on Technology, Communication and Knowledge ({ICTCK})}} (2015-11). \bibinfo{pages}{347--351}.
\newblock
\href{https://doi.org/10.1109/ICTCK.2015.7582694}{doi:\nolinkurl{10.1109/ICTCK.2015.7582694}}


\bibitem[Feng et~al\mbox{.}(2022)]%
        {feng_language-agnostic_2022}
\bibfield{author}{\bibinfo{person}{Fangxiaoyu Feng}, \bibinfo{person}{Yinfei Yang}, \bibinfo{person}{Daniel Cer}, \bibinfo{person}{Naveen Arivazhagan}, {and} \bibinfo{person}{Wei Wang}.} \bibinfo{year}{2022}\natexlab{}.
\newblock \showarticletitle{Language-agnostic {BERT} {Sentence} {Embedding}}. In \bibinfo{booktitle}{\emph{Proceedings of the 60th {Annual} {Meeting} of the {Association} for {Computational} {Linguistics} ({Volume} 1: {Long} {Papers})}}. \bibinfo{publisher}{Association for Computational Linguistics}, \bibinfo{address}{Dublin, Ireland}, \bibinfo{pages}{878--891}.
\newblock
\href{https://doi.org/10.18653/v1/2022.acl-long.62}{doi:\nolinkurl{10.18653/v1/2022.acl-long.62}}


\bibitem[Feng et~al\mbox{.}(2021a)]%
        {feng_heterogeneity-aware_2021}
\bibfield{author}{\bibinfo{person}{Shangbin Feng}, \bibinfo{person}{Zhaoxuan Tan}, \bibinfo{person}{Rui Li}, {and} \bibinfo{person}{Minnan Luo}.} \bibinfo{year}{2021}\natexlab{a}.
\newblock \bibinfo{title}{Heterogeneity-aware Twitter Bot Detection with Relational Graph Transformers}.
\newblock
\href{https://doi.org/10.48550/arXiv.2109.02927}{doi:\nolinkurl{10.48550/arXiv.2109.02927}}
\showeprint[arxiv]{2109.02927 [cs]}


\bibitem[Feng et~al\mbox{.}(2021c)]%
        {feng_satar_2021}
\bibfield{author}{\bibinfo{person}{Shangbin Feng}, \bibinfo{person}{Herun Wan}, \bibinfo{person}{Ningnan Wang}, \bibinfo{person}{Jundong Li}, {and} \bibinfo{person}{Minnan Luo}.} \bibinfo{year}{2021}\natexlab{c}.
\newblock \showarticletitle{{SATAR}: A Self-supervised Approach to Twitter Account Representation Learning and its Application in Bot Detection}. In \bibinfo{booktitle}{\emph{Proceedings of the 30th {ACM} International Conference on Information \& Knowledge Management}} (2021-10-26). \bibinfo{pages}{3808--3817}.
\newblock
\href{https://doi.org/10.1145/3459637.3481949}{doi:\nolinkurl{10.1145/3459637.3481949}}
\showeprint[arxiv]{2106.13089 [cs]}


\bibitem[Feng et~al\mbox{.}(2021)]%
        {feng2021twibot}
\bibfield{author}{\bibinfo{person}{Shangbin Feng}, \bibinfo{person}{Herun Wan}, \bibinfo{person}{Ningnan Wang}, \bibinfo{person}{Jundong Li}, {and} \bibinfo{person}{Minnan Luo}.} \bibinfo{year}{2021}\natexlab{}.
\newblock \showarticletitle{Twibot-20: A comprehensive twitter bot detection benchmark}. In \bibinfo{booktitle}{\emph{Proceedings of the 30th ACM international conference on information \& knowledge management}}. \bibinfo{pages}{4485--4494}.
\newblock


\bibitem[Feng et~al\mbox{.}(2021b)]%
        {feng_botrgcn_2021}
\bibfield{author}{\bibinfo{person}{Shangbin Feng}, \bibinfo{person}{Herun Wan}, \bibinfo{person}{Ningnan Wang}, {and} \bibinfo{person}{Minnan Luo}.} \bibinfo{year}{2021}\natexlab{b}.
\newblock \showarticletitle{{BotRGCN}: Twitter bot detection with relational graph convolutional networks}. In \bibinfo{booktitle}{\emph{Proceedings of the 2021 {IEEE}/{ACM} International Conference on Advances in Social Networks Analysis and Mining}} (Virtual Event Netherlands, 2021-11-08). \bibinfo{publisher}{{ACM}}, \bibinfo{pages}{236--239}.
\newblock
\showISBNx{978-1-4503-9128-3}
\href{https://doi.org/10.1145/3487351.3488336}{doi:\nolinkurl{10.1145/3487351.3488336}}


\bibitem[Feng et~al\mbox{.}(2024)]%
        {feng2024does}
\bibfield{author}{\bibinfo{person}{Shangbin Feng}, \bibinfo{person}{Herun Wan}, \bibinfo{person}{Ningnan Wang}, \bibinfo{person}{Zhaoxuan Tan}, \bibinfo{person}{Minnan Luo}, {and} \bibinfo{person}{Yulia Tsvetkov}.} \bibinfo{year}{2024}\natexlab{}.
\newblock \showarticletitle{What does the bot say? opportunities and risks of large language models in social media bot detection}.
\newblock \bibinfo{journal}{\emph{arXiv preprint arXiv:2402.00371}} (\bibinfo{year}{2024}).
\newblock


\bibitem[Ferrara et~al\mbox{.}(2016)]%
        {ferrara_rise_2016}
\bibfield{author}{\bibinfo{person}{Emilio Ferrara}, \bibinfo{person}{Onur Varol}, \bibinfo{person}{Clayton Davis}, \bibinfo{person}{Filippo Menczer}, {and} \bibinfo{person}{Alessandro Flammini}.} \bibinfo{year}{2016}\natexlab{}.
\newblock \showarticletitle{The Rise of Social Bots}.
\newblock  \bibinfo{volume}{59}, \bibinfo{number}{7} (\bibinfo{year}{2016}), \bibinfo{pages}{96--104}.
\newblock
\showISSN{0001-0782, 1557-7317}
\href{https://doi.org/10.1145/2818717}{doi:\nolinkurl{10.1145/2818717}}
\showeprint[arxiv]{1407.5225 [cs]}


\bibitem[Ferreira Dos~Santos et~al\mbox{.}(2019)]%
        {ferreira_dos_santos_uncovering_2019}
\bibfield{author}{\bibinfo{person}{Eric Ferreira Dos~Santos}, \bibinfo{person}{Danilo Carvalho}, \bibinfo{person}{Livia Ruback}, {and} \bibinfo{person}{Jonice Oliveira}.} \bibinfo{year}{2019}\natexlab{}.
\newblock \showarticletitle{Uncovering Social Media Bots: a Transparency-focused Approach}. In \bibinfo{booktitle}{\emph{Companion Proceedings of The 2019 World Wide Web Conference}} (San Francisco {USA}, 2019-05-13). \bibinfo{publisher}{{ACM}}, \bibinfo{pages}{545--552}.
\newblock
\showISBNx{978-1-4503-6675-5}
\href{https://doi.org/10.1145/3308560.3317599}{doi:\nolinkurl{10.1145/3308560.3317599}}


\bibitem[Fonseca~Abreu et~al\mbox{.}(2020)]%
        {fonseca_abreu_twitter_2020}
\bibfield{author}{\bibinfo{person}{Jefferson~Viana Fonseca~Abreu}, \bibinfo{person}{Celia Ghedini~Ralha}, {and} \bibinfo{person}{Joao~Jose Costa~Gondim}.} \bibinfo{year}{2020}\natexlab{}.
\newblock \showarticletitle{Twitter Bot Detection with Reduced Feature Set}. In \bibinfo{booktitle}{\emph{2020 {IEEE} International Conference on Intelligence and Security Informatics ({ISI})}} (Arlington, {VA}, {USA}, 2020-11-09). \bibinfo{publisher}{{IEEE}}, \bibinfo{pages}{1--6}.
\newblock
\showISBNx{978-1-72818-800-3}
\href{https://doi.org/10.1109/ISI49825.2020.9280525}{doi:\nolinkurl{10.1109/ISI49825.2020.9280525}}


\bibitem[Gansky and McDonald(2022)]%
        {gansky_counterfacctual_2022}
\bibfield{author}{\bibinfo{person}{Ben Gansky} {and} \bibinfo{person}{Sean McDonald}.} \bibinfo{year}{2022}\natexlab{}.
\newblock \showarticletitle{{CounterFAccTual}: {How} {FAccT} {Undermines} {Its} {Organizing} {Principles}}. In \bibinfo{booktitle}{\emph{Proceedings of the 2022 {ACM} {Conference} on {Fairness}, {Accountability}, and {Transparency}}} \emph{(\bibinfo{series}{{FAccT} '22})}. \bibinfo{publisher}{Association for Computing Machinery}, \bibinfo{address}{New York, NY, USA}, \bibinfo{pages}{1982--1992}.
\newblock
\showISBNx{978-1-4503-9352-2}
\href{https://doi.org/10.1145/3531146.3533241}{doi:\nolinkurl{10.1145/3531146.3533241}}


\bibitem[Gillespie(2018)]%
        {gillespie2018custodians}
\bibfield{author}{\bibinfo{person}{Tarleton Gillespie}.} \bibinfo{year}{2018}\natexlab{}.
\newblock \bibinfo{booktitle}{\emph{Custodians of the Internet: Platforms, content moderation, and the hidden decisions that shape social media}}.
\newblock \bibinfo{publisher}{Yale University Press}.
\newblock


\bibitem[Givens and Morris(2020)]%
        {givens_centering_2020}
\bibfield{author}{\bibinfo{person}{Alexandra~Reeve Givens} {and} \bibinfo{person}{Meredith~Ringel Morris}.} \bibinfo{year}{2020}\natexlab{}.
\newblock \showarticletitle{Centering disability perspectives in algorithmic fairness, accountability, \& transparency}. In \bibinfo{booktitle}{\emph{Proceedings of the 2020 {Conference} on {Fairness}, {Accountability}, and {Transparency}}} \emph{(\bibinfo{series}{{FAT}* '20})}. \bibinfo{publisher}{Association for Computing Machinery}, \bibinfo{address}{New York, NY, USA}, \bibinfo{pages}{684}.
\newblock
\href{https://doi.org/10.1145/3351095.3375686}{doi:\nolinkurl{10.1145/3351095.3375686}}


\bibitem[Gray and Witt(2021)]%
        {gray_feminist_2021}
\bibfield{author}{\bibinfo{person}{Joanne Gray} {and} \bibinfo{person}{Alice Witt}.} \bibinfo{year}{2021}\natexlab{}.
\newblock \showarticletitle{A feminist data ethics of care for machine learning: {The} what, why, who and how}.
\newblock \bibinfo{journal}{\emph{First Monday}} (\bibinfo{date}{Dec.} \bibinfo{year}{2021}).
\newblock
\href{https://doi.org/10.5210/fm.v26i12.11833}{doi:\nolinkurl{10.5210/fm.v26i12.11833}}


\bibitem[Guo et~al\mbox{.}(2021)]%
        {guo_social_2021}
\bibfield{author}{\bibinfo{person}{Qinglang Guo}, \bibinfo{person}{Haiyong Xie}, \bibinfo{person}{Yangyang Li}, \bibinfo{person}{Wen Ma}, {and} \bibinfo{person}{Chao Zhang}.} \bibinfo{year}{2021}\natexlab{}.
\newblock \showarticletitle{Social Bots Detection via Fusing {BERT} and Graph Convolutional Networks}.
\newblock  \bibinfo{volume}{14}, \bibinfo{number}{1} (\bibinfo{year}{2021}), \bibinfo{pages}{30}.
\newblock
\showISSN{2073-8994}
\href{https://doi.org/10.3390/sym14010030}{doi:\nolinkurl{10.3390/sym14010030}}


\bibitem[Hagendorff(2020)]%
        {hagendorff_ethics_2020}
\bibfield{author}{\bibinfo{person}{Thilo Hagendorff}.} \bibinfo{year}{2020}\natexlab{}.
\newblock \showarticletitle{The {Ethics} of {AI} {Ethics}: {An} {Evaluation} of {Guidelines}}.
\newblock \bibinfo{journal}{\emph{Minds and Machines}} \bibinfo{volume}{30}, \bibinfo{number}{1} (\bibinfo{date}{March} \bibinfo{year}{2020}), \bibinfo{pages}{99--120}.
\newblock
\href{https://doi.org/10.1007/s11023-020-09517-8}{doi:\nolinkurl{10.1007/s11023-020-09517-8}}


\bibitem[Hajli et~al\mbox{.}(2022)]%
        {hajli2022social}
\bibfield{author}{\bibinfo{person}{Nick Hajli}, \bibinfo{person}{Usman Saeed}, \bibinfo{person}{Mina Tajvidi}, {and} \bibinfo{person}{Farid Shirazi}.} \bibinfo{year}{2022}\natexlab{}.
\newblock \showarticletitle{Social bots and the spread of disinformation in social media: the challenges of artificial intelligence}.
\newblock \bibinfo{journal}{\emph{British Journal of Management}} \bibinfo{volume}{33}, \bibinfo{number}{3} (\bibinfo{year}{2022}), \bibinfo{pages}{1238--1253}.
\newblock


\bibitem[Hardt et~al\mbox{.}(2016)]%
        {hardt_equality_2016}
\bibfield{author}{\bibinfo{person}{Moritz Hardt}, \bibinfo{person}{Eric Price}, \bibinfo{person}{Eric Price}, {and} \bibinfo{person}{Nati Srebro}.} \bibinfo{year}{2016}\natexlab{}.
\newblock \showarticletitle{Equality of {Opportunity} in {Supervised} {Learning}}. In \bibinfo{booktitle}{\emph{Advances in {Neural} {Information} {Processing} {Systems}}}, Vol.~\bibinfo{volume}{29}. \bibinfo{publisher}{Curran Associates, Inc.}
\newblock
\urldef\tempurl%
\url{https://papers.nips.cc/paper_files/paper/2016/hash/9d2682367c3935defcb1f9e247a97c0d-Abstract.html}
\showURL{%
\tempurl}


\bibitem[Hayawi et~al\mbox{.}(2022)]%
        {hayawi_deeprobot_2022}
\bibfield{author}{\bibinfo{person}{Kadhim Hayawi}, \bibinfo{person}{Sujith Mathew}, \bibinfo{person}{Neethu Venugopal}, \bibinfo{person}{Mohammad~M. Masud}, {and} \bibinfo{person}{Pin-Han Ho}.} \bibinfo{year}{2022}\natexlab{}.
\newblock \showarticletitle{{DeeProBot}: a hybrid deep neural network model for social bot detection based on user profile data}.
\newblock  \bibinfo{volume}{12}, \bibinfo{number}{1} (\bibinfo{year}{2022}), \bibinfo{pages}{43}.
\newblock
\showISSN{1869-5469}
\href{https://doi.org/10.1007/s13278-022-00869-w}{doi:\nolinkurl{10.1007/s13278-022-00869-w}}


\bibitem[Huang and Vishnoi(2019)]%
        {huang_stable_2019}
\bibfield{author}{\bibinfo{person}{Lingxiao Huang} {and} \bibinfo{person}{Nisheeth Vishnoi}.} \bibinfo{year}{2019}\natexlab{}.
\newblock \showarticletitle{Stable and {Fair} {Classification}}. In \bibinfo{booktitle}{\emph{Proceedings of the 36th {International} {Conference} on {Machine} {Learning}}}. \bibinfo{publisher}{PMLR}, \bibinfo{pages}{2879--2890}.
\newblock
\urldef\tempurl%
\url{https://proceedings.mlr.press/v97/huang19e.html}
\showURL{%
\tempurl}


\bibitem[Hui et~al\mbox{.}(2019)]%
        {hui_botslayer_2019}
\bibfield{author}{\bibinfo{person}{Pik-Mai Hui}, \bibinfo{person}{Kai-Cheng Yang}, \bibinfo{person}{Christopher Torres-Lugo}, \bibinfo{person}{Zachary Monroe}, \bibinfo{person}{Marc {McCarty}}, \bibinfo{person}{Benjamin Serrette}, \bibinfo{person}{Valentin Pentchev}, {and} \bibinfo{person}{Filippo Menczer}.} \bibinfo{year}{2019}\natexlab{}.
\newblock \showarticletitle{{BotSlayer}: real-time detection of bot amplification on Twitter}.
\newblock  \bibinfo{volume}{4}, \bibinfo{number}{42} (\bibinfo{year}{2019}), \bibinfo{pages}{1706}.
\newblock
\showISSN{2475-9066}
\href{https://doi.org/10.21105/joss.01706}{doi:\nolinkurl{10.21105/joss.01706}}


\bibitem[Jacobs et~al\mbox{.}(2024)]%
        {jacobs2024det}
\bibfield{author}{\bibinfo{person}{Charity~S Jacobs}, \bibinfo{person}{Hui Xian~Lynnette Ng}, {and} \bibinfo{person}{Kathleen~M Carley}.} \bibinfo{year}{2024}\natexlab{}.
\newblock \bibinfo{booktitle}{\emph{DET: Detection Evasion Techniques of State-Sponsored Accounts}}.
\newblock \bibinfo{type}{{T}echnical {R}eport}. \bibinfo{institution}{Center for Open Science}.
\newblock


\bibitem[Jacobs et~al\mbox{.}(2023)]%
        {jacobs_tracking_2023}
\bibfield{author}{\bibinfo{person}{Charity~S. Jacobs}, \bibinfo{person}{Lynnette Hui~Xian Ng}, {and} \bibinfo{person}{Kathleen~M. Carley}.} \bibinfo{year}{2023}\natexlab{}.
\newblock \showarticletitle{Tracking {China}’s {Cross}-{Strait} {Bot} {Networks} {Against} {Taiwan}}. In \bibinfo{booktitle}{\emph{Social, {Cultural}, and {Behavioral} {Modeling}}}, \bibfield{editor}{\bibinfo{person}{Robert Thomson}, \bibinfo{person}{Samer Al-khateeb}, \bibinfo{person}{Annetta Burger}, \bibinfo{person}{Patrick Park}, {and} \bibinfo{person}{Aryn A.~Pyke}} (Eds.). \bibinfo{publisher}{Springer Nature Switzerland}, \bibinfo{address}{Cham}, \bibinfo{pages}{115--125}.
\newblock
\href{https://doi.org/10.1007/978-3-031-43129-6_12}{doi:\nolinkurl{10.1007/978-3-031-43129-6_12}}


\bibitem[Jakesch et~al\mbox{.}(2022)]%
        {10.1145/3531146.3533097}
\bibfield{author}{\bibinfo{person}{Maurice Jakesch}, \bibinfo{person}{Zana Bu\c{c}inca}, \bibinfo{person}{Saleema Amershi}, {and} \bibinfo{person}{Alexandra Olteanu}.} \bibinfo{year}{2022}\natexlab{}.
\newblock \showarticletitle{How Different Groups Prioritize Ethical Values for Responsible AI}. In \bibinfo{booktitle}{\emph{2022 ACM Conference on Fairness, Accountability, and Transparency}} (Seoul, Republic of Korea) \emph{(\bibinfo{series}{FAccT '22})}. \bibinfo{publisher}{Association for Computing Machinery}, \bibinfo{address}{New York, NY, USA}, \bibinfo{pages}{310–323}.
\newblock
\href{https://doi.org/10.1145/3531146.3533097}{doi:\nolinkurl{10.1145/3531146.3533097}}


\bibitem[Javed et~al\mbox{.}(2024)]%
        {javed_ethical_2024}
\bibfield{author}{\bibinfo{person}{Haseeb Javed}, \bibinfo{person}{Hafiz~Abdul Muqeet}, \bibinfo{person}{Tahir Javed}, \bibinfo{person}{Atiq~Ur Rehman}, {and} \bibinfo{person}{Rizwan Sadiq}.} \bibinfo{year}{2024}\natexlab{}.
\newblock \showarticletitle{Ethical {Frameworks} for {Machine} {Learning} in {Sensitive} {Healthcare} {Applications}}.
\newblock \bibinfo{journal}{\emph{IEEE Access}}  \bibinfo{volume}{12} (\bibinfo{year}{2024}), \bibinfo{pages}{16233--16254}.
\newblock
\href{https://doi.org/10.1109/ACCESS.2023.3340884}{doi:\nolinkurl{10.1109/ACCESS.2023.3340884}}


\bibitem[Jobin et~al\mbox{.}(2019)]%
        {jobin_global_2019}
\bibfield{author}{\bibinfo{person}{Anna Jobin}, \bibinfo{person}{Marcello Ienca}, {and} \bibinfo{person}{Effy Vayena}.} \bibinfo{year}{2019}\natexlab{}.
\newblock \showarticletitle{The global landscape of {AI} ethics guidelines}.
\newblock \bibinfo{journal}{\emph{Nature Machine Intelligence}} \bibinfo{volume}{1}, \bibinfo{number}{9} (\bibinfo{date}{Sept.} \bibinfo{year}{2019}), \bibinfo{pages}{389--399}.
\newblock
\href{https://doi.org/10.1038/s42256-019-0088-2}{doi:\nolinkurl{10.1038/s42256-019-0088-2}}


\bibitem[Kamiran and Calders(2009)]%
        {kamiran_classifying_2009}
\bibfield{author}{\bibinfo{person}{Faisal Kamiran} {and} \bibinfo{person}{Toon Calders}.} \bibinfo{year}{2009}\natexlab{}.
\newblock \showarticletitle{Classifying without discriminating}. In \bibinfo{booktitle}{\emph{Control and {Communication} 2009 2nd {International} {Conference} on {Computer}}}. \bibinfo{pages}{1--6}.
\newblock
\href{https://doi.org/10.1109/IC4.2009.4909197}{doi:\nolinkurl{10.1109/IC4.2009.4909197}}


\bibitem[Kamiran and Calders(2012)]%
        {kamiran_data_2012}
\bibfield{author}{\bibinfo{person}{Faisal Kamiran} {and} \bibinfo{person}{Toon Calders}.} \bibinfo{year}{2012}\natexlab{}.
\newblock \showarticletitle{Data preprocessing techniques for classification without discrimination}.
\newblock \bibinfo{journal}{\emph{Knowledge and Information Systems}} \bibinfo{volume}{33}, \bibinfo{number}{1} (\bibinfo{date}{Oct.} \bibinfo{year}{2012}), \bibinfo{pages}{1--33}.
\newblock
\href{https://doi.org/10.1007/s10115-011-0463-8}{doi:\nolinkurl{10.1007/s10115-011-0463-8}}


\bibitem[Kantepe and Ganiz(2017)]%
        {kantepe_preprocessing_2017}
\bibfield{author}{\bibinfo{person}{Mücahit Kantepe} {and} \bibinfo{person}{Murat~Can Ganiz}.} \bibinfo{year}{2017}\natexlab{}.
\newblock \showarticletitle{Preprocessing framework for Twitter bot detection}. In \bibinfo{booktitle}{\emph{2017 International Conference on Computer Science and Engineering ({UBMK})}} (2017-10). \bibinfo{pages}{630--634}.
\newblock
\href{https://doi.org/10.1109/UBMK.2017.8093483}{doi:\nolinkurl{10.1109/UBMK.2017.8093483}}


\bibitem[Karpathy et~al\mbox{.}(2014)]%
        {karpathy2014large}
\bibfield{author}{\bibinfo{person}{Andrej Karpathy}, \bibinfo{person}{George Toderici}, \bibinfo{person}{Sanketh Shetty}, \bibinfo{person}{Thomas Leung}, \bibinfo{person}{Rahul Sukthankar}, {and} \bibinfo{person}{Li Fei-Fei}.} \bibinfo{year}{2014}\natexlab{}.
\newblock \showarticletitle{Large-scale video classification with convolutional neural networks}. In \bibinfo{booktitle}{\emph{Proceedings of the IEEE conference on Computer Vision and Pattern Recognition}}. \bibinfo{pages}{1725--1732}.
\newblock


\bibitem[Kasakowskij et~al\mbox{.}(2020)]%
        {kasakowskij2020network}
\bibfield{author}{\bibinfo{person}{Thomas Kasakowskij}, \bibinfo{person}{Julia F{\"u}rst}, \bibinfo{person}{Jan Fischer}, {and} \bibinfo{person}{Kaja~J Fietkiewicz}.} \bibinfo{year}{2020}\natexlab{}.
\newblock \showarticletitle{Network enforcement as denunciation endorsement? A critical study on legal enforcement in social media}.
\newblock \bibinfo{journal}{\emph{Telematics and Informatics}}  \bibinfo{volume}{46} (\bibinfo{year}{2020}), \bibinfo{pages}{101317}.
\newblock


\bibitem[Keyes et~al\mbox{.}(2019)]%
        {keyes_mulching_2019}
\bibfield{author}{\bibinfo{person}{Os Keyes}, \bibinfo{person}{Jevan Hutson}, {and} \bibinfo{person}{Meredith Durbin}.} \bibinfo{year}{2019}\natexlab{}.
\newblock \showarticletitle{A {Mulching} {Proposal}: {Analysing} and {Improving} an {Algorithmic} {System} for {Turning} the {Elderly} into {High}-{Nutrient} {Slurry}}. In \bibinfo{booktitle}{\emph{Extended {Abstracts} of the 2019 {CHI} {Conference} on {Human} {Factors} in {Computing} {Systems}}} \emph{(\bibinfo{series}{{CHI} {EA} '19})}. \bibinfo{publisher}{Association for Computing Machinery}, \bibinfo{address}{New York, NY, USA}, \bibinfo{pages}{1--11}.
\newblock
\showISBNx{978-1-4503-5971-9}
\href{https://doi.org/10.1145/3290607.3310433}{doi:\nolinkurl{10.1145/3290607.3310433}}


\bibitem[Kiene and Hill(2020)]%
        {kiene2020uses}
\bibfield{author}{\bibinfo{person}{Charles Kiene} {and} \bibinfo{person}{Benjamin~Mako Hill}.} \bibinfo{year}{2020}\natexlab{}.
\newblock \showarticletitle{Who uses bots? A statistical analysis of bot usage in moderation teams}. In \bibinfo{booktitle}{\emph{Extended abstracts of the 2020 CHI conference on human factors in computing systems}}. \bibinfo{pages}{1--8}.
\newblock


\bibitem[Kim and Aravindan(2018)]%
        {kim_aravindan_2018}
\bibfield{author}{\bibinfo{person}{Jack Kim} {and} \bibinfo{person}{Aradhana Aravindan}.} \bibinfo{year}{2018}\natexlab{}.
\newblock \bibinfo{title}{Facebook refuses Singapore request to remove post after critical website blocked}.
\newblock
\urldef\tempurl%
\url{https://www.reuters.com/article/us-singapore-politics-malaysia-scandal/facebook-refuses-singapore-request-to-remove-post-after-critical-website-blocked-idUSKCN1NF05T}
\showURL{%
\tempurl}


\bibitem[Knauth(2019)]%
        {knauth_language-agnostic_2019}
\bibfield{author}{\bibinfo{person}{Jürgen Knauth}.} \bibinfo{year}{2019}\natexlab{}.
\newblock \showarticletitle{Language-Agnostic Twitter-Bot Detection}. In \bibinfo{booktitle}{\emph{Proceedings of the International Conference on Recent Advances in Natural Language Processing ({RANLP} 2019)}} (Varna, Bulgaria, 2019-09), \bibfield{editor}{\bibinfo{person}{Ruslan Mitkov} {and} \bibinfo{person}{Galia Angelova}} (Eds.). \bibinfo{publisher}{{INCOMA} Ltd.}, \bibinfo{pages}{550--558}.
\newblock
\href{https://doi.org/10.26615/978-954-452-056-4_065}{doi:\nolinkurl{10.26615/978-954-452-056-4_065}}


\bibitem[Kohli et~al\mbox{.}(2018)]%
        {kohli_translation_2018}
\bibfield{author}{\bibinfo{person}{Nitin Kohli}, \bibinfo{person}{Renata Barreto}, {and} \bibinfo{person}{Joshua~A Kroll}.} \bibinfo{year}{2018}\natexlab{}.
\newblock \showarticletitle{Translation {Tutorial}: {A} {Shared} {Lexicon} for {Research} and {Practice} in {Human}-{Centered} {Software} {Systems}}. In \bibinfo{booktitle}{\emph{1st {Conference} on {Fairness}, {Accountability}, and {Transparancy}}}. \bibinfo{address}{New York, NY, USA}.
\newblock


\bibitem[Kotek et~al\mbox{.}(2024)]%
        {kotek2024protected}
\bibfield{author}{\bibinfo{person}{Hadas Kotek}, \bibinfo{person}{David~Q Sun}, \bibinfo{person}{Zidi Xiu}, \bibinfo{person}{Margit Bowler}, {and} \bibinfo{person}{Christopher Klein}.} \bibinfo{year}{2024}\natexlab{}.
\newblock \showarticletitle{Protected group bias and stereotypes in Large Language Models}.
\newblock \bibinfo{journal}{\emph{arXiv preprint arXiv:2403.14727}} (\bibinfo{year}{2024}).
\newblock


\bibitem[Kouvela et~al\mbox{.}(2020)]%
        {kouvela_bot-detective_2020}
\bibfield{author}{\bibinfo{person}{Maria Kouvela}, \bibinfo{person}{Ilias Dimitriadis}, {and} \bibinfo{person}{Athena Vakali}.} \bibinfo{year}{2020}\natexlab{}.
\newblock \showarticletitle{Bot-Detective: An explainable Twitter bot detection service with crowdsourcing functionalities}. In \bibinfo{booktitle}{\emph{Proceedings of the 12th International Conference on Management of Digital {EcoSystems}}} (Virtual Event United Arab Emirates, 2020-11-02). \bibinfo{publisher}{{ACM}}, \bibinfo{pages}{55--63}.
\newblock
\showISBNx{978-1-4503-8115-4}
\href{https://doi.org/10.1145/3415958.3433075}{doi:\nolinkurl{10.1145/3415958.3433075}}


\bibitem[Kudugunta and Ferrara(2018)]%
        {kudugunta_deep_2018}
\bibfield{author}{\bibinfo{person}{Sneha Kudugunta} {and} \bibinfo{person}{Emilio Ferrara}.} \bibinfo{year}{2018}\natexlab{}.
\newblock \showarticletitle{Deep Neural Networks for Bot Detection}.
\newblock   \bibinfo{volume}{467} (\bibinfo{year}{2018}), \bibinfo{pages}{312--322}.
\newblock
\showISSN{00200255}
\href{https://doi.org/10.1016/j.ins.2018.08.019}{doi:\nolinkurl{10.1016/j.ins.2018.08.019}}
\showeprint[arxiv]{1802.04289 [cs]}


\bibitem[Kudugunta and Ferrara(2018)]%
        {kudugunta2018deep}
\bibfield{author}{\bibinfo{person}{Sneha Kudugunta} {and} \bibinfo{person}{Emilio Ferrara}.} \bibinfo{year}{2018}\natexlab{}.
\newblock \showarticletitle{Deep neural networks for bot detection}.
\newblock \bibinfo{journal}{\emph{Information Sciences}}  \bibinfo{volume}{467} (\bibinfo{year}{2018}), \bibinfo{pages}{312--322}.
\newblock


\bibitem[Kusner et~al\mbox{.}(2017)]%
        {kusner2017counterfactual}
\bibfield{author}{\bibinfo{person}{Matt~J Kusner}, \bibinfo{person}{Joshua Loftus}, \bibinfo{person}{Chris Russell}, {and} \bibinfo{person}{Ricardo Silva}.} \bibinfo{year}{2017}\natexlab{}.
\newblock \showarticletitle{Counterfactual fairness}.
\newblock \bibinfo{journal}{\emph{Advances in neural information processing systems}}  \bibinfo{volume}{30} (\bibinfo{year}{2017}).
\newblock


\bibitem[Laufer et~al\mbox{.}(2022)]%
        {laufer_four_2022}
\bibfield{author}{\bibinfo{person}{Benjamin Laufer}, \bibinfo{person}{Sameer Jain}, \bibinfo{person}{A.~Feder Cooper}, \bibinfo{person}{Jon Kleinberg}, {and} \bibinfo{person}{Hoda Heidari}.} \bibinfo{year}{2022}\natexlab{}.
\newblock \showarticletitle{Four {Years} of {FAccT}: {A} {Reflexive}, {Mixed}-{Methods} {Analysis} of {Research} {Contributions}, {Shortcomings}, and {Future} {Prospects}}. In \bibinfo{booktitle}{\emph{Proceedings of the 2022 {ACM} {Conference} on {Fairness}, {Accountability}, and {Transparency}}} \emph{(\bibinfo{series}{{FAccT} '22})}. \bibinfo{publisher}{Association for Computing Machinery}, \bibinfo{address}{New York, NY, USA}, \bibinfo{pages}{401--426}.
\newblock
\showISBNx{978-1-4503-9352-2}
\href{https://doi.org/10.1145/3531146.3533107}{doi:\nolinkurl{10.1145/3531146.3533107}}


\bibitem[Lee et~al\mbox{.}(2021)]%
        {lee_seven_2021}
\bibfield{author}{\bibinfo{person}{Kyumin Lee}, \bibinfo{person}{Brian Eoff}, {and} \bibinfo{person}{James Caverlee}.} \bibinfo{year}{2021}\natexlab{}.
\newblock \showarticletitle{Seven Months with the Devils: A Long-Term Study of Content Polluters on Twitter}.
\newblock  \bibinfo{volume}{5}, \bibinfo{number}{1} (\bibinfo{year}{2021}), \bibinfo{pages}{185--192}.
\newblock
\showISSN{2334-0770, 2162-3449}
\href{https://doi.org/10.1609/icwsm.v5i1.14106}{doi:\nolinkurl{10.1609/icwsm.v5i1.14106}}


\bibitem[Lee et~al\mbox{.}(2011)]%
        {lee2011seven}
\bibfield{author}{\bibinfo{person}{Kyumin Lee}, \bibinfo{person}{Brian Eoff}, {and} \bibinfo{person}{James Caverlee}.} \bibinfo{year}{2011}\natexlab{}.
\newblock \showarticletitle{Seven months with the devils: A long-term study of content polluters on twitter}. In \bibinfo{booktitle}{\emph{Proceedings of the international AAAI conference on web and social media}}, Vol.~\bibinfo{volume}{5}. \bibinfo{pages}{185--192}.
\newblock


\bibitem[Leurs(2017)]%
        {leurs_feminist_2017}
\bibfield{author}{\bibinfo{person}{Koen Leurs}.} \bibinfo{year}{2017}\natexlab{}.
\newblock \showarticletitle{Feminist {Data} {Studies}: {Using} {Digital} {Methods} for {Ethical}, {Reflexive} and {Situated} {Socio}-{Cultural} {Research}}.
\newblock \bibinfo{journal}{\emph{Feminist Review}} \bibinfo{volume}{115}, \bibinfo{number}{1} (\bibinfo{date}{March} \bibinfo{year}{2017}), \bibinfo{pages}{130--154}.
\newblock
\href{https://doi.org/10.1057/s41305-017-0043-1}{doi:\nolinkurl{10.1057/s41305-017-0043-1}}


\bibitem[Liedke and Wang(2023)]%
        {walker_matsa_2023}
\bibfield{author}{\bibinfo{person}{Jacob Liedke} {and} \bibinfo{person}{Luxuan Wang}.} \bibinfo{year}{2023}\natexlab{}.
\newblock \bibinfo{title}{Social Media and News Factsheet}.
\newblock
\urldef\tempurl%
\url{https://www.pewresearch.org/journalism/fact-sheet/social-media-and-news-fact-sheet/}
\showURL{%
\tempurl}


\bibitem[Likarchuk et~al\mbox{.}(2023)]%
        {likarchuk2023manipulation}
\bibfield{author}{\bibinfo{person}{Nataliia Likarchuk}, \bibinfo{person}{Zoriana Velychko}, \bibinfo{person}{Olha Andrieieva}, \bibinfo{person}{Raisa Lenda}, {and} \bibinfo{person}{Hanna Vusyk}.} \bibinfo{year}{2023}\natexlab{}.
\newblock \showarticletitle{Manipulation as an element of the political process in social networks.}
\newblock \bibinfo{journal}{\emph{Cuestiones Pol{\'\i}ticas}} \bibinfo{volume}{41}, \bibinfo{number}{76} (\bibinfo{year}{2023}).
\newblock


\bibitem[Lim(2023)]%
        {lim2023activist}
\bibfield{author}{\bibinfo{person}{Merlyna Lim}.} \bibinfo{year}{2023}\natexlab{}.
\newblock \showarticletitle{From activist media to algorithmic politics: The Internet, social media, and civil society in Southeast Asia}.
\newblock In \bibinfo{booktitle}{\emph{Routledge handbook of civil and uncivil society in Southeast Asia}}. \bibinfo{publisher}{Routledge}, \bibinfo{pages}{25--44}.
\newblock


\bibitem[Liu et~al\mbox{.}(2023)]%
        {liu_botmoe_2023}
\bibfield{author}{\bibinfo{person}{Yuhan Liu}, \bibinfo{person}{Zhaoxuan Tan}, \bibinfo{person}{Heng Wang}, \bibinfo{person}{Shangbin Feng}, \bibinfo{person}{Qinghua Zheng}, {and} \bibinfo{person}{Minnan Luo}.} \bibinfo{year}{2023}\natexlab{}.
\newblock \showarticletitle{{BotMoE}: Twitter Bot Detection with Community-Aware Mixtures of Modal-Specific Experts}. In \bibinfo{booktitle}{\emph{Proceedings of the 46th International {ACM} {SIGIR} Conference on Research and Development in Information Retrieval}} (Taipei Taiwan, 2023-07-19). \bibinfo{publisher}{{ACM}}, \bibinfo{pages}{485--495}.
\newblock
\showISBNx{978-1-4503-9408-6}
\href{https://doi.org/10.1145/3539618.3591646}{doi:\nolinkurl{10.1145/3539618.3591646}}


\bibitem[Lopez-Joya et~al\mbox{.}(2024)]%
        {lopez-joya_exploring_2024}
\bibfield{author}{\bibinfo{person}{Salvador Lopez-Joya}, \bibinfo{person}{Jose~A. Diaz-Garcia}, \bibinfo{person}{M.~Dolores Ruiz}, {and} \bibinfo{person}{Maria~J. Martin-Bautista}.} \bibinfo{year}{2024}\natexlab{}.
\newblock \bibinfo{title}{Exploring social bots: A feature-based approach to improve bot detection in social networks}.
\newblock
\href{https://doi.org/10.48550/arXiv.2411.06626}{doi:\nolinkurl{10.48550/arXiv.2411.06626}}
\showeprint[arxiv]{2411.06626 [cs]}


\bibitem[Lu and Lee(2024)]%
        {lu2024agents}
\bibfield{author}{\bibinfo{person}{Hsiu-Chi Lu} {and} \bibinfo{person}{Hsuan-wei Lee}.} \bibinfo{year}{2024}\natexlab{}.
\newblock \showarticletitle{Agents of Discord: Modeling the Impact of Political Bots on Opinion Polarization in Social Networks}.
\newblock \bibinfo{journal}{\emph{Social Science Computer Review}} (\bibinfo{year}{2024}), \bibinfo{pages}{08944393241270382}.
\newblock


\bibitem[Lundberg and Lee(2017)]%
        {NIPS2017_7062}
\bibfield{author}{\bibinfo{person}{Scott~M Lundberg} {and} \bibinfo{person}{Su-In Lee}.} \bibinfo{year}{2017}\natexlab{}.
\newblock \showarticletitle{A Unified Approach to Interpreting Model Predictions}.
\newblock In \bibinfo{booktitle}{\emph{Advances in Neural Information Processing Systems 30}}, \bibfield{editor}{\bibinfo{person}{I.~Guyon}, \bibinfo{person}{U.~V. Luxburg}, \bibinfo{person}{S.~Bengio}, \bibinfo{person}{H.~Wallach}, \bibinfo{person}{R.~Fergus}, \bibinfo{person}{S.~Vishwanathan}, {and} \bibinfo{person}{R.~Garnett}} (Eds.). \bibinfo{publisher}{Curran Associates, Inc.}, \bibinfo{pages}{4765--4774}.
\newblock
\urldef\tempurl%
\url{http://papers.nips.cc/paper/7062-a-unified-approach-to-interpreting-model-predictions.pdf}
\showURL{%
\tempurl}


\bibitem[Macy et~al\mbox{.}(2003)]%
        {macy2003polarization}
\bibfield{author}{\bibinfo{person}{Michael~W Macy}, \bibinfo{person}{James~A Kitts}, \bibinfo{person}{Andreas Flache}, {and} \bibinfo{person}{Steve Benard}.} \bibinfo{year}{2003}\natexlab{}.
\newblock \showarticletitle{Polarization in dynamic networks: A Hopfield model of emergent structure}.
\newblock  (\bibinfo{year}{2003}).
\newblock


\bibitem[Magelinski et~al\mbox{.}(2019)]%
        {magelinski_graph-hist_2019}
\bibfield{author}{\bibinfo{person}{Thomas Magelinski}, \bibinfo{person}{David Beskow}, {and} \bibinfo{person}{Kathleen~M. Carley}.} \bibinfo{year}{2019}\natexlab{}.
\newblock \bibinfo{title}{Graph-Hist: Graph Classification from Latent Feature Histograms With Application to Bot Detection}.
\newblock
\href{https://doi.org/10.48550/arXiv.1910.01180}{doi:\nolinkurl{10.48550/arXiv.1910.01180}}
\showeprint[arxiv]{1910.01180 [cs]}


\bibitem[Magelinski et~al\mbox{.}(2020)]%
        {magelinski2020graph}
\bibfield{author}{\bibinfo{person}{Thomas Magelinski}, \bibinfo{person}{David Beskow}, {and} \bibinfo{person}{Kathleen~M Carley}.} \bibinfo{year}{2020}\natexlab{}.
\newblock \showarticletitle{Graph-hist: Graph classification from latent feature histograms with application to bot detection}. In \bibinfo{booktitle}{\emph{Proceedings of the AAAI Conference on Artificial Intelligence}}, Vol.~\bibinfo{volume}{34}. \bibinfo{pages}{5134--5141}.
\newblock


\bibitem[Malik et~al\mbox{.}(2025)]%
        {malik2024textit}
\bibfield{author}{\bibinfo{person}{Ananya Malik}, \bibinfo{person}{Kartik Sharma}, \bibinfo{person}{Shaily Bhatt}, {and} \bibinfo{person}{Lynnette Hui~Xian Ng}.} \bibinfo{year}{2025}\natexlab{}.
\newblock \showarticletitle{Who Speaks Matters: Analysing the Influence of the Speaker’s Linguistic Identity on Hate Classification}. In \bibinfo{booktitle}{\emph{Findings of the Association for Computational Linguistics: EMNLP 2025}}. \bibinfo{pages}{24927--24937}.
\newblock


\bibitem[Manvi et~al\mbox{.}(2024)]%
        {manvi2024large}
\bibfield{author}{\bibinfo{person}{Rohin Manvi}, \bibinfo{person}{Samar Khanna}, \bibinfo{person}{Marshall Burke}, \bibinfo{person}{David Lobell}, {and} \bibinfo{person}{Stefano Ermon}.} \bibinfo{year}{2024}\natexlab{}.
\newblock \showarticletitle{Large language models are geographically biased}.
\newblock \bibinfo{journal}{\emph{arXiv preprint arXiv:2402.02680}} (\bibinfo{year}{2024}).
\newblock


\bibitem[Mariotti et~al\mbox{.}(2021)]%
        {mariotti_framework_2021}
\bibfield{author}{\bibinfo{person}{Ettore Mariotti}, \bibinfo{person}{Jose~M. Alonso}, {and} \bibinfo{person}{Roberto Confalonieri}.} \bibinfo{year}{2021}\natexlab{}.
\newblock \showarticletitle{A {Framework} for {Analyzing} {Fairness}, {Accountability}, {Transparency} and {Ethics}: {A} {Use}-case in {Banking} {Services}}. In \bibinfo{booktitle}{\emph{2021 {IEEE} {International} {Conference} on {Fuzzy} {Systems} ({FUZZ}-{IEEE})}}. \bibinfo{pages}{1--6}.
\newblock
\href{https://doi.org/10.1109/FUZZ45933.2021.9494481}{doi:\nolinkurl{10.1109/FUZZ45933.2021.9494481}}


\bibitem[Mazza et~al\mbox{.}(2019a)]%
        {mazza2019rtbust}
\bibfield{author}{\bibinfo{person}{Michele Mazza}, \bibinfo{person}{Stefano Cresci}, \bibinfo{person}{Marco Avvenuti}, \bibinfo{person}{Walter Quattrociocchi}, {and} \bibinfo{person}{Maurizio Tesconi}.} \bibinfo{year}{2019}\natexlab{a}.
\newblock \showarticletitle{Rtbust: Exploiting temporal patterns for botnet detection on twitter}. In \bibinfo{booktitle}{\emph{Proceedings of the 10th ACM conference on web science}}. \bibinfo{pages}{183--192}.
\newblock


\bibitem[Mazza et~al\mbox{.}(2019b)]%
        {mazza_rtbust_2019}
\bibfield{author}{\bibinfo{person}{Michele Mazza}, \bibinfo{person}{Stefano Cresci}, \bibinfo{person}{Marco Avvenuti}, \bibinfo{person}{Walter Quattrociocchi}, {and} \bibinfo{person}{Maurizio Tesconi}.} \bibinfo{year}{2019}\natexlab{b}.
\newblock \showarticletitle{{RTbust}: {Exploiting} {Temporal} {Patterns} for {Botnet} {Detection} on {Twitter}}. In \bibinfo{booktitle}{\emph{Proceedings of the 10th {ACM} {Conference} on {Web} {Science}}} \emph{(\bibinfo{series}{{WebSci} '19})}. \bibinfo{publisher}{Association for Computing Machinery}, \bibinfo{address}{New York, NY, USA}, \bibinfo{pages}{183--192}.
\newblock
\href{https://doi.org/10.1145/3292522.3326015}{doi:\nolinkurl{10.1145/3292522.3326015}}


\bibitem[Mehrabi et~al\mbox{.}(2021)]%
        {mehrabi_survey_2021}
\bibfield{author}{\bibinfo{person}{Ninareh Mehrabi}, \bibinfo{person}{Fred Morstatter}, \bibinfo{person}{Nripsuta Saxena}, \bibinfo{person}{Kristina Lerman}, {and} \bibinfo{person}{Aram Galstyan}.} \bibinfo{year}{2021}\natexlab{}.
\newblock \showarticletitle{A {Survey} on {Bias} and {Fairness} in {Machine} {Learning}}.
\newblock \bibinfo{journal}{\emph{Comput. Surveys}} \bibinfo{volume}{54}, \bibinfo{number}{6} (\bibinfo{date}{July} \bibinfo{year}{2021}), \bibinfo{pages}{115:1--115:35}.
\newblock
\showISSN{0360-0300}
\href{https://doi.org/10.1145/3457607}{doi:\nolinkurl{10.1145/3457607}}


\bibitem[Milano et~al\mbox{.}(2020)]%
        {milano2020recommender}
\bibfield{author}{\bibinfo{person}{Silvia Milano}, \bibinfo{person}{Mariarosaria Taddeo}, {and} \bibinfo{person}{Luciano Floridi}.} \bibinfo{year}{2020}\natexlab{}.
\newblock \showarticletitle{Recommender systems and their ethical challenges}.
\newblock \bibinfo{journal}{\emph{AI \& Society}} \bibinfo{volume}{35}, \bibinfo{number}{4} (\bibinfo{year}{2020}), \bibinfo{pages}{957--967}.
\newblock


\bibitem[Miller and Redhead(2019)]%
        {miller_beyond_2019}
\bibfield{author}{\bibinfo{person}{Hannah Miller} {and} \bibinfo{person}{Robin Redhead}.} \bibinfo{year}{2019}\natexlab{}.
\newblock \showarticletitle{Beyond ‘rights-based approaches’? {Employing} a process and outcomes framework}.
\newblock \bibinfo{journal}{\emph{The International Journal of Human Rights}} \bibinfo{volume}{23}, \bibinfo{number}{5} (\bibinfo{date}{May} \bibinfo{year}{2019}), \bibinfo{pages}{699--718}.
\newblock
\href{https://doi.org/10.1080/13642987.2019.1607210}{doi:\nolinkurl{10.1080/13642987.2019.1607210}}


\bibitem[Mitchell et~al\mbox{.}(2019)]%
        {mitchell2019model}
\bibfield{author}{\bibinfo{person}{Margaret Mitchell}, \bibinfo{person}{Simone Wu}, \bibinfo{person}{Andrew Zaldivar}, \bibinfo{person}{Parker Barnes}, \bibinfo{person}{Lucy Vasserman}, \bibinfo{person}{Ben Hutchinson}, \bibinfo{person}{Elena Spitzer}, \bibinfo{person}{Inioluwa~Deborah Raji}, {and} \bibinfo{person}{Timnit Gebru}.} \bibinfo{year}{2019}\natexlab{}.
\newblock \showarticletitle{Model cards for model reporting}. In \bibinfo{booktitle}{\emph{Proceedings of the conference on fairness, accountability, and transparency}}. \bibinfo{pages}{220--229}.
\newblock


\bibitem[Mittelstadt(2019)]%
        {mittelstadt_principles_2019}
\bibfield{author}{\bibinfo{person}{Brent Mittelstadt}.} \bibinfo{year}{2019}\natexlab{}.
\newblock \showarticletitle{Principles alone cannot guarantee ethical {AI}}.
\newblock \bibinfo{journal}{\emph{Nature Machine Intelligence}} \bibinfo{volume}{1}, \bibinfo{number}{11} (\bibinfo{date}{Nov.} \bibinfo{year}{2019}), \bibinfo{pages}{501--507}.
\newblock
\showISSN{2522-5839}
\href{https://doi.org/10.1038/s42256-019-0114-4}{doi:\nolinkurl{10.1038/s42256-019-0114-4}}
\newblock
\shownote{Publisher: Nature Publishing Group}.


\bibitem[Morstatter et~al\mbox{.}(2016)]%
        {morstatter_new_2016}
\bibfield{author}{\bibinfo{person}{Fred Morstatter}, \bibinfo{person}{Liang Wu}, \bibinfo{person}{Tahora~H. Nazer}, \bibinfo{person}{Kathleen~M. Carley}, {and} \bibinfo{person}{Huan Liu}.} \bibinfo{year}{2016}\natexlab{}.
\newblock \showarticletitle{A new approach to bot detection: Striking the balance between precision and recall}. In \bibinfo{booktitle}{\emph{2016 {IEEE}/{ACM} International Conference on Advances in Social Networks Analysis and Mining ({ASONAM})}} (2016-08). \bibinfo{pages}{533--540}.
\newblock
\href{https://doi.org/10.1109/ASONAM.2016.7752287}{doi:\nolinkurl{10.1109/ASONAM.2016.7752287}}


\bibitem[Mukherjee(2019)]%
        {mukherjee2019jio}
\bibfield{author}{\bibinfo{person}{Rahul Mukherjee}.} \bibinfo{year}{2019}\natexlab{}.
\newblock \showarticletitle{Jio sparks Disruption 2.0: infrastructural imaginaries and platform ecosystems in ‘Digital India’}.
\newblock \bibinfo{journal}{\emph{Media, Culture \& Society}} \bibinfo{volume}{41}, \bibinfo{number}{2} (\bibinfo{year}{2019}), \bibinfo{pages}{175--195}.
\newblock


\bibitem[Ng et~al\mbox{.}(2023)]%
        {ng2023digital}
\bibfield{author}{\bibinfo{person}{Lynnette~HX Ng}, \bibinfo{person}{Abigail~CM Lim}, \bibinfo{person}{Adrian~XW Lim}, {and} \bibinfo{person}{Araz Taeihagh}.} \bibinfo{year}{2023}\natexlab{}.
\newblock \showarticletitle{Digital ethics for biometric applications in a smart city}.
\newblock \bibinfo{journal}{\emph{Digital Government: Research and Practice}} \bibinfo{volume}{4}, \bibinfo{number}{4} (\bibinfo{year}{2023}), \bibinfo{pages}{1--6}.
\newblock


\bibitem[Ng and Carley(2022)]%
        {ng_botbuster_2022}
\bibfield{author}{\bibinfo{person}{Lynnette Hui~Xian Ng} {and} \bibinfo{person}{Kathleen~M. Carley}.} \bibinfo{year}{2022}\natexlab{}.
\newblock \bibinfo{title}{{BotBuster}: Multi-platform Bot Detection Using A Mixture of Experts}.
\newblock
\href{https://doi.org/10.48550/arXiv.2207.13658}{doi:\nolinkurl{10.48550/arXiv.2207.13658}}
\showeprint[arxiv]{2207.13658 [cs]}


\bibitem[Ng and Carley(2022)]%
        {ng2022my}
\bibfield{author}{\bibinfo{person}{Lynnette Hui~Xian Ng} {and} \bibinfo{person}{Kathleen~M Carley}.} \bibinfo{year}{2022}\natexlab{}.
\newblock \showarticletitle{Is my stance the same as your stance? A cross validation study of stance detection datasets}.
\newblock \bibinfo{journal}{\emph{Information Processing \& Management}} \bibinfo{volume}{59}, \bibinfo{number}{6} (\bibinfo{year}{2022}), \bibinfo{pages}{103070}.
\newblock


\bibitem[Ng and Carley(2023)]%
        {ng2022botbuster}
\bibfield{author}{\bibinfo{person}{Lynnette Hui~Xian Ng} {and} \bibinfo{person}{Kathleen~M Carley}.} \bibinfo{year}{2023}\natexlab{}.
\newblock \showarticletitle{Botbuster: Multi-platform bot detection using a mixture of experts}. In \bibinfo{booktitle}{\emph{Proceedings of the international AAAI conference on web and social media}}, Vol.~\bibinfo{volume}{17}. \bibinfo{pages}{686--697}.
\newblock


\bibitem[Ng and Carley(2024)]%
        {ng_assembling_2024}
\bibfield{author}{\bibinfo{person}{Lynnette Hui~Xian Ng} {and} \bibinfo{person}{Kathleen~M. Carley}.} \bibinfo{year}{2024}\natexlab{}.
\newblock \showarticletitle{Assembling a multi-platform ensemble social bot detector with applications to {US} 2020 elections}.
\newblock \bibinfo{journal}{\emph{Social Network Analysis and Mining}} \bibinfo{volume}{14}, \bibinfo{number}{1} (\bibinfo{date}{Feb.} \bibinfo{year}{2024}), \bibinfo{pages}{45}.
\newblock
\href{https://doi.org/10.1007/s13278-024-01211-2}{doi:\nolinkurl{10.1007/s13278-024-01211-2}}


\bibitem[Ng and Carley(2025a)]%
        {ng2025dual}
\bibfield{author}{\bibinfo{person}{Lynnette Hui~Xian Ng} {and} \bibinfo{person}{Kathleen~M Carley}.} \bibinfo{year}{2025}\natexlab{a}.
\newblock \showarticletitle{The Dual Personas of Social Media Bots}.
\newblock \bibinfo{journal}{\emph{arXiv preprint arXiv:2504.12498}} (\bibinfo{year}{2025}).
\newblock


\bibitem[Ng and Carley(2025b)]%
        {ng2025global}
\bibfield{author}{\bibinfo{person}{Lynnette Hui~Xian Ng} {and} \bibinfo{person}{Kathleen~M Carley}.} \bibinfo{year}{2025}\natexlab{b}.
\newblock \showarticletitle{A global comparison of social media bot and human characteristics}.
\newblock \bibinfo{journal}{\emph{Scientific Reports}} \bibinfo{volume}{15}, \bibinfo{number}{1} (\bibinfo{year}{2025}), \bibinfo{pages}{10973}.
\newblock


\bibitem[Ng and Carley(2025c)]%
        {ng2025social}
\bibfield{author}{\bibinfo{person}{Lynnette Hui~Xian Ng} {and} \bibinfo{person}{Kathleen~M Carley}.} \bibinfo{year}{2025}\natexlab{c}.
\newblock \showarticletitle{Social Cyber Geographical Worldwide Inventory of Bots}.
\newblock \bibinfo{journal}{\emph{arXiv preprint arXiv:2501.18839}} (\bibinfo{year}{2025}).
\newblock


\bibitem[Ng et~al\mbox{.}(2024a)]%
        {ng2024exploratory}
\bibfield{author}{\bibinfo{person}{Lynnette Hui~Xian Ng}, \bibinfo{person}{Ian Kloo}, \bibinfo{person}{Samantha Clark}, {and} \bibinfo{person}{Kathleen~M Carley}.} \bibinfo{year}{2024}\natexlab{a}.
\newblock \showarticletitle{An exploratory analysis of COVID bot vs human disinformation dissemination stemming from the Disinformation Dozen on Telegram}.
\newblock \bibinfo{journal}{\emph{Journal of Computational Social Science}} \bibinfo{volume}{7}, \bibinfo{number}{1} (\bibinfo{year}{2024}), \bibinfo{pages}{695--720}.
\newblock


\bibitem[Ng et~al\mbox{.}(2022)]%
        {ng_stabilizing_2022}
\bibfield{author}{\bibinfo{person}{Lynnette Hui~Xian Ng}, \bibinfo{person}{Dawn~C. Robertson}, {and} \bibinfo{person}{Kathleen~M. Carley}.} \bibinfo{year}{2022}\natexlab{}.
\newblock \showarticletitle{Stabilizing a supervised bot detection algorithm: How much data is needed for consistent predictions?}
\newblock   \bibinfo{volume}{28} (\bibinfo{year}{2022}), \bibinfo{pages}{100198}.
\newblock
\showISSN{24686964}
\href{https://doi.org/10.1016/j.osnem.2022.100198}{doi:\nolinkurl{10.1016/j.osnem.2022.100198}}


\bibitem[Ng et~al\mbox{.}(2024b)]%
        {ng2024cyborgs}
\bibfield{author}{\bibinfo{person}{Lynnette Hui~Xian Ng}, \bibinfo{person}{Dawn~C Robertson}, {and} \bibinfo{person}{Kathleen~M Carley}.} \bibinfo{year}{2024}\natexlab{b}.
\newblock \showarticletitle{Cyborgs for strategic communication on social media}.
\newblock \bibinfo{journal}{\emph{Big Data \& Society}} \bibinfo{volume}{11}, \bibinfo{number}{1} (\bibinfo{year}{2024}), \bibinfo{pages}{20539517241231275}.
\newblock


\bibitem[Nogara et~al\mbox{.}(2022)]%
        {nogara2022disinformation}
\bibfield{author}{\bibinfo{person}{Gianluca Nogara}, \bibinfo{person}{Padinjaredath~Suresh Vishnuprasad}, \bibinfo{person}{Felipe Cardoso}, \bibinfo{person}{Omran Ayoub}, \bibinfo{person}{Silvia Giordano}, {and} \bibinfo{person}{Luca Luceri}.} \bibinfo{year}{2022}\natexlab{}.
\newblock \showarticletitle{The disinformation dozen: An exploratory analysis of covid-19 disinformation proliferation on twitter}. In \bibinfo{booktitle}{\emph{Proceedings of the 14th ACM web science conference 2022}}. \bibinfo{pages}{348--358}.
\newblock


\bibitem[of~Sciences et~al\mbox{.}(2019)]%
        {national2019decadal}
\bibfield{author}{\bibinfo{person}{National~Academies of Sciences}, \bibinfo{person}{Medicine}, \bibinfo{person}{Division of Behavioral}, \bibinfo{person}{Social Sciences}, \bibinfo{person}{Board on Behavioral}, \bibinfo{person}{Sensory Sciences}, \bibinfo{person}{Committee on~a Decadal Survey~of Social}, {and} \bibinfo{person}{Behavioral~Sciences for Applications~to National~Security}.} \bibinfo{year}{2019}\natexlab{}.
\newblock \showarticletitle{A decadal survey of the social and behavioral sciences: A research agenda for advancing intelligence analysis}.
\newblock  (\bibinfo{year}{2019}).
\newblock


\bibitem[Olanrewaju et~al\mbox{.}(2018)]%
        {olanrewaju2018influence}
\bibfield{author}{\bibinfo{person}{Abdus-Samad~Temitope Olanrewaju}, \bibinfo{person}{Naomi Whiteside}, \bibinfo{person}{Mohammad~Alamgir Hossain}, {and} \bibinfo{person}{Paul Mercieca}.} \bibinfo{year}{2018}\natexlab{}.
\newblock \showarticletitle{The influence of social media on entrepreneur motivation and marketing strategies in a developing country}. In \bibinfo{booktitle}{\emph{Conference on e-Business, e-Services and e-Society}}. Springer, \bibinfo{pages}{355--364}.
\newblock


\bibitem[Olteanu et~al\mbox{.}(2021)]%
        {olteanu2021facts}
\bibfield{author}{\bibinfo{person}{Alexandra Olteanu}, \bibinfo{person}{Jean Garcia-Gathright}, \bibinfo{person}{Maarten de Rijke}, \bibinfo{person}{Michael~D Ekstrand}, \bibinfo{person}{Adam Roegiest}, \bibinfo{person}{Aldo Lipani}, \bibinfo{person}{Alex Beutel}, \bibinfo{person}{Alexandra Olteanu}, \bibinfo{person}{Ana Lucic}, \bibinfo{person}{Ana-Andreea Stoica}, {et~al\mbox{.}}} \bibinfo{year}{2021}\natexlab{}.
\newblock \showarticletitle{FACTS-IR: fairness, accountability, confidentiality, transparency, and safety in information retrieval}. In \bibinfo{booktitle}{\emph{ACM SIGIR Forum}}, Vol.~\bibinfo{volume}{53}. ACM New York, NY, USA, \bibinfo{pages}{20--43}.
\newblock


\bibitem[Oneto et~al\mbox{.}(2019)]%
        {oneto_taking_2019}
\bibfield{author}{\bibinfo{person}{Luca Oneto}, \bibinfo{person}{Michele Doninini}, \bibinfo{person}{Amon Elders}, {and} \bibinfo{person}{Massimiliano Pontil}.} \bibinfo{year}{2019}\natexlab{}.
\newblock \showarticletitle{Taking {Advantage} of {Multitask} {Learning} for {Fair} {Classification}}. In \bibinfo{booktitle}{\emph{Proceedings of the 2019 {AAAI}/{ACM} {Conference} on {AI}, {Ethics}, and {Society}}} \emph{(\bibinfo{series}{{AIES} '19})}. \bibinfo{publisher}{Association for Computing Machinery}, \bibinfo{address}{New York, NY, USA}, \bibinfo{pages}{227--237}.
\newblock
\href{https://doi.org/10.1145/3306618.3314255}{doi:\nolinkurl{10.1145/3306618.3314255}}


\bibitem[Ormazabal et~al\mbox{.}(2019)]%
        {ormazabal2019analyzing}
\bibfield{author}{\bibinfo{person}{Aitor Ormazabal}, \bibinfo{person}{Mikel Artetxe}, \bibinfo{person}{Gorka Labaka}, \bibinfo{person}{Aitor Soroa}, {and} \bibinfo{person}{Eneko Agirre}.} \bibinfo{year}{2019}\natexlab{}.
\newblock \showarticletitle{Analyzing the Limitations of Cross-lingual Word Embedding Mappings}. In \bibinfo{booktitle}{\emph{Proceedings of the 57th Annual Meeting of the Association for Computational Linguistics}}. \bibinfo{pages}{4990--4995}.
\newblock


\bibitem[Pinch(2012)]%
        {pinch2012social}
\bibfield{author}{\bibinfo{person}{Trevor Pinch}.} \bibinfo{year}{2012}\natexlab{}.
\newblock \showarticletitle{The social construction of technology: A review}.
\newblock \bibinfo{journal}{\emph{Technological change}} (\bibinfo{year}{2012}), \bibinfo{pages}{17--35}.
\newblock


\bibitem[Raji et~al\mbox{.}(2020)]%
        {raji2020closing}
\bibfield{author}{\bibinfo{person}{Inioluwa~Deborah Raji}, \bibinfo{person}{Andrew Smart}, \bibinfo{person}{Rebecca~N White}, \bibinfo{person}{Margaret Mitchell}, \bibinfo{person}{Timnit Gebru}, \bibinfo{person}{Ben Hutchinson}, \bibinfo{person}{Jamila Smith-Loud}, \bibinfo{person}{Daniel Theron}, {and} \bibinfo{person}{Parker Barnes}.} \bibinfo{year}{2020}\natexlab{}.
\newblock \showarticletitle{Closing the AI accountability gap: Defining an end-to-end framework for internal algorithmic auditing}. In \bibinfo{booktitle}{\emph{Proceedings of the 2020 conference on fairness, accountability, and transparency}}. \bibinfo{pages}{33--44}.
\newblock


\bibitem[Rauchfleisch and Kaiser(2020)]%
        {rauchfleisch2020false}
\bibfield{author}{\bibinfo{person}{Adrian Rauchfleisch} {and} \bibinfo{person}{Jonas Kaiser}.} \bibinfo{year}{2020}\natexlab{}.
\newblock \showarticletitle{The false positive problem of automatic bot detection in social science research}.
\newblock \bibinfo{journal}{\emph{PloS one}} \bibinfo{volume}{15}, \bibinfo{number}{10} (\bibinfo{year}{2020}), \bibinfo{pages}{e0241045}.
\newblock


\bibitem[Sayyadiharikandeh et~al\mbox{.}(2020)]%
        {sayyadiharikandeh_detection_2020}
\bibfield{author}{\bibinfo{person}{Mohsen Sayyadiharikandeh}, \bibinfo{person}{Onur Varol}, \bibinfo{person}{Kai-Cheng Yang}, \bibinfo{person}{Alessandro Flammini}, {and} \bibinfo{person}{Filippo Menczer}.} \bibinfo{year}{2020}\natexlab{}.
\newblock \showarticletitle{Detection of Novel Social Bots by Ensembles of Specialized Classifiers}. In \bibinfo{booktitle}{\emph{Proceedings of the 29th {ACM} International Conference on Information \& Knowledge Management}} (2020-10-19). \bibinfo{pages}{2725--2732}.
\newblock
\href{https://doi.org/10.1145/3340531.3412698}{doi:\nolinkurl{10.1145/3340531.3412698}}
\showeprint[arxiv]{2006.06867 [cs]}


\bibitem[Sayyadiharikandeh et~al\mbox{.}(2020)]%
        {sayyadiharikandeh2020detection}
\bibfield{author}{\bibinfo{person}{Mohsen Sayyadiharikandeh}, \bibinfo{person}{Onur Varol}, \bibinfo{person}{Kai-Cheng Yang}, \bibinfo{person}{Alessandro Flammini}, {and} \bibinfo{person}{Filippo Menczer}.} \bibinfo{year}{2020}\natexlab{}.
\newblock \showarticletitle{Detection of novel social bots by ensembles of specialized classifiers}. In \bibinfo{booktitle}{\emph{Proceedings of the 29th ACM International Conference on Information \& Knowledge Management}}. \bibinfo{pages}{2725--2732}.
\newblock


\bibitem[Schuchard and Crooks(2021)]%
        {schuchard2021insights}
\bibfield{author}{\bibinfo{person}{Ross~J Schuchard} {and} \bibinfo{person}{Andrew~T Crooks}.} \bibinfo{year}{2021}\natexlab{}.
\newblock \showarticletitle{Insights into elections: An ensemble bot detection coverage framework applied to the 2018 US midterm elections}.
\newblock \bibinfo{journal}{\emph{Plos one}} \bibinfo{volume}{16}, \bibinfo{number}{1} (\bibinfo{year}{2021}), \bibinfo{pages}{e0244309}.
\newblock


\bibitem[Shao et~al\mbox{.}(2019)]%
        {shao2019does}
\bibfield{author}{\bibinfo{person}{Yong Shao}, \bibinfo{person}{Chenchen Zhang}, \bibinfo{person}{Jing Zhou}, \bibinfo{person}{Ting Gu}, {and} \bibinfo{person}{Yuan Yuan}.} \bibinfo{year}{2019}\natexlab{}.
\newblock \showarticletitle{How does culture shape creativity? A mini-review}.
\newblock \bibinfo{journal}{\emph{Frontiers in psychology}}  \bibinfo{volume}{10} (\bibinfo{year}{2019}), \bibinfo{pages}{1219}.
\newblock


\bibitem[Sharma et~al\mbox{.}(2022)]%
        {sharma2022characterizing}
\bibfield{author}{\bibinfo{person}{Karishma Sharma}, \bibinfo{person}{Emilio Ferrara}, {and} \bibinfo{person}{Yan Liu}.} \bibinfo{year}{2022}\natexlab{}.
\newblock \showarticletitle{Characterizing Online Engagement with Disinformation and Conspiracies in the 2020 US Presidential Election}. In \bibinfo{booktitle}{\emph{Proceedings of the International AAAI Conference on Web and Social Media}}, Vol.~\bibinfo{volume}{16}. \bibinfo{pages}{908--919}.
\newblock


\bibitem[Shukla et~al\mbox{.}(2023)]%
        {shukla_social_2023}
\bibfield{author}{\bibinfo{person}{Anant Shukla}, \bibinfo{person}{Martin Jurecek}, {and} \bibinfo{person}{Mark Stamp}.} \bibinfo{year}{2023}\natexlab{}.
\newblock \bibinfo{title}{Social Media Bot Detection using Dropout-{GAN}}.
\newblock
\href{https://doi.org/10.48550/arXiv.2311.05079}{doi:\nolinkurl{10.48550/arXiv.2311.05079}}
\showeprint[arxiv]{2311.05079 [cs]}


\bibitem[Siegel(2022)]%
        {siegel_2022}
\bibfield{author}{\bibinfo{person}{Tatiana Siegel}.} \bibinfo{year}{2022}\natexlab{}.
\newblock \bibinfo{title}{Exclusive: Fake accounts fueled the 'Snyder cut' online army}.
\newblock
\urldef\tempurl%
\url{https://www.rollingstone.com/tv-movies/tv-movie-features/justice-league-the-snyder-cut-bots-fans-1384231/}
\showURL{%
\tempurl}


\bibitem[Singh et~al\mbox{.}(2020)]%
        {singh2020multidimensional}
\bibfield{author}{\bibinfo{person}{Maneet Singh}, \bibinfo{person}{Rishemjit Kaur}, {and} \bibinfo{person}{SRS Iyengar}.} \bibinfo{year}{2020}\natexlab{}.
\newblock \showarticletitle{Multidimensional analysis of fake news spreaders on Twitter}. In \bibinfo{booktitle}{\emph{International Conference on Computational Data and Social Networks}}. Springer, \bibinfo{pages}{354--365}.
\newblock


\bibitem[Smith and Topin(2016)]%
        {smith2016deep}
\bibfield{author}{\bibinfo{person}{Leslie~N Smith} {and} \bibinfo{person}{Nicholay Topin}.} \bibinfo{year}{2016}\natexlab{}.
\newblock \showarticletitle{Deep convolutional neural network design patterns}.
\newblock \bibinfo{journal}{\emph{arXiv preprint arXiv:1611.00847}} (\bibinfo{year}{2016}).
\newblock


\bibitem[Speith(2022)]%
        {speith2022review}
\bibfield{author}{\bibinfo{person}{Timo Speith}.} \bibinfo{year}{2022}\natexlab{}.
\newblock \showarticletitle{A review of taxonomies of explainable artificial intelligence (XAI) methods}. In \bibinfo{booktitle}{\emph{2022 ACM Conference on Fairness, Accountability, and Transparency}}. \bibinfo{pages}{2239--2250}.
\newblock


\bibitem[Starbird et~al\mbox{.}(2019)]%
        {starbird2019disinformation}
\bibfield{author}{\bibinfo{person}{Kate Starbird}, \bibinfo{person}{Ahmer Arif}, {and} \bibinfo{person}{Tom Wilson}.} \bibinfo{year}{2019}\natexlab{}.
\newblock \showarticletitle{Disinformation as collaborative work: Surfacing the participatory nature of strategic information operations}.
\newblock \bibinfo{journal}{\emph{Proceedings of the ACM on human-computer interaction}} \bibinfo{volume}{3}, \bibinfo{number}{CSCW} (\bibinfo{year}{2019}), \bibinfo{pages}{1--26}.
\newblock


\bibitem[Subrahmanian et~al\mbox{.}(2016)]%
        {subrahmanian2016darpa}
\bibfield{author}{\bibinfo{person}{Venkatramanan~S Subrahmanian}, \bibinfo{person}{Amos Azaria}, \bibinfo{person}{Skylar Durst}, \bibinfo{person}{Vadim Kagan}, \bibinfo{person}{Aram Galstyan}, \bibinfo{person}{Kristina Lerman}, \bibinfo{person}{Linhong Zhu}, \bibinfo{person}{Emilio Ferrara}, \bibinfo{person}{Alessandro Flammini}, {and} \bibinfo{person}{Filippo Menczer}.} \bibinfo{year}{2016}\natexlab{}.
\newblock \showarticletitle{The DARPA Twitter bot challenge}.
\newblock \bibinfo{journal}{\emph{Computer}} \bibinfo{volume}{49}, \bibinfo{number}{6} (\bibinfo{year}{2016}), \bibinfo{pages}{38--46}.
\newblock


\bibitem[Subrahmanian et~al\mbox{.}(2016)]%
        {subrahmanian_darpa_2016}
\bibfield{author}{\bibinfo{person}{V.~S. Subrahmanian}, \bibinfo{person}{Amos Azaria}, \bibinfo{person}{Skylar Durst}, \bibinfo{person}{Vadim Kagan}, \bibinfo{person}{Aram Galstyan}, \bibinfo{person}{Kristina Lerman}, \bibinfo{person}{Linhong Zhu}, \bibinfo{person}{Emilio Ferrara}, \bibinfo{person}{Alessandro Flammini}, \bibinfo{person}{Filippo Menczer}, \bibinfo{person}{Andrew Stevens}, \bibinfo{person}{Alexander Dekhtyar}, \bibinfo{person}{Shuyang Gao}, \bibinfo{person}{Tad Hogg}, \bibinfo{person}{Farshad Kooti}, \bibinfo{person}{Yan Liu}, \bibinfo{person}{Onur Varol}, \bibinfo{person}{Prashant Shiralkar}, \bibinfo{person}{Vinod Vydiswaran}, \bibinfo{person}{Qiaozhu Mei}, {and} \bibinfo{person}{Tim Hwang}.} \bibinfo{year}{2016}\natexlab{}.
\newblock \showarticletitle{The {DARPA} Twitter Bot Challenge}.
\newblock  \bibinfo{volume}{49}, \bibinfo{number}{6} (\bibinfo{year}{2016}), \bibinfo{pages}{38--46}.
\newblock
\showISSN{0018-9162}
\href{https://doi.org/10.1109/MC.2016.183}{doi:\nolinkurl{10.1109/MC.2016.183}}
\showeprint[arxiv]{1601.05140 [cs]}


\bibitem[Suchman(2007)]%
        {suchman2007human}
\bibfield{author}{\bibinfo{person}{Lucille~Alice Suchman}.} \bibinfo{year}{2007}\natexlab{}.
\newblock \bibinfo{booktitle}{\emph{Human-machine reconfigurations: Plans and situated actions}}.
\newblock \bibinfo{publisher}{Cambridge university press}.
\newblock


\bibitem[Suzor et~al\mbox{.}(2019)]%
        {suzor_what_2019}
\bibfield{author}{\bibinfo{person}{Nicolas~P. Suzor}, \bibinfo{person}{Sarah~Myers West}, \bibinfo{person}{Andrew Quodling}, {and} \bibinfo{person}{Jillian York}.} \bibinfo{year}{2019}\natexlab{}.
\newblock \showarticletitle{What {Do} {We} {Mean} {When} {We} {Talk} {About} {Transparency}? {Toward} {Meaningful} {Transparency} in {Commercial} {Content} {Moderation}}.
\newblock \bibinfo{journal}{\emph{International Journal of Communication}} \bibinfo{volume}{13}, \bibinfo{number}{0} (\bibinfo{date}{March} \bibinfo{year}{2019}), \bibinfo{pages}{18}.
\newblock
\urldef\tempurl%
\url{https://ijoc.org/index.php/ijoc/article/view/9736}
\showURL{%
\tempurl}


\bibitem[Team(2006)]%
        {global2006world}
\bibfield{author}{\bibinfo{person}{Global Deception~Research Team}.} \bibinfo{year}{2006}\natexlab{}.
\newblock \showarticletitle{A world of lies}.
\newblock \bibinfo{journal}{\emph{Journal of cross-cultural psychology}} \bibinfo{volume}{37}, \bibinfo{number}{1} (\bibinfo{year}{2006}), \bibinfo{pages}{60--74}.
\newblock


\bibitem[Varol et~al\mbox{.}(2017)]%
        {varol2017online}
\bibfield{author}{\bibinfo{person}{Onur Varol}, \bibinfo{person}{Emilio Ferrara}, \bibinfo{person}{Clayton Davis}, \bibinfo{person}{Filippo Menczer}, {and} \bibinfo{person}{Alessandro Flammini}.} \bibinfo{year}{2017}\natexlab{}.
\newblock \showarticletitle{Online human-bot interactions: Detection, estimation, and characterization}. In \bibinfo{booktitle}{\emph{Proceedings of the International AAAI Conference on Web and Social Media}}, Vol.~\bibinfo{volume}{11}. \bibinfo{pages}{280--289}.
\newblock


\bibitem[Varol et~al\mbox{.}(2017)]%
        {varol_online_2017}
\bibfield{author}{\bibinfo{person}{Onur Varol}, \bibinfo{person}{Emilio Ferrara}, \bibinfo{person}{Clayton~A. Davis}, \bibinfo{person}{Filippo Menczer}, {and} \bibinfo{person}{Alessandro Flammini}.} \bibinfo{year}{2017}\natexlab{}.
\newblock \bibinfo{title}{Online Human-Bot Interactions: Detection, Estimation, and Characterization}.
\newblock
\href{https://doi.org/10.48550/arXiv.1703.03107}{doi:\nolinkurl{10.48550/arXiv.1703.03107}}
\showeprint[arxiv]{1703.03107 [cs]}


\bibitem[Veale and Binns(2017)]%
        {veale_fairer_2017}
\bibfield{author}{\bibinfo{person}{Michael Veale} {and} \bibinfo{person}{Reuben Binns}.} \bibinfo{year}{2017}\natexlab{}.
\newblock \showarticletitle{Fairer machine learning in the real world: {Mitigating} discrimination without collecting sensitive data}.
\newblock \bibinfo{journal}{\emph{Big Data \& Society}} \bibinfo{volume}{4}, \bibinfo{number}{2} (\bibinfo{date}{Dec.} \bibinfo{year}{2017}), \bibinfo{pages}{2053951717743530}.
\newblock
\href{https://doi.org/10.1177/2053951717743530}{doi:\nolinkurl{10.1177/2053951717743530}}


\bibitem[Vese(2022)]%
        {vese2022governing}
\bibfield{author}{\bibinfo{person}{Donato Vese}.} \bibinfo{year}{2022}\natexlab{}.
\newblock \showarticletitle{Governing fake news: the regulation of social media and the right to freedom of expression in the era of emergency}.
\newblock \bibinfo{journal}{\emph{European Journal of Risk Regulation}} \bibinfo{volume}{13}, \bibinfo{number}{3} (\bibinfo{year}{2022}), \bibinfo{pages}{477--513}.
\newblock


\bibitem[Viswanath et~al\mbox{.}(2010)]%
        {viswanath_analysis_2010}
\bibfield{author}{\bibinfo{person}{Bimal Viswanath}, \bibinfo{person}{Ansley Post}, \bibinfo{person}{Krishna~P. Gummadi}, {and} \bibinfo{person}{Alan Mislove}.} \bibinfo{year}{2010}\natexlab{}.
\newblock \showarticletitle{An analysis of social network-based Sybil defenses}. In \bibinfo{booktitle}{\emph{Proceedings of the {ACM} {SIGCOMM} 2010 conference}} (New Delhi India, 2010-08-30). \bibinfo{publisher}{{ACM}}, \bibinfo{pages}{363--374}.
\newblock
\showISBNx{978-1-4503-0201-2}
\href{https://doi.org/10.1145/1851182.1851226}{doi:\nolinkurl{10.1145/1851182.1851226}}


\bibitem[Wagner et~al\mbox{.}(2020)]%
        {wagner2020regulating}
\bibfield{author}{\bibinfo{person}{Ben Wagner}, \bibinfo{person}{Krisztina Rozgonyi}, \bibinfo{person}{Marie-Therese Sekwenz}, \bibinfo{person}{Jennifer Cobbe}, {and} \bibinfo{person}{Jatinder Singh}.} \bibinfo{year}{2020}\natexlab{}.
\newblock \showarticletitle{Regulating transparency? Facebook, twitter and the German network enforcement act}. In \bibinfo{booktitle}{\emph{Proceedings of the 2020 conference on fairness, accountability, and transparency}}. \bibinfo{pages}{261--271}.
\newblock


\bibitem[Wallach and Allen(2008)]%
        {wallach_moral_2008}
\bibfield{author}{\bibinfo{person}{Wendell Wallach} {and} \bibinfo{person}{Colin Allen}.} \bibinfo{year}{2008}\natexlab{}.
\newblock \bibinfo{booktitle}{\emph{Moral {Machines}: {Teaching} {Robots} {Right} from {Wrong}}}.
\newblock \bibinfo{publisher}{Oxford University Press}.
\newblock


\bibitem[Wang et~al\mbox{.}(2013)]%
        {wang_you_nodate}
\bibfield{author}{\bibinfo{person}{Gang Wang}, \bibinfo{person}{Tristan Konolige}, \bibinfo{person}{Christo Wilson}, \bibinfo{person}{Xiao Wang}, \bibinfo{person}{Haitao Zheng}, {and} \bibinfo{person}{Ben~Y Zhao}.} \bibinfo{year}{2013}\natexlab{}.
\newblock \showarticletitle{You are how you click: Clickstream analysis for sybil detection}. In \bibinfo{booktitle}{\emph{22nd USENIX security symposium (USENIX Security 13)}}. \bibinfo{pages}{241--256}.
\newblock


\bibitem[Wang et~al\mbox{.}({[n.\,d.]})]%
        {wang_social_nodate}
\bibfield{author}{\bibinfo{person}{Gang Wang}, \bibinfo{person}{Manish Mohanlal}, \bibinfo{person}{Christo Wilson}, \bibinfo{person}{Xiao Wang}, \bibinfo{person}{Miriam Metzger}, \bibinfo{person}{Haitao Zheng}, {and} \bibinfo{person}{Ben~Y Zhao}.} \bibinfo{year}{[n.\,d.]}\natexlab{}.
\newblock \showarticletitle{Social Turing Tests: Crowdsourcing Sybil Detection}.
\newblock  (\bibinfo{year}{[n.\,d.]}).
\newblock


\bibitem[Wang et~al\mbox{.}(2012)]%
        {wang2012social}
\bibfield{author}{\bibinfo{person}{Gang Wang}, \bibinfo{person}{Manish Mohanlal}, \bibinfo{person}{Christo Wilson}, \bibinfo{person}{Xiao Wang}, \bibinfo{person}{Miriam Metzger}, \bibinfo{person}{Haitao Zheng}, {and} \bibinfo{person}{Ben~Y Zhao}.} \bibinfo{year}{2012}\natexlab{}.
\newblock \showarticletitle{Social turing tests: Crowdsourcing sybil detection}.
\newblock \bibinfo{journal}{\emph{arXiv preprint arXiv:1205.3856}} (\bibinfo{year}{2012}).
\newblock


\bibitem[Wei and Nguyen(2020)]%
        {wei_twitter_2020}
\bibfield{author}{\bibinfo{person}{Feng Wei} {and} \bibinfo{person}{Uyen~Trang Nguyen}.} \bibinfo{year}{2020}\natexlab{}.
\newblock \bibinfo{title}{Twitter Bot Detection Using Bidirectional Long Short-term Memory Neural Networks and Word Embeddings}.
\newblock
\href{https://doi.org/10.48550/arXiv.2002.01336}{doi:\nolinkurl{10.48550/arXiv.2002.01336}}
\showeprint[arxiv]{2002.01336 [cs]}


\bibitem[Wexler et~al\mbox{.}(2019)]%
        {wexler2019if}
\bibfield{author}{\bibinfo{person}{James Wexler}, \bibinfo{person}{Mahima Pushkarna}, \bibinfo{person}{Tolga Bolukbasi}, \bibinfo{person}{Martin Wattenberg}, \bibinfo{person}{Fernanda Vi{\'e}gas}, {and} \bibinfo{person}{Jimbo Wilson}.} \bibinfo{year}{2019}\natexlab{}.
\newblock \showarticletitle{The what-if tool: Interactive probing of machine learning models}.
\newblock \bibinfo{journal}{\emph{IEEE transactions on visualization and computer graphics}} \bibinfo{volume}{26}, \bibinfo{number}{1} (\bibinfo{year}{2019}), \bibinfo{pages}{56--65}.
\newblock


\bibitem[Widder et~al\mbox{.}(2022)]%
        {widder_limits_2022}
\bibfield{author}{\bibinfo{person}{David~Gray Widder}, \bibinfo{person}{Dawn Nafus}, \bibinfo{person}{Laura Dabbish}, {and} \bibinfo{person}{James Herbsleb}.} \bibinfo{year}{2022}\natexlab{}.
\newblock \showarticletitle{Limits and {Possibilities} for “{Ethical} {AI}” in {Open} {Source}: {A} {Study} of {Deepfakes}}. In \bibinfo{booktitle}{\emph{Proceedings of the 2022 {ACM} {Conference} on {Fairness}, {Accountability}, and {Transparency}}} \emph{(\bibinfo{series}{{FAccT} '22})}. \bibinfo{publisher}{Association for Computing Machinery}, \bibinfo{address}{New York, NY, USA}, \bibinfo{pages}{2035--2046}.
\newblock
\showISBNx{978-1-4503-9352-2}
\href{https://doi.org/10.1145/3531146.3533779}{doi:\nolinkurl{10.1145/3531146.3533779}}


\bibitem[Widder et~al\mbox{.}(2023)]%
        {widder_its_2023}
\bibfield{author}{\bibinfo{person}{David~Gray Widder}, \bibinfo{person}{Derrick Zhen}, \bibinfo{person}{Laura Dabbish}, {and} \bibinfo{person}{James Herbsleb}.} \bibinfo{year}{2023}\natexlab{}.
\newblock \showarticletitle{It’s about power: {What} ethical concerns do software engineers have, and what do they (feel they can) do about them?}. In \bibinfo{booktitle}{\emph{Proceedings of the 2023 {ACM} {Conference} on {Fairness}, {Accountability}, and {Transparency}}} \emph{(\bibinfo{series}{{FAccT} '23})}. \bibinfo{publisher}{Association for Computing Machinery}, \bibinfo{address}{New York, NY, USA}, \bibinfo{pages}{467--479}.
\newblock
\showISBNx{9798400701924}
\href{https://doi.org/10.1145/3593013.3594012}{doi:\nolinkurl{10.1145/3593013.3594012}}


\bibitem[Winner(2020)]%
        {winner2020whale}
\bibfield{author}{\bibinfo{person}{Langdon Winner}.} \bibinfo{year}{2020}\natexlab{}.
\newblock \bibinfo{booktitle}{\emph{The whale and the reactor: A search for limits in an age of high technology}}.
\newblock \bibinfo{publisher}{University of Chicago Press}.
\newblock


\bibitem[Woolley and Howard(2016)]%
        {woolley2016automation}
\bibfield{author}{\bibinfo{person}{Samuel~C. Woolley} {and} \bibinfo{person}{Philip~N. Howard}.} \bibinfo{year}{2016}\natexlab{}.
\newblock \showarticletitle{Automation, algorithms, and politics| Political communication, computational propaganda, and autonomous agents—Introduction}.
\newblock \bibinfo{journal}{\emph{International Journal of Communication}}  \bibinfo{volume}{10} (\bibinfo{year}{2016}), \bibinfo{pages}{4882--4890}.
\newblock


\bibitem[X(2024)]%
        {xrules}
\bibfield{author}{\bibinfo{person}{X}.} \bibinfo{year}{2024}\natexlab{}.
\newblock \bibinfo{title}{Abuse and Harassment}.
\newblock \bibinfo{howpublished}{\url{https://help.x.com/en/rules-and-policies/abusive-behavior}}.
\newblock
\newblock
\shownote{[Accessed 03-03-2025]}.


\bibitem[Yang et~al\mbox{.}(2013)]%
        {yang_empirical1_2013}
\bibfield{author}{\bibinfo{person}{Chao Yang}, \bibinfo{person}{Robert Harkreader}, {and} \bibinfo{person}{Guofei Gu}.} \bibinfo{year}{2013}\natexlab{}.
\newblock \showarticletitle{Empirical Evaluation and New Design for Fighting Evolving Twitter Spammers}.
\newblock  \bibinfo{volume}{8}, \bibinfo{number}{8} (\bibinfo{year}{2013}), \bibinfo{pages}{1280--1293}.
\newblock
\showISSN{1556-6021}
\href{https://doi.org/10.1109/TIFS.2013.2267732}{doi:\nolinkurl{10.1109/TIFS.2013.2267732}}
\newblock
\shownote{Conference Name: {IEEE} Transactions on Information Forensics and Security}.


\bibitem[Yang and Menczer(2023)]%
        {yang2023anatomy}
\bibfield{author}{\bibinfo{person}{Kai-Cheng Yang} {and} \bibinfo{person}{Filippo Menczer}.} \bibinfo{year}{2023}\natexlab{}.
\newblock \showarticletitle{Anatomy of an AI-powered malicious social botnet}.
\newblock \bibinfo{journal}{\emph{arXiv preprint arXiv:2307.16336}} (\bibinfo{year}{2023}).
\newblock


\bibitem[Yang et~al\mbox{.}(2019)]%
        {yang2019arming}
\bibfield{author}{\bibinfo{person}{Kai-Cheng Yang}, \bibinfo{person}{Onur Varol}, \bibinfo{person}{Clayton~A Davis}, \bibinfo{person}{Emilio Ferrara}, \bibinfo{person}{Alessandro Flammini}, {and} \bibinfo{person}{Filippo Menczer}.} \bibinfo{year}{2019}\natexlab{}.
\newblock \showarticletitle{Arming the public with artificial intelligence to counter social bots}.
\newblock \bibinfo{journal}{\emph{Human Behavior and Emerging Technologies}} \bibinfo{volume}{1}, \bibinfo{number}{1} (\bibinfo{year}{2019}), \bibinfo{pages}{48--61}.
\newblock


\bibitem[Yang et~al\mbox{.}(2020)]%
        {yang_scalable_2020}
\bibfield{author}{\bibinfo{person}{Kai-Cheng Yang}, \bibinfo{person}{Onur Varol}, \bibinfo{person}{Pik-Mai Hui}, {and} \bibinfo{person}{Filippo Menczer}.} \bibinfo{year}{2020}\natexlab{}.
\newblock \showarticletitle{Scalable and Generalizable Social Bot Detection through Data Selection}.
\newblock  \bibinfo{volume}{34}, \bibinfo{number}{1} (\bibinfo{year}{2020}), \bibinfo{pages}{1096--1103}.
\newblock
\showISSN{2374-3468, 2159-5399}
\href{https://doi.org/10.1609/aaai.v34i01.5460}{doi:\nolinkurl{10.1609/aaai.v34i01.5460}}
\showeprint[arxiv]{1911.09179 [cs]}


\bibitem[Yang et~al\mbox{.}(2020)]%
        {yang2020scalable}
\bibfield{author}{\bibinfo{person}{Kai-Cheng Yang}, \bibinfo{person}{Onur Varol}, \bibinfo{person}{Pik-Mai Hui}, {and} \bibinfo{person}{Filippo Menczer}.} \bibinfo{year}{2020}\natexlab{}.
\newblock \showarticletitle{Scalable and generalizable social bot detection through data selection}. In \bibinfo{booktitle}{\emph{Proceedings of the AAAI Conference on Artificial Intelligence}}, Vol.~\bibinfo{volume}{34}. \bibinfo{pages}{1096--1103}.
\newblock


\bibitem[Zhao and Jiang(2011)]%
        {zhao2011cultural}
\bibfield{author}{\bibinfo{person}{Chen Zhao} {and} \bibinfo{person}{Gonglue Jiang}.} \bibinfo{year}{2011}\natexlab{}.
\newblock \showarticletitle{Cultural differences on visual self-presentation through social networking site profile images}. In \bibinfo{booktitle}{\emph{Proceedings of the SIGCHI Conference on Human Factors in Computing Systems}}. \bibinfo{pages}{1129--1132}.
\newblock


\bibitem[Zhou et~al\mbox{.}(2024)]%
        {zhou2024lgb}
\bibfield{author}{\bibinfo{person}{Ming Zhou}, \bibinfo{person}{Dan Zhang}, \bibinfo{person}{Yuandong Wang}, \bibinfo{person}{Yangli-ao Geng}, \bibinfo{person}{Yuxiao Dong}, {and} \bibinfo{person}{Jie Tang}.} \bibinfo{year}{2024}\natexlab{}.
\newblock \showarticletitle{Lgb: Language model and graph neural network-driven social bot detection}.
\newblock \bibinfo{journal}{\emph{arXiv preprint arXiv:2406.08762}} (\bibinfo{year}{2024}).
\newblock


\end{thebibliography}

\newpage

\appendix
\renewcommand{\thetable}{\thesection.\arabic{table}}

\section{Performance of Bot Detection Algorithms}
\setcounter{table}{0}
\autoref{tab:english_perc_full} shows the difference in performance of bot detection algorithms for English-based users and non-English-based users. The language the users write in were taken from X language detectors.

\begin{table}[h]
    \centering
    \small
    \begin{tabular}{p{2.2cm}p{1.5cm}p{1.8cm}p{1.8cm}p{1.8cm}p{1.8cm}}
    \toprule
        \textbf{Dataset} & \textbf{Total Users} & \textbf{Users that write in English (\%)} & \textbf{Accuracy identifying bots from users that primarily write in English (\%)} &  \textbf{Users that do not write in English (\%)} & \textbf{Accuracy identifying bots from users that do not primarily write in English (\%)} \\ \midrule 
        \textbf{astroturf} \cite{sayyadiharikandeh2020detection} & 3782 & 1408 (37.2) & 98.8 & 2374 (62.8) & 98.6 \\ \hline 
        \textbf{botometer-feedback-2019} \cite{yang2019arming} & 4975 & 2364 (47.5) & 99.2 & 2611 (52.5) & 98.6 \\ \hline 
        \textbf{botwiki-2019} \cite{yang2020scalable} & 1066 & 459 (43.1) & 97.2 & 607 (56.9) & 95.9 \\ \hline 
        \textbf{cresci-rtbust-2019} \cite{mazza2019rtbust} & 3580 & 341 (9.5) & 98.8 & 3239 (90.5) & 94.9 \\ \hline 
        \textbf{cresci-stock-2018} \cite{cresci2018fake} & 58283 & 23902 (41.0) & 98.8 & 34381 (59.0) & 92.9 \\ \hline 
        \textbf{gilani-2017} \cite{diesner2017proceedings} & 20776 & 5607 (27.0) & 97.9 & 15169 (73.0) & 97.4 \\ \hline 
        \textbf{midterm-2018} \cite{yang2020scalable} & 68145 & 38740 (56.8) & 99.6 & 29405 (43.2) & 98.6 \\ \hline 
        \textbf{varol-2018} \cite{varol2017online} & 15719 & 7925 (50.4) & 98.2 & 7794 (49.6) & 97.7 \\ \hline 
        \textbf{verified-2019} \cite{yang2020scalable} & 27737 & 13949 (50.3) & 99.9 & 13788 (49.7) & 99.9 \\ \hline
        \textbf{Average} & ~ & 40.3\% & 98.7 & 59.7\% & 97.2 \\ \bottomrule 
    \end{tabular}
    \caption{Difference in bot detection performance for English and Non-English based tweets.}
    \label{tab:english_perc_full}
\end{table}

\section{Reddit User Discussions that Account is Flagged as Bot}
\setcounter{table}{0}
We searched three subreddits (r/Instagram, r/Facebook, r/Twitter) in May 2024 for the phrase ``I'm a bot'', commonly mentioned in posts about social media users being misclassified as bots. We manually examined the top 20 resulting threads for each subreddit (60 in total) and labeled if the thread was relevant, the consequences that the user faced, and the appeal action that the user took. \autoref{tab:reddit_table} presents the full table of the user responses after they were flagged as bots.

    
\newgeometry{left=1cm, right=1cm}
\begin{center}
\small 
\begin{longtable}{@{}p{1cm}p{2.5cm}p{1.2cm}p{1.2cm}p{2.0cm}p{2.0cm}@{}}
    \hline
    \textbf{ID} & \textbf{Title} & \textbf{Subreddit} & \textbf{Relevant} & \textbf{Consequences} & \textbf{Appeal Actions} \\
    \hline
    \endfirsthead

    \multicolumn{6}{c}%
    {{\bfseries Table \thetable\ (continued)}} \\
    \hline
    \textbf{ID} & \textbf{Title} & \textbf{Subreddit} & \textbf{Relevant} & \textbf{Consequences} & \textbf{Appeal Actions} \\
    \hline
    \endhead

    \hline
    \caption{Reddit user responses to account being flagged as bot} \label{tab:reddit_table} \\
    \multicolumn{6}{r}{{(Continued on next page)}} \\
    \endfoot

    \endlastfoot

\hline
15ii09c & why is ig so sure im a bot  & Instagram & TRUE & Forced password changes, login stress & ~ \\ \hline
13eiplz & instagram thinks im a bot for just engaging with  & Instagram & TRUE & Can't engage and grow nature photography posting hobby & ~ \\ \hline
132an8j & instagram is claiming im a bot and wont let me  & Instagram & TRUE & Can't interact with friends & Won't let them appeal \\ \hline
ccwyo3 & instagram thinks im a bot  & Instagram & TRUE & ~ & ~ \\ \hline
17m59lc & everytime i login have to verify my identity and  & Instagram & TRUE & login annoyance and stress & ~ \\ \hline
owant5 & instagram thinks im a bot  & Instagram & TRUE & ~ & ~ \\ \hline
jhw877 & instagram thinks im a bot any suggestions for  & Instagram & TRUE & Can't share art & No response \\ \hline
18mr8d3 & apparently i liked too many posts and instagram  & Instagram & TRUE & ~ & ~ \\ \hline
ulz57p & proofed im not a bot still deleted  & Instagram & TRUE & Account removed & No response \\ \hline
i5889e & instagram thinks im a bot  & Instagram & TRUE & ~ & No response \\ \hline
hwkif4 & did i screw my art account into seeming like a bot  & Instagram & TRUE & Trying to do certain things to ""prove"" not a bot & ~ \\ \hline
19cvwps & instagram thinks im a robot even after i proved  & Instagram & TRUE & Account disabled & ~ \\ \hline
rxzq5n & please help i un followed 2k people in a week and  & Instagram & TRUE & ~ & ~ \\ \hline
1avba4b & instagrams been on something first my account  & Instagram & TRUE & ~ & ~ \\ \hline
cmyhcq & action blocked when following someone has been  & Instagram & TRUE & Blocking actions like following & No response \\ \hline
1895ekp & facebook thinks im a bot and im not sure how i  & Facebook & TRUE & Banned for 6 days. Can't grow Facebook group & Doesn't know how to appeal \\ \hline
ybnm2r & any ideas does the ai think im a bot  & Facebook & TRUE & ~ & ~ \\ \hline
ur4q70 & facebook thinks im a bot and put me in facebook  & Facebook & TRUE & Messaging restrictions & ~ \\ \hline
ur4q70 & facebook thinks im a bot and put me in facebook  & Facebook & TRUE & Messaging restrictions & ~ \\ \hline
o9lmnw & why did my account get permanently banned as soon  & Facebook & TRUE & losing past games & ~ \\ \hline
rmpz6b & why am i being restricted for an account i just  & Facebook & TRUE & feature restrictions & ~ \\ \hline
n25mj6 & facebook bot saw the worst no way to appeal  & Facebook & TRUE & ~ & No way to appeal \\ \hline
17vgzvf & twitter thinks im a bot  & X & TRUE & action restrictions & Can't appeal \\ \hline
14ov8wb & twitter thinks im a bot and wont let me comment  & X & TRUE & action restrictions & ~ \\ \hline
145zcvi & why does twitter think im a bot  & X & TRUE & login stress & ~ \\ \hline
1bgv5ia & even after verifying every 5 minutes the past few  & X & TRUE & login stress & ~ \\ \hline
iz11p9 & twitter thinks im a bot please help  & X & TRUE & no access to account & ~ \\ \hline
17bqcah & if im only getting likes from those prn bots will  & X & TRUE & just stress in thinking might be a bot & ~ \\ \hline
itthxe & account locked because twitter ai thinks im a bot  & X & TRUE & locked out of account unless give intrusive personal info & ~ \\ \hline
tdcn0q & twitter ai thinks im a bot  & X & TRUE & ~ & ~ \\ \hline
jik6zc & twitter thinks im a bot  & X & TRUE & ~ & ~ \\ \hline
18046aj & i was going about my business on the app and was  & X & TRUE & Difficult captchas to regain account & ~ \\ \hline
vx5qoh & twitter keeps locking my account thinking im a bot  & X & TRUE & locked out of account & ~ \\ \hline
c27kog & help account locked because twitter thinks im a  & X & TRUE & locked out of account & ~ \\ \hline
gj5cm2 & twitter thinks im a bot  & X & TRUE & Can't like & ~ \\ \hline
1bpgp2i & twitter thinks my new account is spambot  & X & TRUE & action restrictions & ~ \\ \hline
8aoqzh & my account was locked indefinitely because it  & X & TRUE & locked out of account & ~ \\ \hline
1137gdj & does anyone know how to prove im not a bot my  & X & TRUE & tweets don't show up for people, locked out of account & ~ \\ \hline
jtn5h4 & i started a new twitter account last week and im  & X & TRUE & Can't grow influence & ~ \\ \hline
14otjh8 & with all thats been going on with the limiting i  & X & TRUE & ~ & No idea how to appeal \\ \hline
1bh5y7v & label put on my account  & X & TRUE & Label about being a harmful account & ~ \\ \hline
8cygk0 & can someone explain me how they got that amount  & X & FALSE & ~ & ~ \\ \hline
51cm0w & im getting hacked by a porn bot i think  & Instagram & FALSE & Not relevant (hacked by porn bot) & ~ \\ \hline
11yj3g9 & im suddenly getting tons of these bot accounts  & Instagram & FALSE & Not relevant (followed by bots) & ~ \\ \hline
1al2lob & my account got banned for the dumbest reason  & Instagram & FALSE & Not relevant (moderation for impersonation) & ~ \\ \hline
1bfg4vf & looks like im not the only one pissed about this  & Instagram & FALSE & Not relevant & ~ \\ \hline
x7oi6t & im so tired of getting these comments from these  & Instagram & FALSE & ~ & ~ \\ \hline
1bezpre & so im a facebook group moderator and i need to  & Facebook & FALSE & ~ & ~ \\ \hline
1alyn1s & facebook password reset issue they think im a bot  & Facebook & FALSE & Login issues & ~ \\ \hline
1bezpo4 & so im a facebook group moderator and i need to  & Facebook & FALSE & ~ & ~ \\ \hline
1aggdev & i jus tried to buy a ps4 but he said it was sold  & Facebook & FALSE & ~ & ~ \\ \hline
fv1xfk & im getting a lot of seeming fake bot messages to  & Facebook & FALSE & ~ & ~ \\ \hline
jyqf5o & i create this account yesterday and this happen  & Facebook & FALSE & Can't create account & ~ \\ \hline
1bmu0nq & whats with the influx of spam bots on some pages  & Facebook & FALSE & ~ & ~ \\ \hline
n0uyvf & throughout the day and mostly during late morning  & Facebook & FALSE & hacked & ~ \\ \hline
1c9hbnl & im deactivating and deleting zucks social media  & Facebook & FALSE & ~ & ~ \\ \hline
1bn1zsy & butthurtbook censoring me for replying normally  & Facebook & FALSE & community standards issues & ~ \\ \hline
up0ws8 & noticed a major increase in the number of spam  & Facebook & FALSE & ~ & ~ \\ \hline
hxcp7h & when i post on facebook in a certain group it  & Facebook & FALSE & community standards issues & ~ \\ \hline
u8lekr & made a facebook account a few weeks ago and im  & Facebook & FALSE & ~ & ~ \\ \hline
\end{longtable}
\end{center}
\restoregeometry

\newgeometry{left=1cm, right=1cm}
\begin{center}
\small
\begin{longtable}{@{}p{1.2cm}p{2cm}p{1.3cm}p{4.5cm}p{1.1cm}@{}}
\hline
\textbf{Platform} & \textbf{Event} & \textbf{Language} & \textbf{Algorithm/Approach} & \textbf{Reference} \\
\hline
\endfirsthead

\multicolumn{5}{c}%
{{\bfseries Table \thetable\ (continued)}} \\
\hline
\textbf{Platform} & \textbf{Event} & \textbf{Language} & \textbf{Algorithm/Approach} & \textbf{Reference} \\
\hline
\endhead

\hline
\caption{Summary of Bot Detection Systems} \label{tab:bot-detection} \\
\multicolumn{5}{r}{{Continued on next page}} \\
\endfoot

\endlastfoot

Twitter, Reddit & US elections & English & Mixture of Experts (MoE) combining classifiers for username, metadata, and posts & \cite{ng_botbuster_2022} \\ \hline
Twitter & NULL & Italian & unsupervised feature extraction and clustering of retweet temporal patterns & \cite{mazza_rtbust_2019} \\ \hline
Twitter & NULL & English & Graph-based classification with 1D CNN using latent feature histograms & \cite{magelinski_graph-hist_2019} \\ \hline
Twitter & NULL & English & Contextual LSTM deep neural network that exploits both content and metadata & \cite{kudugunta_deep_2018} \\ \hline
Twitter & NULL & English, German & Ensemble of Random Forests trained on distinct bot classes & \cite{sayyadiharikandeh_detection_2020} \\ \hline
Twitter & NULL & English & Supervised ML (1,000+ features, e.g., posting frequency, URL ratios) & \cite{varol_online_2017} \\ \hline
Twitter & COVID-19 discussions (2020) 2020 U.S. Elections 2019 NATO exercises & English & Supervised ML with data size sensitivity testing & \cite{ng_stabilizing_2022} \\ \hline
Twitter & NULL & English & logistic regression and rule-based heuristics & \cite{hui_botslayer_2019} \\ \hline
Twitter & NULL & English & Community-aware Mixture-of-Experts (MoE) with multimodal inputs & \cite{liu_botmoe_2023} \\ \hline
Twitter & Includes data from the 2018 U.S. Midterm Elections & English & Random Forest classifier with minimal user metadata features for scalable and generalizable bot detection & \cite{yang_scalable_2020} \\ \hline
Twitter & NULL & English & Bidirectional LSTM with word embeddings for contextual tweet analysis & \cite{wei_twitter_2020} \\ \hline
Twitter & NULL & English & Self-supervised learning with hierarchical RNNs and attention mechanisms for semantic, property, and neighborhood representation & \cite{feng_satar_2021} \\ \hline
Twitter & NULL & English & Relational Graph Convolutional Networks with multi-modal user feature encoding on heterogeneous graphs without feature engineering & \cite{feng_botrgcn_2021} \\ \hline
Twitter & NULL & English & Relational Graph Transformers with Semantic Attention Networks on heterogeneous graphs & \cite{feng_heterogeneity-aware_2021} \\ \hline
Twitter & Honeypot dataset (36,000 content polluters collected over seven months) & English & Random Forest classifier with features from user metadata, friendship networks, content, and historical behavior, trained using social honeypots & \cite{lee_seven_2021} \\ \hline
Twitter & NULL & English & DenStream clustering algorithm for real-time spam tweet detection using user metadata, content, and time-based features & \cite{eshraqi_detecting_2015} \\ \hline
Twitter & NULL & English & Graph Convolutional Neural Networks (GCNN) with inductive representation learning leveraging user profile features and social network structure & \cite{ali_alhosseini_detect_2019} \\ \hline
Twitter & NULL & English & Random Forest on engineered features, uses SHAP for interpretability, and clusters misclassified accounts with UMAP and DBSCAN & \cite{lopez-joya_exploring_2024} \\ \hline
Twitter & NULL & English & Utilizes graph neural networks with a community-aware contrastive learning module to mine hard positive/negative samples & \cite{chen_cacl_2024} \\ \hline
Twitter & NULL & English & Proposes a GAN-based approach with multiple discriminators and data augmentation to overcome mode collapse and improve detection accuracy & \cite{shukla_social_2023} \\ \hline
Twitter & NULL & English & Feature-based supervised learning system that extracts highly predictive behavioral and meta-data features to distinguish bots from humans & \cite{ferrara_rise_2016} \\ \hline
Twitter & NULL & English & Graph-based detection relying on the assumption that sybil accounts have few connections to legitimate users (identifies densely interconnected groups of sybils) & \cite{viswanath_analysis_2010} \\ \hline
General & NULL & English & Detects group attacks by spotting lockstep behavior in social networks & \cite{beutel_copycatch_2013} \\ \hline
Twitter/ General & NULL & English & Identifies coordinated (synchronized) behavior among accounts to flag groups of bots & \cite{cao_uncovering_2014} \\ \hline
Renren & NULL & Chinese & Clickstream modeling using k-gram based sequence similarity metrics with graph partitioning for unsupervised Sybil detection & \cite{wang_you_nodate} \\ \hline
Facebook, Renren & NULL & Chinese + English & Crowdsourced Sybil detection leveraging human judgment with multi-tiered filtering, vote aggregation, and \revision{mechanical} turker selection for scalable OSN profile verification & \cite{wang2012social} \\ \hline
Twitter & NULL & Multiple (all languages) & Warped Correlation with Dynamic Time Warping (DTW) and Lag-sensitive Hashing & \cite{chavoshi_debot_2016} \\ \hline
Twitter & NULL & Multiple (all languages) & Temporal Pattern Mining, including Motif Discovery, Discord Discovery, Subsequence Join, Bursts, and Dynamic Clusters & \cite{chavoshi_temporal_2017} \\ \hline
Twitter & NULL & English & Supervised classification using 20+ account features (e.g., links, hashtags, follower ratios) to detect spam bots & \cite{benevenuto_detecting_nodate} \\ \hline
Twitter & NULL & English & Random Forest-based ensemble leveraging corrected entropy measures, Bayesian spam detection, and account property features for classifying Twitter accounts as human, bot, or cyborg & \cite{chu_detecting_2012} \\ \hline
Facebook & NULL & English & Coordinated programmable socialbots that exploit social engineering and triadic closure principles to infiltrate OSNs and harvest user data & \cite{boshmaf_design_2013} \\ \hline
Twitter & NULL & English & Robust Twitter spam detection using newly designed graph-, neighbor-, automation-, and timing-based features integrated into a supervised machine learning framework & \cite{yang_empirical1_2013} \\ \hline
Twitter & NULL & English & Sentiment-aware ensemble classifier combining tweet sentiment, linguistic cues, and contextual network features for distinguishing bots from humans on Twitter & \cite{dickerson_using_2014} \\ \hline
Twitter & 2015 influence campaign & English & Robust multi-model ensemble integrating content-based and network-based classifiers, as showcased in the DARPA Twitter Bot Challenge & \cite{subrahmanian2016darpa} \\ \hline
Twitter & NULL & English & A boosting-based bot detection framework that fuses topic modeling with heuristic features to optimize the F1 score by striking a careful balance between precision and recall & \cite{morstatter_new_2016} \\ \hline
Twitter & NULL & English & A deep learning framework combining CNN-LSTM based temporal text analysis with network embedding of user behaviors for scalable and robust bot detection in social media & \cite{cai_behavior_2017} \\ \hline
Twitter & NULL & English & Preprocessing framework integrating multi-stage feature engineering---including metrics like hashtag similarity, session duration, and tweeting entropy---with rigorous feature selection and labeled training set creation for scalable Twitter bot detection & \cite{kantepe_preprocessing_2017} \\ \hline
Twitter & NULL & English & Supervised ML pipeline fusing diverse profile and tweet features into a binary feature matrix, preprocessed by logistic regression and classified with SVM & \cite{efthimion_supervised_2018} \\ \hline
Twitter & NULL & English & Tiered supervised machine learning with event-based annotation that fuses features from tweet text, account metadata, timeline dynamics, and network interactions to train scalable classifiers for Twitter bot detection & \cite{beskow_bot-hunter_nodate} \\ \hline
Twitter & NULL & English & Random string detection leveraging character n-gram TF-IDF features, normalized case and numeric counts, and Shannon entropy, with logistic regression to filter 15-character random screen names for Twitter bot labeling & \cite{beskow_its_2019} \\ \hline
Twitter & NULL & English, Multilingual & AdaBoost with SMOTE-ENN on a compact, language-agnostic feature set extracted from account metadata, tweet behavior, and temporal patterns for high-accuracy Twitter bot detection & \cite{knauth_language-agnostic_2019} \\ \hline
Twitter & FIFA World Cup 2014 & English / Portuguese & Leverages a simplified CART decision tree trained on normalized textual features from topic-aligned Twitter messages, providing both bot detection and transparent, interpretable decision paths for user guidance & \cite{ferreira_dos_santos_uncovering_2019} \\ \hline
Twitter & NULL & English & Leverages a reduced set of five Twitter profile counters---statuses, followers, friends, favorites, and listed counts---to train supervised ML models (RF, SVM, NB, and ocSVM) for efficient bot detection with performance on par with state-of-the-art approaches & \cite{fonseca_abreu_twitter_2020} \\ \hline
Twitter & NULL & English & Leverages BERT contextualized tweet embeddings augmented with emoji2vec and categorical features (retweets, URLs, mentions, hashtags) to train shallow classifiers for effective bot detection & \cite{dukic_are_2020} \\ \hline
Twitter & NULL & English & Random Forest ensemble trained on 36 tweet-based features extracted via the Twitter API, combined with a LIME explainability module and integrated crowdsourcing feedback to deliver interpretable bot detection scores & \cite{kouvela_bot-detective_2020} \\ \hline
Twitter & NULL & English & Bipartite Graph Self-supervised Representation Distillation combining graph neural networks and semantic analysis & \cite{guo_social_2021} \\ \hline
Twitter & NULL & English & Utilizes classical and structural graph embedding techniques to automatically extract unsupervised features from Twitter's social network, which are then employed in node classification to detect bot accounts & \cite{dehghan_detecting_2023} \\ \hline
Twitter & NULL & English & Deep neural network combining profile text (bio) with structured features to detect suspicious account patterns & \cite{hayawi_deeprobot_2022} \\ \hline
Instagram & NULL & English & Random Forest classifier using profile, activity, and content features to classify accounts as real, fake, or bot & \cite{lopez-joya_exploring_2024} \\ \hline
\end{longtable}

\restoregeometry
\end{center}

\end{document}